%% file: main.tex
\documentclass[sigconf]{acmart}

\AtBeginDocument{%
  \providecommand\BibTeX{{%
    \normalfont B\kern-0.5em{\scshape i\kern-0.25em b}\kern-0.8em\TeX}}}

\emergencystretch=\maxdimen
\copyrightyear{2025}
\acmYear{2025}
\setcopyright{acmlicensed}\acmConference[DIS '25]{Designing Interactive Systems Conference}{July 5--9, 2025}{Funchal, Portugal}
\acmBooktitle{Designing Interactive Systems Conference (DIS '25), July 5--9, 2025, Funchal, Portugal}
\acmDOI{10.1145/3715336.3735810}
\acmISBN{979-8-4007-1485-6/2025/07}

\author{Ananya Bhattacharjee}
\affiliation{%
  \institution{Computer Science, University of Toronto}
  \city{Toronto}
  \state{Ontario}
  \country{Canada}
}

\author{Sarah Yi Xu}
\affiliation{%
  \institution{Computer Science, University of Toronto}
  \city{Toronto}
  \state{Ontario}
  \country{Canada}
}
\author{Pranav Rao}
\affiliation{%
  \institution{Computer Science, University of Toronto}
  \city{Toronto}
  \state{Ontario}
  \country{Canada}
}
\author{Yuchen Zeng}
\affiliation{%
  \institution{Computer Science, University of Toronto}
  \city{Toronto}
  \state{Ontario}
  \country{Canada}
}

\author{Jonah Meyerhoff}
\affiliation{%
  \institution{Preventive Medicine, Northwestern University}
  \city{Chicago}
  \state{Illinois}
  \country{USA}
}
\author{Syed Ishtiaque Ahmed}
\affiliation{%
  \institution{Computer Science, University of Toronto}
  \city{Toronto}
  \state{Ontario}
  \country{Canada}
}

\author{David C Mohr}
\affiliation{%
  \institution{Preventive Medicine, Northwestern University}
  \city{Chicago}
  \state{Illinois}
  \country{USA}
}

\author{Michael Liut}
\affiliation{%
  \institution{Mathematical and Computational Sciences, University of Toronto Mississauga}
  \city{Mississauga}
  \state{Ontario}
  \country{Canada}
}
\author{Alex Mariakakis}
\affiliation{%
  \institution{Computer Science, University of Toronto}
  \city{Toronto}
  \state{Ontario}
  \country{Canada}
}
\author{Rachel Kornfield}
\affiliation{%
  \institution{Preventive Medicine, Northwestern University}
  \city{Chicago}
  \state{Illinois}
  \country{USA}
}

\author{Joseph Jay Williams}
\affiliation{%
  \institution{Computer Science, University of Toronto}
  \city{Toronto}
  \state{Ontario}
  \country{Canada}
}

\newcommand{\revision}[1]{\textcolor{black}{#1}}

\usepackage{hyperref}
\usepackage{booktabs}
\usepackage{caption}
\usepackage{subcaption}
\usepackage{gensymb}
\usepackage{xspace}
\usepackage{tabularx}
\usepackage{footmisc}
\usepackage{enumitem}
\usepackage{mdframed}
\usepackage{bbding}
\usepackage{longtable}
\usepackage{graphicx}
\usepackage{tabu}
\newcommand{\promptcomment}[1]{\vspace{0.0cm}\begin{mdframed}[backgroundcolor=gray!20]#1\end{mdframed}\vspace{0.4cm}}

\usepackage{mdframed}
\usepackage[most]{tcolorbox}
\usepackage{pdflscape} 
\usepackage{adjustbox}
\usepackage{graphicx}
\usepackage{booktabs} 
\usepackage{tabularx} 
\usepackage{geometry} 
\usepackage{multirow}
\usepackage{colortbl}
\usepackage{hhline}
\usepackage{cuted}
\newcommand{\italquote}[1]{\begin{quote}``\textit{#1}''\end{quote}}
\newcommand{\DIS}[1]{\textcolor{black}{#1}}
\newenvironment{DISenv}
  {\color{black}} 
  {} 

\begin{document}

\title[User Perceptions of Personalized LLM-Enhanced Narrative Interventions]{\textit{Perfectly to a Tee}: Understanding User Perceptions of Personalized LLM-Enhanced Narrative Interventions}


\renewcommand{\shortauthors}{Ananya Bhattacharjee et al.}

\begin{abstract}
\input{Section/abstract}

\end{abstract}

\ccsdesc[500]{Human-centered computing~Empirical studies in HCI}

\keywords{Story, Narrative, Digital Mental Health, LLM, Intervention}
\begin{teaserfigure}
    \centering
    \includegraphics[width=0.8\linewidth]{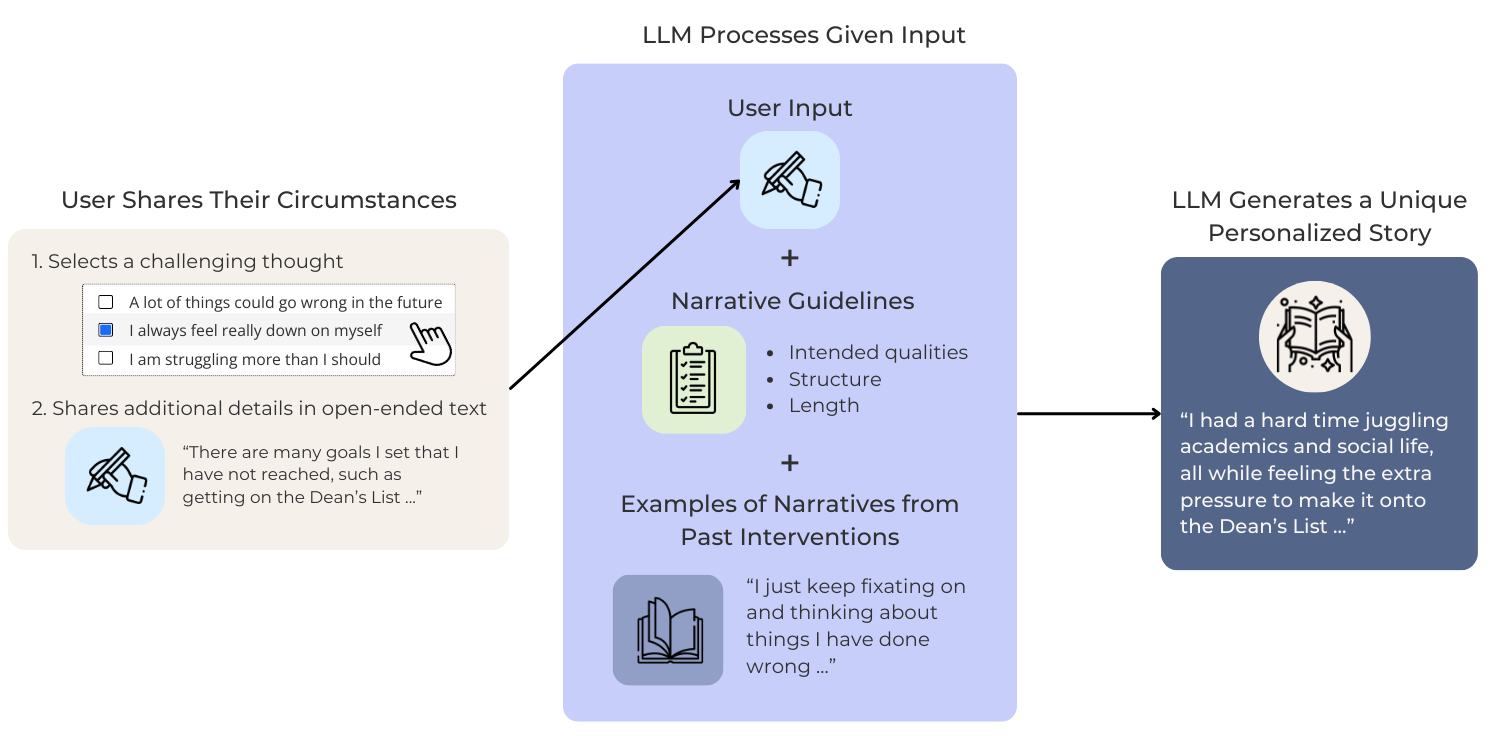}
    \caption{Process of Generating Personalized LLM-Enhanced Narratives: User inputs their circumstances through structured and open-ended responses, which are combined with narrative guidelines and examples from past interventions, allowing the LLM to generate unique, personalized stories for each user.}
    \label{fig:diagram}
    \Description{Process of Generating Personalized LLM-Enhanced Narratives: User inputs their circumstances through structured and open-ended responses, which are combined with narrative guidelines and examples from past interventions, allowing the LLM to generate unique, personalized stories for each user.}
\end{teaserfigure}


\maketitle

\input{Section/intro}
\input{Section/related-work}
\input{Section/design}
\input{Section/results}
\input{Section/discussion}

\input{Section/conclusion}

\bibliographystyle{ACM-Reference-Format}
\bibliography{main}
\clearpage
\input{Section/appendix}

\end{document}

%% file: Section/abstract.tex
Stories about overcoming personal struggles can effectively illustrate the application of psychological theories in real life, yet they may fail to resonate with individuals' experiences. In this work, we employ large language models (LLMs) to create tailored narratives that acknowledge and address unique challenging thoughts and situations faced by individuals. Our study, involving 346 young adults across two settings, demonstrates that personalized LLM-enhanced stories were perceived to be better than human-written ones in conveying key takeaways, promoting reflection, and reducing belief in negative thoughts. These stories were not only seen as more relatable but also similarly authentic to human-written ones, highlighting the potential of LLMs in helping young adults manage their struggles. The findings of this work provide crucial design considerations for future narrative-based digital mental health interventions, such as the need to maintain relatability without veering into implausibility and refining the wording and tone of AI-enhanced content.

%% file: Section/intro.tex
\section{Introduction}



\noindent
Stories about individuals overcoming personal struggles can serve as a powerful medium for promoting wellbeing \cite{bhattacharjee2022kind, anderson2015digital, llewellyn2019characteristics, kornfield2022involving}. When individuals engage with narratives that reflect their own experiences, they can draw parallels between their lives and those of the characters, creating opportunities for self-reflection \cite{ferguson2000teaching}. For instance, a narrative about someone overcoming disappointment from job search rejection could provide valuable insights to others facing similar setbacks. This reflection process helps readers identify patterns in their thoughts or behaviors, leading to practical insights that empower them to make positive changes \cite{bhattacharjee2023design, bhattacharjee2022kind}. Furthermore, engaging with such stories can help validate and normalize mental health challenges \cite{kornfield2022meeting}. Given these substantial benefits, stories are increasingly being integrated into digital mental health (DMH) interventions \cite{meyerhoff2024small, bhattacharjee2022kind, baumel2018digital}. 



\DIS{However, many DMH interventions face significant challenges in delivering engaging and applicable content \cite{hornstein2023personalization, bhattacharjee2023investigating}.  A key issue is the lack of tailoring to users' unique struggles \cite{matthews2024personalisation}, which can make the content feel generic and disconnected from their immediate needs.} Such perceptions not only increase the likelihood that the content will be overlooked but also risk dismissing users' genuine struggles \cite{bhattacharjee2023investigating}. Consequently, seemingly irrelevant narratives may undermine the potential effectiveness of a DMH intervention \cite{bhattacharjee2022kind}. \DIS{Although some interventions have attempted to address this issue by using pre-defined rules for basic content adjustments, these strategies are inherently limited in the complexity of inputs they can handle and the range of personalization that can be applied during live interactions.}


\DIS{
Large language models (LLMs) offer promise in addressing these challenges by leveraging their ability to process vast amounts of data and adapt to diverse user inputs.} They are rapidly gaining recognition for their capabilities as personalized intervention tools and supportive scaffolds in a variety of settings \cite{bhattacharjee2024understanding, kim2023mindfuldiary, abd2023large}. 
Given their flexibility, LLMs can interpret users' descriptions of their situations in order to create stories specifically tailored to their circumstances. We posit that such capability of dynamic personalization -- adapting content in real-time based on user inputs \cite{mahmood2014dynamic} -- will significantly enhance the relevance and impact of digital narrative interventions. 
Revisiting the previous example of job search rejection, a user might gain more benefits from a narrative tailored to reflect the competitive nature of their field or environment.
By connecting the story to specific circumstances and concerns, a narrative might provide deeper insights and more appropriate strategies, effectively addressing the user's comprehensive experience.

\DIS{Although LLMs show potential for generating personalized stories, they are not inherently trained to adhere to established guidelines or structures \cite{bhattacharjee2024understanding}. Given the sensitivity of mental health challenges, implementing an LLM-driven intervention requires careful measures to prevent generating harmful or insensitive content \cite{demszky2023using}. To address these risks, LLMs can be preloaded with narratives and mental health content from prior DMH interventions, constraining them to generate content within those boundaries. This approach would allow personalization to adapt the story's context to match an individual's circumstances while ensuring the core takeaways and guidance remain consistent with validated methods.}






\DIS{In this work, we developed an intervention designed to dynamically personalize stories to support individuals experiencing challenging thoughts. The intervention was designed to ensure that each participant received a story uniquely tailored to their specific circumstances. As shown in \autoref{fig:diagram}, the intervention achieves dynamic personalization by incorporating user input into the prompt, where participants describe their challenging thoughts and provide open-ended descriptions of the difficulties they are facing. The LLM is also prompted to follow narrative intervention guidelines to ensure the generated story aligns with the intended format and qualities. Our goal was to understand how well this personalization approach can adhere to the intended qualities of the narratives while also assessing their impact as an intervention on specific outcomes, such as reducing belief in challenging thoughts.}


\DIS{Our research focused on young adults aged 18--25, a demographic particularly susceptible to mental health issues due to the inherent transitional nature of this life stage, encompassing significant changes in education, career, and personal relationships \cite{kornfield2022meeting, park2018social}. Statistically, this age group exhibits a much higher prevalence of mental health concerns compared to other adult groups \cite{nimh2020mentalhealth}. These factors make them a relevant group for our research, where we aim to design stories that can help manage common challenges they face.}

Our work was guided by the following research questions:

\begin{itemize} 

\item \textbf{RQ1:} How do young adults perceive the key narrative qualities of personalized LLM-enhanced stories, such as authenticity and the ability to communicate takeaways, compared to human-written stories from past DMH interventions?

\item \textbf{RQ2:} How effective are personalized LLM-enhanced stories as interventions for managing negative thoughts, such as reducing belief in negative thoughts and inspiring solutions, compared to human-written stories from past DMH interventions? 
\end{itemize}

\noindent
To address these research questions, we conducted a randomized experiment with two groups of young adult participants: crowdworkers ($N=174$) and students ($N=172$). The intervention was implemented as a single-session intervention (SSI) \cite{schleider2017little}, where participants received and responded to a single story. Participants were randomly assigned to engage with either a human-written story from past interventions (Non-LLM story)  \cite{meyerhoff2024small} or a personalized LLM-enhanced story (LLM story). \DIS{We evaluated the impact that LLM stories had on participants' beliefs in their negative thoughts and associated emotions relative to Non-LLM stories. We also gathered feedback on participants' perceptions of the narratives, focusing on attributes such as the authenticity of the stories and the clarity with which key takeaways were communicated.}  Our findings revealed that LLM stories were perceived as comparably authentic to Non-LLM stories, yet they performed better on average in communicating key takeaways, promoting reflection, reducing belief in negative thoughts, and inspiring actionable solutions. Our results suggest the promise of LLMs to create custom narratives supporting personal wellbeing, and inspire design considerations for personalized LLM-enhanced narrative interventions, such as the need for a careful balance between relatability and avoiding implausibility and the importance of refining the tone and wording in AI-generated content.

\begin{DISenv}    
Our contributions include:  
\begin{itemize}  
    \item Design of an intervention for delivering personalized LLM-enhanced narratives that align with users' challenging thoughts and circumstances.  
    \item A randomized experiment conducted in two distinct settings, comparing personalized LLM-enhanced narratives with narratives from previous DMH interventions.  
    \item Design implications to guide future DMH interventions in utilizing LLMs for personalizing narrative-based support.  
\end{itemize}  
\end{DISenv}





%% file: Section/related-work.tex
\section{Related Work}

For the review of related work, we begin by discussing how story-based interventions can promote psychological wellbeing. Following this, we talk about the need for dynamic personalization in DMH interventions, focusing on the challenges and potential of LLMs in this context.

\subsection{Promoting Psychological Wellbeing Through Stories}


Storytelling has long been a traditional method for imparting morals and lessons, as evidenced through various mediums such as fables, allegories, parables and poems~\cite{sarlej2012representing,gibbs2002aesop}. Among the most famous examples of moral storytelling are Aesop's Fables~\cite{aesopfable}, famous for concluding with succinct moral takeaways like ``look before you leap'' or ``honesty is the best policy.'' Recent research underscores the role of storytelling in supporting psychological wellbeing, highlighting how narratives can make challenging experiences more relatable and offer paths to recovery~\cite{nurser2018personal, llewellyn2020not, llewellyn2019characteristics, anderson2015digital}. Through these stories, individuals can find inspiration to mirror the story characters' resilience and recovery strategies in their own lives~\cite{shaw2015effect,nurser2018personal, williams2018recovery, lipsey2020evaluation}. Furthermore, storytelling can encourage self-reflection, helping individuals see parallels between themselves and the characters, which can lead to a deeper understanding of their mental state and encourage constructive changes in thought patterns and behaviors~\cite{dym2019coming, payne2006narrative, hinyard2007using, bhattacharjee2022kind}. This reflective practice is also crucial in various therapeutic settings where sharing and discussing personal stories about one's journey through adversity helps individuals learn from each other~\cite{payne2006narrative, lipsey2020evaluation}. 

Storytelling serves as one method of fostering ``diffuse sociality,'' \cite{burgess2019think} a low-effort way for individuals to connect with others and share experiences without direct or face-to-face interaction. This form of connection can significantly reduce feelings of isolation by revealing that others face similar struggles \cite{burgess2019think, o2017design}. Such a sense of connectedness can be especially beneficial in DMH interventions, where users may prefer to engage independently. Research shows that individuals can feel a strong connection to the characters of a story, seeing their own challenges reflected in the narratives, and finding reassurance in shared human experiences \cite{bhattacharjee2022kind, akram2020exploratory, lipsey2020evaluation}. 

There is growing evidence supporting the benefits of narrative approaches in promoting psychological wellbeing and behavioral change \cite{bhattacharjee2022kind, saksono2021storymap, grimes2008eatwell, park2021wrote, collins2022covid, davis2020understand}.  Research has shown that storytelling interventions can reduce symptoms of depression across various populations \cite{vromans2011narrative, rodriguez2014mindfulness}; for instance, \citet{rodriguez2014mindfulness} have utilized mindfulness-based narrative therapy as a therapeutic intervention for treating depression in cancer patients. Behavioral change interventions have leveraged stories to subtly guide audiences toward specific viewpoints or actions by immersing them in relatable narratives \cite{bilandzic2013narrative}. Some interventions embed educational content within an entertainment framework containing narratives, engaging audiences while delivering important messages \cite{singhal2012entertainment}. Stories have been used successfully to encourage physical activity \cite{saksono2021storymap}, promote healthy eating habits \cite{grimes2008eatwell}, and counter stigma against mental health conditions~\cite{deconstructing, evans2014effect, walkinourshoes} and sexual harassment victims \cite{dimond2013hollaback}. This broad applicability underscores the power of stories to inspire behavioral change and foster psychological wellbeing in diverse contexts.

\DIS{Narrative interventions can be delivered in multiple formats. In our work, we utilize them as SSIs \cite{schleider2020acceptability, schleider2017little} — structured, standalone sessions designed to provide focused support that has immediate effects on specific outcomes. While sustained engagement over a longitudinal intervention can have enhanced benefits, we chose this format for our exploratory study since our research questions center around the impact of LLM integration. The intervention was designed to prompt participants to reflect on their challenging thoughts by reading a story about someone experiencing a similar struggle and making progress, encouraging them to reevaluate their beliefs. Similar to prior research that has evaluated the short-term benefits of SSIs \cite{sharma2023cognitive, bhattacharjee2024exploring}, we evaluate the utility of our intervention by examining the immediate impact it has on participants' belief in their challenging thoughts and the associated emotional intensity.}

\subsection{Dynamic Personalization in DMH Interventions}
Because mental health challenges are deeply personal and varied, static or pre-generated DMH content often fails to resonate with users \cite{bhattacharjee2024exploring}. \DIS{The ability to dynamically personalize content in real time based on users' specific input and evolving needs has therefore become a highly sought-after feature, as users often prefer intervention content that reflects their specific challenges \cite{jahedi2024personalization, garrigos2003modelling}.  In the context of DMH, real-time personalization can enhance user engagement, increase intervention effectiveness, and improve overall user satisfaction \cite{bhattacharjee2023investigating, baumel2019objective}. Attaining such benefits not only requires an intervention to be able to process and respond to user input, but also to understand how to change the original content without deviating too far from its original materials. As described below, implementing these affordances remains a significant challenge.}


\subsubsection{Challenges of Dynamic Personalization:} Early DMH interventions, such as rule-based chatbots, relied on pre-scripted responses and decision trees for support \cite{weizenbaum1966eliza, abd2019overview}. While these systems have been effective at generating structured conversations, they have been inherently limited in their capacity for dynamic content generation. Since responses were predefined, the systems struggled to adapt to the unique and evolving contexts of individual users' life experiences \cite{abd2019overview, bhattacharjee2023investigating}. As a result, the interactions often felt static and generic, lacking the personalization necessary to address users' specific needs in real time.

Some prior studies \cite{ma2023contextbot, morris2014crowd, smith2021effective, saksono2023evaluating, morris2015efficacy} explored supporting user wellbeing through dynamic content generated via crowdsourcing. For instance, Panoply \cite{morris2014crowd, morris2015efficacy}, a crowd-powered system, allowed users to submit descriptions of their stressors. These descriptions were then sent to crowdworkers who crafted supportive messages in response. Similarly, Flip*Doubt was a crowd-powered web application that provided users with cognitive reappraisals of negative thoughts \cite{smith2021effective}.  Additionally, many online peer support groups have been established with a similar motivation \cite{gatos2021hci, baumel2018digital, wang2023metrics, collins2022covid}. These platforms aim to foster wellbeing by connecting individuals facing similar challenges, enabling them to share stories, learn from each other, and engage in discussions that promote reflection. However, these platforms heavily depend on the availability and responsiveness of participants, which is not always guaranteed. Moreover, the varying levels of comfort among crowdworkers and peers in sharing their experiences could limit the effectiveness of the support provided. Furthermore, crowdworkers and peers may also require training to ensure they deliver appropriate and empathetic support \cite{chen2021scaffolding}.

The advent of machine learning and deep learning brought some advancements in dynamic personalization for DMH interventions. These models began to harness personalized data, such as user history, mood patterns, and behavioral cues, to generate responses that were more attuned to each user \cite{torous2018mental, adler2021identifying, fitzpatrick2017delivering}. Numerous conversational chatbots were developed during this era, using AI algorithms to choose from pre-set branches \cite{fitzpatrick2017delivering, meyerhoff2024small, aguilera2020mhealth}, but their capacity to process open-ended input from users and adapt accordingly was still very limited. Natural language processing (NLP) techniques, like information retrieval or word embedding \cite{carpenter2016seeing}, played important roles in improving how systems interpreted free-form user input, enabling more relevant and context-sensitive outputs \cite{le2021machine}. For instance, \citet{morris2018towards} have employed these techniques to repurpose responses from existing online peer support databases, presenting the content that most align with their specific concerns. However, despite these advancements, the ability to dynamically generate new content still remained limited. These systems excelled at pattern recognition but fell short in the creativity needed to produce truly customized content without large-scale training datasets. As a result, most systems continued to rely on selecting responses from a predefined content bank rather than creating entirely new content \cite{abd2019overview}.

\subsubsection{Potential of LLMs in Dynamic Personalization for DMH Interventions}
Recent advancements in the development of LLMs like GPT and Gemini represent a leap forward for dynamic personalization. These models have the potential to dynamically generate coherent and rich text, offering a significant improvement over earlier systems that relied on selecting from predefined content \cite{bhattacharjee2024understanding}. These capabilities can extend to various domains, particularly in DMH interventions, where understanding nuanced user needs is crucial. LLMs have the potential to tailor interventions to an individual's unique experience that might involve a challenge they face or a recent experience. This is particularly facilitated by the LLMs' ability to process open-ended input, allowing users to express and respond to their situations in rich detail and in any form they choose.

For example, \citet{sharma2023cognitive} developed an LLM-based cognitive reframing tool that dynamically adapts its responses based on the specific challenges faced by users.  Similarly, CLOVA CareCall, a conversational AI, collects data on individuals’ general health and engages as a conversational partner to alleviate loneliness by generating dynamic, human-like questions and responses \cite{jo2023understanding, jo2024understanding}. Additionally, \citet{hedderich2024piece} introduced an LLM-based chatbot designed to assist teachers in providing support to adolescents dealing with cyberbullying, with responses that recognize and adapt to the context of each situation. Studies such as the one by \citet{kim2024diarymate} have utilized LLMs to offer contextual writing support in journal writing. In their work, Kim et al.'s system provided real-time assistance as users expanded their journal entries. Collectively, these applications underscore the promise of LLMs in enabling dynamic personalization that not only acknowledges but also responds appropriately to the nuanced needs of users.

However, adapting LLMs for dynamic personalization in DMH tools is not a straightforward task. Research has shown that LLMs can produce inaccurate or clinically invalid suggestions within the wellbeing context \cite{ma2024evaluating, demszky2023using}. Therefore, interventions must ensure that content is not only accurate but also delivered effectively to support wellbeing. Additionally, it is crucial to determine which intervention elements can be personalized and which must remain standardized. \citet{bhattacharjee2024understanding} emphasize that LLMs should operate within specific evidence-based psychological strategies when providing support in managing emotions. In this domain, LLMs require careful direction regarding the situational context of the user and should adhere strictly to predetermined structures and content guidelines. 

\DIS{In the context of our study, we designed our intervention to maintain desirable qualities such as authenticity and the ability to communicate key takeaways.
We achieve this by not only providing guidance in the prompt, but also by feeding stories from past DMH interventions into the LLMs. We analyzed the quality of personalization by examining how the personalized stories preserved these desirable narrative qualities.}

%% file: Section/design.tex
\section{Study Design}
\label{sec: study_design}

We describe the design of our study and the various steps we employed to execute it below.

\subsection{Procedure}

Participants were recruited for a brief activity, lasting between 10 and 20 minutes, aimed at aiding the development of technology-based narrative interventions for managing negative thoughts and emotions. \revision{The study targeted members of the general population in order to explore diverse experiences with common negative thoughts and emotions.} During recruitment, it was communicated that their participation would contribute to research in this area. The study activity, which included dynamic story selection and generation through the use of LLMs, was conducted on the Qualtrics platform.

After providing affirmative informed consent, participants were asked to select a common challenging thought that they frequently encounter (we label this question as \textbf{Q-Challenging-Thought}). This selection was made from six options identified from previous research on mental health challenges in young adults \cite{meyerhoff2024small}. The options included feelings of being down on oneself, worrying about the future, feelings of struggling, feelings of underachievement compared to others, disappointments in reliance on others, and social anxiety.  Participants then evaluated their belief in, and the emotional intensity of, the selected thought on a scale from 1 (`very low') to 7 (`very high'), inspired by questions from a past study \cite{sharma2023cognitive}. They were also requested to provide a detailed description of their thought, circumstances, and emotions to the extent they felt comfortable (we label this question as \textbf{Q-Challenging-Thought-Description}).

Following this initial sequence of questions, participants were randomly assigned to engage with either a story sourced from prior studies (Non-LLM condition) or a personalized LLM-enhanced version of the story (LLM condition). In both groups, stories addressed the same challenging thought endorsed by the participant. \revision{Participants were blinded to their assigned condition in that the story was broadly described as illustrating how someone might navigate a situation similar to the one the participant had mentioned.} Each story concluded with a reflection question, prompting participants to reflect on their behavior and potential strategies to overcome their challenges. The number of words written by participants in this reflection was captured as an indicator of their engagement with the intervention \cite{wang2023metrics}. We detail the design of stories in both conditions in Section~\ref{sec: design_llm_nonllm}.

Subsequently, participants rated four qualities of the stories — such as authenticity and the effectiveness of communicating key takeaways — on a scale from 1 (`very low') to 7 (`very high'). They also provided qualitative feedback on their ratings and additional quantitative evaluations concerning related qualities like perceived relatability and engagement. Qualitative comments on both helpful and unhelpful aspects of the stories were also collected. These questions were also inspired from prior literature \cite{bhattacharjee2022kind, bhattacharjee2024exploring, sharma2023cognitive}. Following best practices in survey design, we also incorporated attention-check questions in the middle of the activity to ensure participant attentiveness and the quality of the responses \cite{hauser2016attentive}.
\input{Table/question}

\autoref{tab: quant_q} highlights the quantitative questions, and we label them for future reference. All questions were asked on a 1–7 scale, where 1 represents the lowest or least favorable response (e.g., ``strongly disagree'' or ``not at all''), and 7 represents the highest or most favorable response (e.g., ``strongly agree'' or ``extremely''). We also highlight the sources of these questions, along with the corresponding research question.


For our quantitative comparison, we hypothesized that participants would perceive the personalized LLM-enhanced stories more favorably across all qualities except for authenticity, compared to those in the Non-LLM condition. This hypothesis was based on the LLMs' potential to improve DMH content across various settings. However, we expected lower scores for authenticity in the LLM condition because these stories, while based on human-written content, were adapted by LLMs, unlike the stories in the Non-LLM condition.

\revision{Although this was not a clinical intervention, the study was designed and carried out in close collaboration with licensed clinical psychologists who were part of the research team throughout the project. They contributed to all major aspects of the intervention, including the creation of stories in the Non-LLM condition (Section~\ref{subsec: non-llm}), the design of prompts for the LLM condition (Section~\ref{subsec: llm}), and the implementation of ethical considerations (Section~\ref{subsec: ethics}).}

\subsection{Participants}

We conducted two separate experiments to compare stories in the LLM and the Non-LLM conditions, recruiting \revision{crowdworkers from the Prolific platform} and students from a large computer science course at a major North American University. All participants were required to be between 18 and 25 years old. For the crowdworkers, an additional requirement was residency in North America. Since our study aimed to help the general population manage common challenging thoughts, we did not impose any criteria related to wellbeing or mental health symptoms.

A total of 180 crowdworkers and 179 students initially participated in our study. However, \revision{data from 6 crowdworkers and 7 students were discarded due to quality issues identified through instructional manipulation checks~\cite{oppenheimer2009instructional}}, resulting in a final sample of 174 crowdworkers and 172 students.

Participants were randomly assigned to either the LLM or the Non-LLM condition. After the assignments, we had 87 crowdworkers in both the Non-LLM and LLM conditions, and 90 students in the Non-LLM condition and 82 in the LLM condition. We refer to them using a combination of the first letters of their group (Crowdworker = C, Student = S) and their condition (Non-LLM = N, LLM = L), followed by a numeric ID indicating their participant number within that condition. For example, CL82 refers to a crowdworker in the LLM condition, while SN56 refers to a student in the Non-LLM condition. Crowdworkers were compensated at a rate of \$10 USD per hour for their participation, while student participants received a 1\% credit towards their course grade. \autoref{tab: demography} details participants' demographic information, including age, gender, and race.

\input{Table/demography}

\subsection{Design of the Intervention in the Study}
\label{sec: design_llm_nonllm}

\DIS{In this section, we describe the design of our intervention in both the Non-LLM and LLM conditions. In both cases, the intervention was implemented as an SSI—a structured, standalone session designed to provide focused support with immediate effects on specific outcomes. SSIs can be highly effective in delivering rapid relief and promoting reflection or reappraisal within a short timeframe (typically 10-20 minutes), making them accessible and adaptable to diverse populations \cite{schleider2017little, schleider2020acceptability}.  They can also serve as an entry point to broader mental health care by helping individuals develop insights or skills that may encourage further engagement with mental health resources \cite{bhattacharjee2024exploring}. Research has demonstrated their success in managing negative thoughts, reducing stress, and alleviating symptoms of anxiety and depression in both clinical and non-clinical settings \cite{sharma2023cognitive, schleider2017little, schleider2020acceptability, bhattacharjee2024exploring}.}


\DIS{Building on prior SSIs for managing negative thoughts \cite{sharma2023cognitive, bhattacharjee2024exploring} and leveraging the potential of narratives to assist individuals in addressing negative thoughts and situational challenges \cite{bhattacharjee2022kind, meyerhoff2024small}, we developed an SSI centered on using stories to support participants in managing common negative thoughts.} Stories in the Non-LLM condition, along with an example LLM-enhanced version for each, are included in Appendix \ref{sec: examples}. It is important to note that, even when using the same negative thoughts and descriptions as inputs, the LLM may generate outputs that differ from the provided examples. We provide a detailed explanation of the design of stories below.

\subsubsection{Design of the Stories in the Non-LLM Condition}
\label{subsec: non-llm}
In the Non-LLM condition, our stories were sourced from \citet{meyerhoff2024small}, who themselves based their stories on the experiences reported by young adult participants according to \citet{kornfield2022meeting}. \revision{To generate stories, a separate group of crowdworkers distinct from those who participated in the main study} were given a list of common challenges to write about, or they were allowed to supply their own. They were then asked how they took a step that helped them better manage the challenge or feel better about it, even if it was a small or incremental step. They were provided with examples of evidence-based psychological strategies they may have used in addressing their challenge, or they were allowed to describe their own approach. Crowdworkers were asked to write true stories in the first person and to include as much detail as they could about their challenge and how they had addressed it, all while omitting identifying information. After clinical psychology and human-computer interaction researchers edited and curated the stories for clarity and to align with the intended psychological strategies, \citet{meyerhoff2024small} eventually deployed these stories as a part of a text messaging program designed to help young adults manage mental health and wellbeing. 

Each story corresponded to one of the six challenging thoughts outlined in Q-Challenging-Thought, illustrating how a young adult navigates and overcomes such challenges. These stories were designed to promote evidence-based psychological strategies for managing negative thoughts, such as practicing self-compassion \cite{allen2010self} or engaging in cognitive restructuring~\cite{clark2013cognitive}. 
The stories featured four essential qualities as outlined by \citet{bhattacharjee2022kind}: authenticity, key takeaways, active participation (i.e., promoting reflection), and a balance between positivity and realistic struggles.  We describe below how these stories were crafted to embody each desired quality:

\begin{itemize}
\item \textbf{Authenticity:} Crafted from a first-person perspective, the stories aimed to endow the narrative character with a distinct voice, facilitating a deeper connection between the character and the reader \cite{delgado1990story}. By incorporating detailed descriptions of the characters and their circumstances, the stories also created vivid and relatable imagery for the audience \cite{rosenthal1971specificity}.  
For instance, instead of generalizing about high school graduates, they mentioned specifics such as a high school graduate feeling overshadowed by peers who have seemingly achieved more or dealing with intense self-criticism while working from home during the pandemic.

\item \textbf{Key Takeaways:} Drawing from the format of Aesop’s Fables \cite{aesopfable}, each story concluded with a clear message for the reader, linking a psychological principle to the narrative’s characters and events. For example, stories included takeaways like ``Being deliberate about the things we tell ourselves can make a difference in how we feel.'' or ``Self-compassion is all about giving the same kindness to ourselves we give to others.''.

\item \textbf{Promoting Reflection:} The stories fostered interactive engagement by including a reflective question at the end of each story, such as ``Have you ever faced a similar challenge?''. This technique was designed to prompt users to evaluate their own experiences in light of the story's message, facilitating personal reflection. This method is akin to the dialogic inquiry used in storytelling, where the storyteller periodically pauses to invite listeners to relate the narrative to their own lives \cite{wells2000dialogic, slovak2016scaffolding}.

\item \textbf{Balance Between Positivity and Realistic Struggles:} Each story ended on a positive note, depicting an individual making progress or gaining a new perspective on their challenges and illustrating the value of change. The conclusions were deliberately realistic; rather than depicting characters who have made enormous progress or solved their problems entirely, the stories emphasized that small, incremental steps can lead to meaningful improvements in one's situation~\cite{bhattacharjee2022kind}.

\end{itemize}

\subsubsection{Design of the Stories in the LLM Condition}
\label{subsec: llm}

\begin{DISenv}
Our implementation of personalized LLM-enhanced narratives was informed by recent research across various contexts, which demonstrated the benefits of adapting content to align with user contexts \cite{bhattacharjee2024understanding, stade2024large}.  For our project, we employed a prompt engineering method. The prompt design was an iterative effort involving faculty and graduate students with expertise in human-computer interaction, LLMs, psychology, cognitive science, and DMH. The prompt design underwent multiple rounds of revision, applying the prompts to a variety of stories from past interventions and refining them iteratively. Following recommended practices of prompt engineering \cite{stade2024large, ekin2023prompt}, we developed a structured prompt using OpenAI's GPT-4 that aimed to achieve the following objectives:

\begin{itemize}
    \item \textbf{Dynamic Personalization:} We aimed to generate stories during the interaction that reflect participants' unique situations and challenges.  This was achieved by incorporating two responses from participants: the \textbf{selected option from Q-Challenging-Thought} and the \textbf{response to the open-ended question Q-Challenging-Thought-Description}.
    \item \textbf{Adherence to Guidelines:} We aimed to ensure that the generated narratives align with the intended qualities in prior DMH interventions.  This also involved integrating prior relevant examples by including the \textbf{corresponding story to the selected option from Q-Selected-Thought}. We included instructions to avoid the introduction of new suggestions or strategies not present in the examples. The stories were designed to maintain a similar length to those in the Non-LLM condition ($\leq$250 words) and concluded with a reflective question.
\end{itemize}

\end{DISenv}



The complete prompt is as follows:

\vspace{2mm}
\promptcomment{Given the user's thought, [[selected choice from Q-Challenging-Thought]],  and their provided description, [[response to the open-ended question Q-Challenging-Thought-Description]], your task is to share a narrative. This narrative should recount the journey of an individual who encountered challenges similar to user's and the steps they took towards making progress, incorporating relevant examples and strategies for improvement. Here are some guidelines to follow:

\noindent
Authenticity: The experience shared should feel genuine and relatable, mirroring real-life emotions and scenarios. The story should be told from a first-person perspective.

\noindent  
Key Takeaways: Clearly articulate lessons or insights that the user can apply to similar situations in their life.

\noindent  
Reflective Question: At the end of the narrative, there should be a reflective question along the lines of ``Have you had a situation where you had to deal with a similar challenge?” or "What strategies might work during such challenging times?''. 
		
\noindent
Balancing Positivity with Realistic Struggles: While maintaining a positive outlook, acknowledge the challenges and struggles inherent to the situation.
		
\noindent
Use the following example as a basis for your narrative. It is important that you do not introduce new suggestions or strategies beyond those mentioned in the example.

\noindent		
[[Corresponding story to the selected choice from Q-Selected-Thought]]

\noindent  
Please generate a first-person narrative that incorporates these guidelines and examples, making it as engaging and helpful as possible for the user to gain new perspectives and coping mechanisms. The text should start with ``Here's someone's experience that might resonate with you: [[newline]]'' and should not contain more than 250 words. As shown in the example, conclude with a clear separator, ``******,'' followed by a prompt for a reflective question that starts with ``Please answer this reflective question:''.}

\subsection{Data Analysis}
We determined the statistical significance of differences between the groups for each measure using one-tailed Welch's t-tests. 
P-values were adjusted using the Benjamini-Hochberg (BH) procedure for multiple comparisons. The null hypothesis (H0) posited that the mean for the hypothesized superior group is less than or equal to the mean for the other condition. Conversely, the alternative hypothesis (Ha) contended that the mean for the hypothesized superior group is greater than that of the baseline condition.


After the data were anonymized and cleaned, we utilized the thematic analysis approach outlined by \citet{clarke2017thematic} to examine the qualitative data. First, two coders from the research team reviewed the data to familiarize themselves with the data. Using an open-coding method \cite{khandkar2009open}, each coder independently developed an initial codebook for the data of 25 participants. \revision{This process reflected an iterative approach that integrated both deductive and inductive techniques. The analysis of RQ1 was guided by prior work on narrative qualities~\cite{bhattacharjee2022kind} and followed a deductive coding approach, whereas RQ2 was approached inductively, allowing themes to emerge directly from the data.}
We then implemented a consensus coding method similar to that outlined by \citet{clarke2017thematic}. During this phase, the coders held several joint sessions to review and integrate their separate codebooks. This consensus coding process, through repeated discussions and refinement, was designed to improve reliability and minimize personal biases. The coders determined common codes, refined the definitions, and removed codes that did not directly relate to the research questions. The updated codebook was then applied iteratively to the data from an additional 25 participants, making adjustments as new findings emerged. After finalizing the codebook, each coder independently applied the established codes to half of the remaining data.

\revision{Our analysis generated 138 codes such as ``new perspective'',  ``matching personal struggles'', ``giving hope'', ``detailed character portrayal'', ``robotic'', ``informative'', and ``clear takeaways''. Coders used axial coding \cite{williams2019art} in subsequent meetings to organize these codes into eight major themes (four per RQ).} Detailed findings structured by these themes are presented in Section~\ref{sec: results}.




\subsection{Ethical Considerations}
\label{subsec: ethics}

Our research activity was approved by the Research Ethics Board at the first author's university. Given the nature of our research, which involved aspects of reflection on negative thoughts and challenging life situations, we were mindful of the ethical considerations at play throughout the study, and we took proactive measures to address potential concerns. All participants provided affirmative informed consent prior to beginning any study procedures. They were informed about the possibility that participation might cause anxiety, sadness, or other forms of distress. Participants were also requested to not include any identifiable information in their responses to ensure confidentiality and privacy. 


Moreover, while we did not specifically ask participants about their experiences with suicide-related thoughts and behaviors, we recognized that suicidal thoughts, behaviors, or other risk-related issues may arise due to the open-ended nature of the questions and assessments as well as the probabalistic nature of LLM responses.
We provided all participants with contact information for crisis services, such as suicide prevention helplines, to ensure they had access to support, if needed. Additionally, \revision{the research team reviewed all stories and participant responses daily during the study period to monitor for any indications of suicide-related risk (e.g., suicidal ideation, self-harm). Team members were trained to assess participants using the Columbia-Suicide Risk Assessment protocol \cite{posner2008columbia} if any potential suicide or self-harm-related responses were identified. However, no such risks emerged during the course of the study, and we did not need to conduct any follow-up assessments.}

%% file: Table/question.tex
\begin{table*}[ht]
\centering
\captionsetup{justification=centering, singlelinecheck=true} 
\caption{Quantitative questions from the survey} 
	\label{tab: quant_q}
    \Description{Quantitative survey questions used to evaluate the stories. Includes pre- and post-story measures such as belief in negative thought and emotion intensity, as well as post-story assessments of narrative qualities such as authenticity, takeaways, reflection, relatability,  and future engagement. Each item is linked to a corresponding research question and literature source.}
	\begin{tabular}{|>{\raggedright\arraybackslash}p{0.3\textwidth}|
		>{\raggedright\arraybackslash}p{0.41\textwidth}|p{2cm}|c|} \hline

		\rowcolor{lightgray!50}\multicolumn{4}{|l|}{\textbf{Questions asked before and after reading the story}}                                                                                                                                                                \\ \hline

		\rowcolor{lightgray!50}\textbf{Question Label}                   & \textbf{Actual Question}                                                                                    & \textbf{Corresponding RQ} & \textbf{Source}                                            \\ \hline
		\multirow{1}{*}{\textbf{Belief in Negative Thought}}                             & How strongly do you believe in your thought?                                                                & \multirow{1}{*}{RQ2}                      & \multirow{1}{*}{\cite{sharma2023cognitive} }               \\ \hline
		\multirow{1}{*}{\textbf{Emotion Intensity}     }                                  & How strong is the emotion associated with this thought?                                                     & \multirow{1}{*}{RQ2}                       & \multirow{1}{*}{\cite{sharma2023cognitive} }                                \\ \hline
		\multicolumn{4}{c}{}                                                                                                                                                                                                                                                    \\ \hline

		\rowcolor{lightgray!50}\rowcolor{lightgray!50}\multicolumn{4}{|l|}{\textbf{Questions asked after reading the story}}                                                                                                                                                    \\ \hline
		\rowcolor{lightgray!50}\textbf{Question Label}                   & \textbf{Actual Question}                                                                                    & \textbf{Corresponding RQ} & \textbf{Source}                                            \\ \hline
		\multirow{1}{*}{\textbf{Perceived Authenticity}}                                 & How would you rate the authenticity of the story in portraying characters, settings, and emotions?          & \multirow{1}{*}{RQ1}                       & \multirow{1}{*}{\cite{bhattacharjee2022kind}}                               \\ \hline
		\textbf{Perceived Ability to Communicate Key Takeaways}          & How effectively do you think the story communicated its key messages or lessons?                            & \multirow{1}{*}{RQ1}                       & \multirow{1}{*}{\cite{bhattacharjee2022kind}}                               \\ \hline
        \multirow{1}{*}{\hspace{-5pt}\begin{tabular}{p{0.3\textwidth}}
		\textbf{Perceived Ability to Promote Reflection} 
		\end{tabular}}                & How much did the story encourage you to reflect on your own thoughts, experiences, or beliefs?              & \multirow{1}{*}{RQ1}                       & \multirow{1}{*}{\cite{bhattacharjee2022kind}}                               \\ \hline
		\multirow{1}{*}{\hspace{-5pt}\begin{tabular}{p{0.3\textwidth}}
		    \textbf{Perceived Balance of Positivity with Realistic Struggle}
		\end{tabular}} & How well did the story balance positive outcomes with realistic portrayals of challenges?                   & \multirow{1}{*}{RQ1}                       & \multirow{1}{*}{\cite{bhattacharjee2022kind}}                               \\ \hline
		\multirow{1}{*}{\textbf{Perceived Relatability}}                                 & Agreement with the statement: `I could relate to the experiences shared in the story.'                      & \multirow{1}{*}{RQ2}                       & \multirow{1}{*}{\cite{meyerhoff2024small}}           \\ \hline
  \multirow{1}{*}{\hspace{-5pt}\begin{tabular}{p{0.3\textwidth}}
		    \textbf{Perceived Capacity of Inspiring Solutions}
		\end{tabular}} & Agreement with the statement: `The story helped me think of potential ways I can manage negative thoughts.' & \multirow{1}{*}{RQ2}      & \multirow{1}{*}{\cite{bhattacharjee2024exploring}} \\ \hline
		\textbf{Likelihood of Future Engagement}                         & Agreement with the statement: `I would love to read similar stories in future.'                             & \multirow{1}{*}{RQ2}           & \multirow{1}{*}{\cite{bhattacharjee2024exploring}}                          \\ \hline
	\end{tabular}
\end{table*}

%% file: Table/demography.tex
\begin{table*}[ht]
	\centering
	\captionsetup{justification=centering, singlelinecheck=true} 
	\caption{{Demographics of the crowdworkers and students}} 
	\label{tab: demography}
    \Description{Demographic characteristics of participants, including mean age, gender distribution, and racial identity for both crowdworkers and students. Highlights demographic differences between the two populations.}
	\begin{tabular}{|>{\raggedright\arraybackslash}p{0.3\textwidth}|
		>{\centering\arraybackslash}p{0.3\textwidth}|
		>{\centering\arraybackslash}p{0.3\textwidth}|}
		\hline
		                                  & \textbf{Crowdworkers (N = 174)} & \textbf{Students (N = 172)} \\
		\hline
		\textbf{Age, mean ± std err}                      & 22.5 ± 0.2                & 20.9 ± 0.1            \\ \hhline{|=|=|=|}
		\multicolumn{3}{|l|}{\textbf{Gender, N (\%)}} \\ \hline
		Women                            & 101 (58.0\%)                  & 25 (14.5\%)               \\ \hline
		Men                              & 63 (36.2\%)                   & 142 (82.6\%)              \\ \hline
		Non-binary/third gender                        & 7 (4.0\%)                     & 0 (0\%)                   \\ \hline
		Self-described                      & 1 (0.6\%)                     & 2 (1.2\%)                 \\ \hline
		Not reported                      & 2 (1.1\%)                     & 3 (1.7\%)                 \\ 
        \hhline{|=|=|=|}
		\multicolumn{3}{|l|}{\textbf{Race, N (\%)}} \\ \hline
		White                             & 68 (39.1\%)                   & 19 (11.0\%)               \\ \hline
		More than one race                & 13 (7.5\%)                    & 6 (3.5\%)                 \\ \hline
		Black or African American         & 13 (7.5\%)                    & 6 (3.5\%)                 \\ \hline
		Asian                             & 65 (37.4\%)                   & 126 (73.3\%)              \\ \hline
		American Indian or Alaskan Native & 2 (1.1\%)                     & 0 (0\%)                   \\ \hline
		Not reported                      & 13 (7.5\%)                    & 15 (8.7\%)                \\ \hline
	\end{tabular}
\end{table*}

%% file: Section/results.tex
\section{Results}
\label{sec: results}

We structure our findings around the research questions we originally set. First, we examine how personalized LLM-enhanced stories were perceived in terms of exhibiting desirable narrative qualities. Subsequently, we explore their effectiveness as interventions, specifically looking at their impact on managing negative thoughts. We primarily explore user perceptions of stories in the LLM condition, occasionally drawing comparisons to the Non-LLM condition to highlight differences and similarities.

\subsection{Personalized LLM-Enhanced Stories in Exhibiting Key Narrative Qualities}

\autoref{tab: stat1} presents the summary statistics of the ratings given by participants regarding the narrative qualities of the stories. Additional details about ratings across different challenging thoughts are provided in Appendix \ref{sec: app_evaluation}. We observed that the stories from the LLM condition were perceived to be superior in all the narrative qualities other than authenticity. The differences between groups in authenticity ratings were not significant. The remaining differences were statistically significant among the crowdworkers, whereas among the students, only the difference in the perceived ability to promote reflection was significant. Below, we support these observations using insights drawn from participants' comments. 

\input{Table/results1}

\subsubsection{Perceived Authenticity}


\revision{Across both populations, perceived authenticity ratings were similar across conditions. Among crowdworkers, the LLM stories ($Mean_{\text{LLM}}$ = 4.92~$\pm$~0.17) received slightly lower ratings than the Non-LLM stories ($Mean_{\text{Non-LLM}}$ = 5.03~$\pm$~0.17), with no significant difference ($p$~=~0.31). Among students, ratings were nearly identical for both conditions ($Mean_{\text{LLM}}$ = 4.29~$\pm$~0.19, $Mean_{\text{Non-LLM}}$ = 4.34~$\pm$~0.18, $p$~=~0.42).}
Participants in both LLM and Non-LLM conditions were drawn to the authenticity of their stories because of their relatability and the sincerity depicted in the characters' journeys. These narratives were appreciated for their candid portrayal of often unspoken feelings, aligning closely with personal and broadly shared human experiences. 
However, for some participants in the Non-LLM condition, the very universality that made the stories relatable also rendered them suspect in terms of authenticity. They perceived these universal themes as too generic, suggesting a lack of specificity that could make the stories feel impersonal.

Participants in the LLM condition provided various insights into the factors contributing to perceived authenticity in the LLM stories. SL473 remarked on the relatability and credibility of their story, stating:

\italquote{I think the setting and thoughts portrayed in the story felt very natural and authentic as it represents a common feeling and scenario that many people face. The fact that the ``journey still wasn't smooth all the time'' helped to make the story feel more realistic.}


\noindent
Other participants resonated with their stories as reflective of their own potential future experiences or as echoing a shared journey with others facing similar challenges. The familiarity of the scenarios and the realistic solutions proposed were key elements that the participants found compelling.

The quality of the writing also played a crucial role in the authenticity of the LLM stories. Descriptive and vivid prose helped participants clearly visualize the scenarios, fostering a deeper connection with the content. CL73 shared, \textit{``I could picture the scenario pretty well in my mind and resonate with it.''} However, several participants criticized the tone and writing style of their stories as being too mechanical and impersonal. Doubts about AI involvement were raised due to perceived shortcomings in the depth of the protagonists and their emotions. In fact, some participants commented that their stories shared patterns similar to the typical responses made by ChatGPT. CL41 noted:

\italquote{Anyone could have written it, and the sentence structure is very even and AI-like. I have used a lot of AI text generators in the past so it gives me the same feeling. \ldots [[The story was]] generic and removed from their own feelings.}

Furthermore, some participants felt that the LLM stories were overly tailored to their specific situations, making the AI's involvement too apparent and diminishing the authenticity of the narratives. This precise mirroring of their inputs led to a perception that the stories were merely processing and regurgitating information, which reduced their impact and made them feel less genuine.

\subsubsection{Perceived Ability to Communicate Key Takeaways}

\revision{LLM stories were rated significantly higher than Non-LLM stories in terms of communicating key takeaways among crowdworkers ($Mean_{\text{LLM}}$ = 6.01~$\pm$~0.10 vs. $Mean_{\text{Non-LLM}}$ = 5.51~$\pm$~0.14, $p$~<~0.01). Among students, the LLM stories also received higher ratings ($Mean_{\text{LLM}}$ = 5.48~$\pm$~0.17 vs. $Mean_{\text{Non-LLM}}$ = 5.27~$\pm$~0.16), though the difference was not statistically significant ($p$~=~0.21).}
Participants in both conditions valued the structured progression within their stories, which clearly delineated the flow from beginning to end. However, some participants in the Non-LLM condition commented that the takeaways were not very evident as they did not relate to their stories. For instance, SN63 acknowledged understanding the message but felt a disconnect from their personal experiences. 

Participants in the LLM condition felt that their stories were crafted to closely mirror their current life situations, which made the messages more poignant and relevant. This led CL34 to say, \textit{"It explains what I am currently going through perfectly to a tee."} Customization helped participants see their stories not just as general advice, but as directly applicable guidance. 
The stories' structure played a crucial role in effectively communicating key takeaways as well. By charting a clear progression from past to future, the stories in the LLM condition provided contextual grounding that guided the reader through the speaker’s process of discovering solutions. CL19 explained:

\italquote{It broke down the story clearly. Explained why this person felt this way and what circumstances were around it. Clarified the feelings they were having and the choices they made to better themselves and finished with how this struggle makes them a better person now.}

Criticisms of the LLM stories in this regard were minimal, mainly focusing on the tone and presentation of the content. The feedback echoed previous sentiments, highlighting that mechanical and impersonal tones detracted from the authenticity of the narratives. Additionally, there was some concern from the students that advice that seemed to be generated by AI might not be taken as seriously by readers, potentially undermining the impact and credibility of the guidance provided.

\subsubsection{Perceived Ability to Promote Reflection}

\revision{LLM stories were rated significantly higher in promoting reflection among both groups. For crowdworkers, the LLM condition received a mean rating of $Mean_{\text{LLM}}$ = 5.31~$\pm$~0.17, compared to $Mean_{\text{Non-LLM}}$ = 4.71~$\pm$~0.20 in the Non-LLM condition ($p$~=~0.02). Among students, the LLM stories also outperformed Non-LLM stories ($Mean_{\text{LLM}}$ = 4.55~$\pm$~0.20 vs. $Mean_{\text{Non-LLM}}$ = 3.96~$\pm$~0.20, $p$~=~0.04).}
Participants in the Non-LLM condition valued how their stories depicted common experiences that reinforced actions and introduced fresh perspectives, but they were sometimes criticized for being too generic. This generality made their stories seem less reliable as catalysts for deep reflection. The lack of specific, detailed scenarios in these stories sometimes led participants to perceive the stories as insufficiently engaging or thought-provoking. 

Participants in the LLM condition were more enthusiastic about the successes that their stories had in promoting reflection. They particularly noted the influence of the recommended behaviors within the stories. This perceived applicability allowed participants to see the advice and strategies presented as both relevant and practical, directly connecting their stories to their own life situations. 

Reflection was triggered in various ways. For participants like CL1, the stories validated the idea that \textit{``everyone has different paths and progressions in their lives''}, helping them appreciate and accept the diversity in personal journeys and perspectives. Additionally, these stories reinforced beliefs or approaches that participants were already exploring or implementing to manage negative thoughts. For some others, the situations depicted were relatable, but the recommended behaviors provided fresh perspectives and potential methods for participants to consider. CL38 expressed:
\italquote{To be fair, the story did give me a step back to look at my situation from a different perspective. It made me think `Maybe I am a little too harsh on myself.'}

While relatability was often beneficial, it sometimes impeded participants' ability to reflect. In certain cases, participants had encountered similar scenarios repeatedly in the past, so the familiarity stripped away the novelty that might have prompted more profound contemplation about their own circumstances. CL65 mentioned that a slightly different situation in the story might have offered a fresh perspective on the problem. These findings highlight how overly familiar situations can diminish the potential for deeper reflection.



\subsubsection{Perceived Balance of Positivity with Realistic Struggles}

\revision{Crowdworkers rated LLM stories significantly higher for balancing positivity with realistic struggles ($Mean_{\text{LLM}}$ = 5.38~$\pm$~0.16) compared to Non-LLM stories ($Mean_{\text{Non-LLM}}$ = 4.80~$\pm$~0.16, $p$~<~0.01). Among students, the LLM condition also received higher ratings ($Mean_{\text{LLM}}$ = 4.62~$\pm$~0.17 vs. $Mean_{\text{Non-LLM}}$ = 4.29~$\pm$~0.18), though the difference was not statistically significant ($p$~=~0.15).}
The structured progression of events in the stories was viewed favorably in both conditions, as it helped maintain a balance between positivity and realistic portrayals of struggles, anchoring the narratives and ensuring they resonated with the audience to some extent. However, participants in the LLM condition were particularly appreciative of the balance achieved between positivity and realistic struggles
They particularly highlighted their stories' empathetic acknowledgment of the protagonist's struggles, which mirrored the challenges faced by the readers themselves. This recognition of shared difficulties established a foundation of empathy, making participants feel understood and more connected to the narrative. Additionally, their stories were commended for illustrating the small, incremental steps protagonists took to address and improve their challenges, enhancing their realism and applicability.

Participants appreciated that their stories did not promise unattainable perfection but instead portrayed a more truthful journey toward improvement. This realistic depiction resonated well with participants like SL86, who stated:

\italquote{I felt like it was an incredibly realistic portrayal. At the end of the story, the storyteller even mentions that not everything went perfect, but they were able to face challenges more effectively. That outcome is realistic and something I hope to obtain in the future.}

\noindent
However, participants noted that both the Non-LLM and LLM stories often adopted a more positive tone than they would have preferred. This sentiment was especially pronounced when participants perceived their situations as beyond their control, leading to a disconnect with the occasional overly optimistic narrative. CL35 expressed this concern:

\italquote{It's not that easy to just focus on what you have in control and the rest will follow in reality. When you have a lot of things on your plate it gets stressful to manage.}

Participants suggested that in some difficult situations, particularly where people may feel they lack control over their situation, portraying progress might be seen as an oversimplification. They recommended acknowledging the challenges more explicitly in such situations and potentially ending the stories in a more open-ended manner, rather than always concluding with progress. 

\subsection{Personalized LLM-Enhanced Stories as an Intervention to Manage Negative Thoughts}

\autoref{tab: stat2} presents the summary statistics of participant ratings on various intervention outcomes. We observed that the LLM stories were perceived to be superior in all the qualities. Additional details about ratings across different challenging thoughts are provided in Appendix \ref{sec: app_evaluation}. Save for the ratings of perceived emotional intensity, the differences in ratings were statistically significant among the crowdworkers. Among the students, only the differences in perceived relatability, perceived capacity of inspiring solutions, and length of written reflection were statistically significant. Below, we supplement these findings with participant comments that highlight both the beneficial aspects and the areas for improvement of the stories.

\input{Table/results2}


\subsubsection{Reevaluating Existing Thoughts and Beliefs}

\revision{Participants who read LLM stories reported significantly greater perceived reduction in belief in their negative thought among crowdworkers ($Mean_{\text{LLM}}$ = 0.64~$\pm$~0.13) compared to those in the Non-LLM condition ($Mean_{\text{Non-LLM}}$ = 0.26~$\pm$~0.10, $p$~=~0.02). Among students, the difference followed the same direction ($Mean_{\text{LLM}}$ = 0.65~$\pm$~0.12 vs. $Mean_{\text{Non-LLM}}$ = 0.49~$\pm$~0.12), though it was not statistically significant ($p$~=~0.21).}

Participants in the Non-LLM condition found that their stories allowed them to organize and reevaluate their thoughts and emotions in a productive manner. Participants recognized the benefits of the reflective writing activity associated with these stories, demonstrating the potential of narratives in facilitating the process of reflection among participants.

Participants in the LLM condition expressed similar sentiments, noting that they were able to reevaluate their negative thought patterns from a more neutral point of view. For instance, CL11 said the following about reading a story related to a person still being in school while their friends had already completed their studies:

\italquote{Now I understand that everybody has their own timing and even if it takes me more time to find out what I want, at least I'll be happy and sure of what I want to do.}

\noindent
The reevaluation process also involved a reflective writing activity associated with the reflection question. 
As shown in \autoref{tab: stat2}, both crowdworkers and students wrote more words on average in response to the LLM stories. 
This activity aimed to provide participants with an opportunity to organize their thoughts and emotions into written form, which may have helped them restructure their thoughts to be more helpful. SL28 expressed one such change, saying:

\italquote{Despite all the negative thoughts I had about myself, I should be proud of the things that I have already achieved which was by completing university and getting that bachelor's degree. \ldots everyone should be enjoying the journey and not just the destination.}

These findings suggest that both Non-LLM and LLM stories could support participants in reevaluating their thoughts and beliefs, offering them new insights and perspectives.

\subsubsection{Potential of Stories in Inspiring Solutions and Actions}


\revision{LLM stories received higher ratings for their ability to inspire potential solutions among both crowdworkers ($Mean_{\text{LLM}}$ = 5.46~$\pm$~0.16) and students ($Mean_{\text{Non-LLM}}$ = 4.76~$\pm$~0.18) compared to the Non-LLM condition ($Mean_{\text{LLM}}$ = 4.74~$\pm$~0.18, $Mean_{\text{Non-LLM}}$ = 4.14~$\pm$~0.19), with both differences reaching statistical significance ($p$~<~0.01, $p$~=~0.04, respectively).}

Participants in the Non-LLM condition generally observed a lack of pragmatic solutions and actions in their stories compared to those in the LLM condition. This subdued response was frequently attributed by some participants to the relatively more generic nature of the pre-scripted stories, which they felt did not adequately reflect the applicability of the narratives to their own personal experiences.

Participants in the LLM condition felt that their stories were more effective at inspiring solutions. They attributed this success to the stories' relatability and the fact that they ended on a positive note, offering a sense of hope and direction. As CL37 explained:

\italquote{Just having a story that reflects mine with a happy ending is something that is important to me and helps provide guidance.}

\noindent
Participants reported learning new techniques through the storytelling process, which helped them reframe their challenges or cope with their emotions more effectively. For instance, CL51 shared how they discovered specific methods for managing anxiety in social situations, particularly when interacting with people with whom they were not comfortable. SL81 echoed the benefits of reading about tangible actions, saying:

\italquote{The reminder that working towards my goals is something that I can do to increase my self-worth was good. It's easy to get lost in those negative emotions, and having something tangible to do about it is better than just hearing that I shouldn't think those things or compare myself to others.}

Participants expressed that this reflection process involved more than just passive reading. The LLM stories helped participants to reflect on their personal struggles through the lens of the story and to actively engage with the suggested solutions. By presenting coping mechanisms embedded within the context of the narrative, participants felt the stories allowed them to see how similar strategies might apply in their own lives. The stories not only illustrated relatable challenges but also modeled potential solutions in a way that felt accessible and actionable. 
SL64 commented:

\italquote{Plainly writing out the advice key message [[by the story]] was helpful since, by relating to the situation, one would feel like they're being helped by someone in real life as well.}


In summary, participants in both groups endorsed the encouragement to actively engage in reflection through writing. However, LLM-enhanced stories were perceived as better at inspiring solutions by depicting progress and providing hope.

\subsubsection{Dual Aspects of Relatability}
Participants in the Non-LLM condition occasionally reported that their stories were not relatable enough to foster a sense of connection with the story characters. 
Participants in the LLM condition, on the other hand, noted that their stories fostered a sense of connection with the story characters, as participants recognized that others were experiencing similar challenges. This sense of shared experience provided reassurance, helping participants feel less isolated in their struggles. For example, SL30 expressed:

\italquote{I like knowing that it's a shared experience and you're not alone in this. \ldots it's reassuring to know that other people struggle with it too.}

However, participants like CL3, CL22, and SN66 mentioned that their stories seemed overly tailored to their specific responses, which created a sense of artificiality. They felt that using the exact examples they had provided actually hampered the impact of the stories since it felt improbable that someone else would have experienced the exact same situation. These participants suggested that, in attempting to customize the stories to fit their context, the narratives may have activated suspicion about the story generation process.
These observations suggest a nuanced balance is necessary: while insufficient relatability can leave participants feeling disconnected, overly personalized narratives may seem improbable or forced.

\subsubsection{Negative Perception of AI Contributing to Reduced Impact}
Participants in the LLM condition also suggested that the potential involvement of AI in the story-generation process might activate some negative associations that could undermine their ability to benefit. Participants like CL86, SL23, and SL57 expressed that they were uncomfortable with the idea of AI-based systems providing support for mental wellbeing and expressed a preference for direct human interaction. Participants even pointed out potential risks regardless of the relatability of the AI-generated story. SL57, for example, highlighted this concern:

\italquote{I don't think an AI-generated response is going to be helpful in any way. It actually just makes me more depressed thinking about how advice and therapy, a very human task, are being delegated to the 1s and 0s.}


Interestingly, some participants in the Non-LLM condition mistakenly believed that AI had been used to generate the stories, and they expressed similar concerns to those in the LLM group. This suggests a broader skepticism among participants about the role of AI in content production.

%% file: Table/results1.tex
\begin{table*}[t]
\centering
\caption{Summary statistics and measures of the stories' narrative qualities with and without LLM enhancement. The results are separated across the two populations that were studied: crowdworkers and university students. 
}
\label{tab: stat1}
\Description{Quantitative comparisons of perceived narrative qualities between LLM and Non-LLM stories, including authenticity, key takeaways, reflection, and balance of positivity with realistic struggles. LLM stories were rated significantly higher on several dimensions, particularly by crowdworkers.}
\resizebox{\textwidth}{!}{%
\begin{tabular}{|>{\columncolor{lightgray!50}}p{35mm}|c|c|c|c|c|c|c|}
\hline

\rowcolor{lightgray!50}
& & \multicolumn{3}{c|}{\textbf{Crowdworkers ($\textbf{n}_{Non-LLM} = \textbf{n}_{LLM} = 87$)}} & \multicolumn{3}{c|}{\textbf{Students ($\textbf{n}_{Non-LLM} = 90, \textbf{n}_{LLM} = 82$)}} \\ 
\hhline{|
>{\arrayrulecolor{lightgray!50}}-
>{\arrayrulecolor{black}}|
>{\arrayrulecolor{lightgray!50}}-
>{\arrayrulecolor{black}}|------
}

\rowcolor{lightgray!50}
& & \multicolumn{2}{c|}{\textbf{Mean}}  &  & \multicolumn{2}{c|}{\textbf{Mean}} & \\  
\hhline{|
>{\arrayrulecolor{lightgray!50}}-
>{\arrayrulecolor{black}}|
>{\arrayrulecolor{lightgray!50}}-
>{\arrayrulecolor{black}}|--|
>{\arrayrulecolor{lightgray!50}}-
>{\arrayrulecolor{black}}|--|
>{\arrayrulecolor{lightgray!50}}-
>{\arrayrulecolor{black}}|
}

\rowcolor{lightgray!50}
\multirow{-3}{*}{\textbf{Quality}} & \multirow{-3}{*}{\begin{tabular}{@{}c@{}} \textbf{Hypothesized}\\ \textbf{ Superior} \\ \textbf{Group}\end{tabular}} & \textbf{Non-LLM} & \textbf{LLM} & \multirow{-2}{*}{\textbf{p-value}}  & \textbf{Non-LLM} & \textbf{LLM} &  \multirow{-2}{*}{\textbf{p-value}}\\ \hhline{*7{|=}|=|}

\textbf{Perceived Authenticity} & Non-LLM & \cellcolor{green!30}5.03 $\pm$ 0.17 & \cellcolor{red!30}4.92 $\pm$ 0.17 & 0.31 & \cellcolor{green!30}4.34 $\pm$ 0.18 & \cellcolor{red!30}4.29 $\pm$ 0.19 & 0.42 \\ \hline

\textbf{Perceived Ability to Communicate Key Takeaways} & \multirow{8}{*}{LLM} & \cellcolor{red!30}\multirow{3}{*}{5.51 $\pm$ 0.14} & \cellcolor{green!30}\multirow{3}{*}{6.01 $\pm$ 0.10} & \multirow{3}{*}{<0.01**}  & \cellcolor{red!30}\multirow{3}{*}{5.27 $\pm$ 0.16} & \cellcolor{green!30}\multirow{3}{*}{5.48 $\pm$ 0.17} & \multirow{3}{*}{0.21}\\ \hhline{|-|~|------} 

\textbf{Perceived Ability to Promote Reflection} & & \cellcolor{red!30}\multirow{2}{*}{4.71 $\pm$ 0.20} & \cellcolor{green!30}\multirow{2}{*}{5.31 $\pm$ 0.17} & \multirow{2}{*}{0.02*}  & \cellcolor{red!30}\multirow{2}{*}{3.96 $\pm$ 0.20} & \cellcolor{green!30}\multirow{2}{*}{4.55 $\pm$ 0.20} & \multirow{2}{*}{0.04*} \\ \hhline{|-|~|------} 

\textbf{Perceived Balance of Positivity with Realistic Struggle} & & \cellcolor{red!30}\multirow{3}{*}{4.80 $\pm$ 0.16} & \cellcolor{green!30}\multirow{3}{*}{5.38 $\pm$ 0.16} & \multirow{3}{*}{<0.01**}  &  \cellcolor{red!30}\multirow{3}{*}{4.29 $\pm$ 0.18} & \cellcolor{green!30}\multirow{3}{*}{4.62 $\pm$ 0.17} & \multirow{3}{*}{0.15} \\ \hline
\end{tabular}%
}
~\\~\\
\footnotesize{* p <.05, ** p < .01, *** p < .001}
\end{table*}

%% file: Table/results2.tex
\begin{table*}[t]
\centering
\caption{Summary statistics and measures of the stories' intervention qualities with and without LLM enhancement. The results are separated across the two populations that were studied: crowdworkers and university students. 
}
\label{tab: stat2}
\Description{Quantitative comparisons of perceived intervention qualities, including reduction in belief in negative thoughts, emotion intensity, relatability, inspiration, likelihood of future engagement, and number of words written for reflection. LLM stories consistently outperformed Non-LLM stories, especially among crowdworkers.}
\resizebox{\textwidth}{!}{%
\begin{tabular}{|>{\columncolor{lightgray!50}}p{33mm}|c|c|c|c|c|c|c|}
\hline

\rowcolor{lightgray!50}
& & \multicolumn{3}{c|}{\textbf{Crowdworkers ($\textbf{n}_{Non-LLM} = \textbf{n}_{LLM} = 87$)}} & \multicolumn{3}{c|}{\textbf{Students ($\textbf{n}_{Non-LLM} = 90, \textbf{n}_{LLM} = 82$)}} \\ 
\hhline{|
>{\arrayrulecolor{lightgray!50}}-
>{\arrayrulecolor{black}}|
>{\arrayrulecolor{lightgray!50}}-
>{\arrayrulecolor{black}}|------
}

\rowcolor{lightgray!50}
& & \multicolumn{2}{c|}{\textbf{Mean}}  &  & \multicolumn{2}{c|}{\textbf{Mean}} & \\  
\hhline{|
>{\arrayrulecolor{lightgray!50}}-
>{\arrayrulecolor{black}}|
>{\arrayrulecolor{lightgray!50}}-
>{\arrayrulecolor{black}}|--|
>{\arrayrulecolor{lightgray!50}}-
>{\arrayrulecolor{black}}|--|
>{\arrayrulecolor{lightgray!50}}-
>{\arrayrulecolor{black}}|
}

\rowcolor{lightgray!50}
\multirow{-3}{*}{\textbf{Quality}} & \multirow{-3}{*}{\begin{tabular}{@{}c@{}} \textbf{Hypothesized}\\ \textbf{ Superior} \\ \textbf{Group}\end{tabular}} & \textbf{Non-LLM} & \textbf{LLM} & \multirow{-2}{*}{\textbf{p-value}}  & \textbf{Non-LLM} & \textbf{LLM} &  \multirow{-2}{*}{\textbf{p-value}}\\ \hhline{*7{|=}|=|}

\textbf{Perceived Reduction in Belief in Negative Thought} & \multirow{12}{*}{LLM} &  \cellcolor{red!30}\multirow{3}{*}{0.26 $\pm$ 0.10} & \cellcolor{green!30}\multirow{3}{*}{0.64 $\pm$ 0.13} & \multirow{3}{*}{0.02*}& \cellcolor{red!30}\multirow{3}{*}{0.49 $\pm$ 0.12} & \cellcolor{green!30}\multirow{3}{*}{0.65 $\pm$ 0.12} & \multirow{3}{*}{0.21} \\ \hhline{|-|~|------} 

\textbf{Perceived Reduction in Emotion Intensity} & & \cellcolor{red!30}\multirow{2}{*}{0.59 $\pm$ 0.13} & \cellcolor{green!30}\multirow{2}{*}{0.89 $\pm$ 0.12} & \multirow{2}{*}{0.06}  & \cellcolor{red!30}\multirow{2}{*}{0.44 $\pm$ 0.10} & \cellcolor{green!30}\multirow{2}{*}{0.65 $\pm$ 0.12} & \multirow{2}{*}{0.15}\\ \hhline{|-|~|------}

\textbf{Perceived Relatability} &  & \cellcolor{red!30}5.16 $\pm$ 0.17 & \cellcolor{green!30}5.75 $\pm$ 0.13 & <0.01** & \cellcolor{red!30}4.48 $\pm$ 0.17 & \cellcolor{green!30}5.05 $\pm$ 0.18 & 0.03* \\ \hhline{|-|~|------} 

\textbf{Perceived Capacity of Inspiring Solutions} &  & \cellcolor{red!30}\multirow{2}{*}{4.74 $\pm$ 0.18} & \cellcolor{green!30}\multirow{2}{*}{5.46 $\pm$ 0.16} & \multirow{2}{*}{<0.01**}  & \cellcolor{red!30}\multirow{2}{*}{4.14 $\pm$ 0.19} & \cellcolor{green!30}\multirow{2}{*}{4.76 $\pm$ 0.18} & \multirow{2}{*}{0.04*}  \\ \hhline{|-|~|------}

\textbf{Likelihood of Future Engagement} & &  \cellcolor{red!30}\multirow{2}{*}{4.53 $\pm$ 0.20} & \cellcolor{green!30}\multirow{2}{*}{5.21 $\pm$ 0.17} & \multirow{2}{*}{<0.01**}  & \cellcolor{red!30}\multirow{2}{*}{3.92 $\pm$ 0.20} & \cellcolor{green!30}\multirow{2}{*}{4.30 $\pm$ 0.20} & \multirow{2}{*}{0.15} \\ \hhline{|-|~|------}

\textbf{Number of Words Written for Reflection} && \cellcolor{red!30}\multirow{2}{*}{31.15 $\pm$ 2.28} & \cellcolor{green!30}\multirow{2}{*}{44.22 $\pm$ 4.04} & \multirow{2}{*}{<0.01**}  & \cellcolor{red!30}\multirow{2}{*}{25.29 $\pm$ 2.42} & \cellcolor{green!30}\multirow{2}{*}{33.56 $\pm$ 2.65} & \multirow{2}{*}{0.04*} \\ \hline

\end{tabular}%
}
~\\~\\
\footnotesize{* p <.05, ** p < .01, *** p < .001}
\end{table*}

%% file: Section/discussion.tex
\section{Discussion}

Our study demonstrates how LLMs can significantly enhance human-generated content so that it is more relatable and impactful to new audiences. We found that these enhancements through dynamic personalization improved the utility of the stories as interventions for managing negative thoughts among young adults. 
These findings add to the growing body of literature on human-LLM collaboration \cite{chan2023mango, denny2023conversing, bhattacharjee2024understanding, guingrich2023chatbots, sharma2023human}, which seeks to utilize and expand upon existing support mechanisms and content, while shining a light on the evaluation and scrutiny required to support the design of such tools for DMH \cite{bhattacharjee2024understanding, korngiebel2021considering, demszky2023using}.

In our discussion, we begin by addressing how our findings address our research questions. We then discuss the design implications of our results, specifically focusing on future DMH interventions and storytelling applications.

\subsection{Key Insights}

\subsubsection{\textbf{RQ1}: Personalized LLM-Enhanced Stories in Exhibiting Narrative Qualities}

Our research highlights the promising capabilities of LLMs to dynamically generate content that can enhance the quality of narratives in DMH interventions. Similar to prior works \cite{sharma2023cognitive, kim2024diarymate, bhattacharjee2024understanding}, the LLM stories in our study processed open-ended inputs from participants and adapted the stories to their unique contexts. This capability underscores the potential of LLMs in crafting personalized stories that can support individuals in managing their specific barriers to psychological wellbeing.

Our findings reveal that the LLM stories in our study enhanced most of the narrative qualities we investigated. \revision{The LLM stories were effective in communicating clear takeaways}, which participants noted as essential for translating the stories’ lessons into their own lives \cite{bhattacharjee2022kind, dyer1982depth}. These stories were perceived to excel in promoting reflection, as explicit suggestions within the narratives encouraged readers to contemplate potential changes in their thinking patterns \cite{bhattacharjee2024exploring, kocielnik2018reflection}. Additionally, the LLM stories balanced positivity with realistic portrayals of struggles. The stories often charted a structured progression from challenging life scenarios that resonated with participants' experiences to outcomes showing small yet meaningful progress \cite{bhattacharjee2022kind}. 

The LLM was able to render all of these benefits while still being comparable in authenticity to human-written stories, which likely reflects that the LLM effectively mirrored participants' life experiences and that it drew on human-written example stories.  However, while not statistically significant, we also note that the trend in the difference between the two conditions, favored the Non-LLM condition. Our qualitative data also revealed concerns about authenticity in the LLM condition, which could potentially reflect specific word choices made by the LLMs, participants' familiarity with typical responses made by popular chatbots \cite{ariyaratne2023comparison}, and the occasional mechanical tone inherent in AI-generated content \cite{cheng2022human}.  Recent studies have also examined text generated by ChatGPT across various contexts, noting an increased and repetitive occurrence of certain words such as `delve', `intricate', and `underscore' \cite{kobak2024delving}. The frequent use of certain relatively formal words or phrases may contribute to compromising the perceived authenticity of LLM-generated stories.


\subsubsection{\textbf{RQ2}: Personalized LLM-Enhanced Stories as an Intervention to Help People Manage Negative Thoughts}

Our findings suggest that LLM stories were relatively more effective in helping individuals reduce their belief in their negative thoughts, aligning with prior literature that underscores the benefits of structured reflection activities for mental health management \cite{o2018suddenly, bhattacharjee2024exploring, bhattacharjee2022kind}. These stories facilitated reflection on personal thought patterns, allowing participants to draw parallels between their own challenges and those faced by the characters within the narratives \cite{akram2020exploratory}. 
The LLM stories appeared to promote diffuse sociality~\cite{burgess2019think}, reducing feelings of isolation by showing that others may also experience similar struggles. Participants noted that this sense of connection reassured them that they were not alone in their challenges. 


The LLM stories also highlight the role of affective empathy in DMH interventions \cite{sharma2023human, sharma2021towards, picard2000affective}. Affective empathy describes an entity's ability to perceive, respond to, and convey emotions \cite{picard2000affective}. The LLM stories demonstrated this capability by dynamically adapting the content to reflect how users felt about the challenges they encountered \cite{motahar2024toward}. Participants particularly valued how the stories acknowledged the characters' struggles, which resonated deeply with their own experiences. They reported significant benefits, such as the ability to reevaluate feelings of failure or explore alternative solutions — outcomes that align with the recognized advantages of empathetic interactions in DMH tools \cite{sharma2023cognitive, ardenghi2024supporting, pereira2019using}.

We also noticed that relatability was often cited as a key factor enhancing the impact of the LLM stories within our study, underscoring the importance of tailoring DMH content to individual contexts \cite{bhattacharjee2023investigating, baumel2019objective, gatos2021hci}. Participants frequently noted that the stories reflected their own experiences, aiding them in visualizing how the advice could be applicable to their own lives \cite{bhattacharjee2022kind}. However, excessive relatability detracted from the story's applicability, as it gave the impression of the story being unrealistically tuned to an individual's unique situations. These comments bear similarities with the uncanny valley effect \cite{seyama2007uncanny} encountered in robotics, which entails discomfort when replicas appear almost, but not quite, human. In the context of DMH, excessively tailored stories can also seem similarly artificial, potentially causing participants to view them as forced and potentially undermining their effectiveness.

\DIS{Finally, we emphasize that LLM-based narrative interventions are not necessarily intended to be implemented in the exact form used in this study, nor are they alone sufficient to address broader wellbeing challenges. Narrative-based approaches are increasingly integrated into interventions targeting specific goals, such as managing depression and anxiety \cite{rodriguez2014mindfulness, meyerhoff2024small}, promoting physical activity \cite{saksono2021storymap}, or encouraging healthy habits \cite{grimes2008eatwell}.} 
\revision{Interventions like ours can function as modular components within a broader framework, addressing specific needs while complementing other strategies. For instance, the current version of our SSI aims to communicate psychological principles for managing commonly experienced negative thoughts, such as feeling down on oneself, worrying about the future, or feeling let down by others. Similar methods could be used to reframe everyday stressors \cite{bhattacharjee2024exploring, meyerhoff2024small} in a way that offers users brief, supportive narratives as a way to regulate immediate emotional responses. These SSIs could be integrated into larger DMH programs, where they can coexist with other approaches such as skill-building exercises or psychoeducational modules. However, deploying LLM-generated narratives for severe conditions, such as post-traumatic stress disorder (PTSD) or suicidal ideation, would likely require additional safeguards such as direct clinician involvement and supervision prior to story delivery.}


\subsection{Design Recommendations}
Our findings guide us to make several design recommendations. We describe them below.

\subsubsection{Balance Between Relatability and Avoiding Implausibility}
\label{subsub: balance}
Our findings highlight a delicate balance in crafting stories that are relatable enough to engage and inspire without crossing into implausibility. Future interventions could address this challenge by creating stories that reflect key aspects of a person’s experience without mirroring their situation too closely, which could otherwise lead to a loss of credibility \cite{seyama2007uncanny}.

One approach could be to offer users a curated set of story options with varying degrees of relatability. For instance, instead of presenting just one narrative, participants might choose from a range of stories that share similar emotional challenges but differ in specifics like context, background, or tone \cite{bhattacharjee2022kind, meyerhoff2024small}. This would allow users to select a narrative that aligns with their preferred level of resonance. Another approach could involve embedding multiple branching points within stories, where users could navigate different pathways as the narrative unfolds \cite{bhattacharjee2022kind}. For example, a user could choose whether the protagonist tackles their challenges by seeking external support or by relying on personal coping strategies. This interactive experience might offer users control over how much the story connects with their own experiences, making the process feel more dynamic and tailored.

\subsubsection{Refining Tone and Wording of AI-generated Content}


While LLM stories demonstrated promising potential as narrative interventions, some participants expressed concerns about the writing style and tone, which they felt diminished the overall impact. This issue arose from the use of certain phrasing or the participants’ familiarity with the way ChatGPT responds, making the stories feel less authentic. The tone of the responses was frequently mentioned, with some participants finding the tone to be too mechanical lacking the personal touch necessary for conversations around managing negative thoughts \cite{cheng2022human}. Others noted that the overly optimistic tone in certain stories did not always align with or acknowledge the severity of their struggles, which, in turn, made them feel that their challenges were being minimized \cite{haugeland2022understanding}. 

To address these issues, future DMH interventions could focus on refining the tone and wording of LLM-generated stories. For example, incorporating more variability in sentence structures and word choice could make the narratives feel less formulaic and more natural \cite{park2023anthropomorphic}. This enhancement could be achieved by explicitly instructing LLMs through customized prompting, experimenting with the temperature parameter to introduce variability, or fine-tuning LLMs using stories that contain diverse structures. Another approach could be to vary the emotional tone of stories depending on the intensity of the challenges being addressed, ensuring the tone is empathetic without being overly upbeat \cite{haugeland2022understanding}. Additionally, allowing for more ambiguity or open-ended conclusions in some stories might prevent them from feeling overly prescriptive, giving participants room to reflect on their own situation rather than feeling as though they are being given an unrealistic solution \cite{bhattacharjee2022kind}. However, all of these refinements should be carefully tested to tailor the tone, structure, and ambiguity to different user needs. 


\subsubsection{Structured Reflection Activities}

Participants in our study expressed appreciation for the reflection activity embedded within the stories, noting that such exercises helped them reevaluate their thoughts and emotions and potentially identify solutions to their problems. However, the reflection activity in our study was limited to a single question. Future studies might consider drawing inspiration from guided reflection activities, which involve a sequence of targeted questions focusing on specific aspects of one's thoughts \cite{o2018suddenly, lee2017designing, bhattacharjee2024actually}. These multi-question formats allow for a more structured presentation of thoughts, capturing nuances that a single question might miss. Incorporating such detailed questioning can enhance the benefits of LLM stories by  providing deeper insights into their thoughts and challenges, and potentially helping find ways of solving them. \cite{bhattacharjee2024exploring, sharma2023cognitive, park2021wrote}. However, it is important to note that answering these more detailed questions requires more time and effort from users. Therefore, careful consideration should be given to the timing and frequency of these questions to maintain a balance between depth of engagement and participant burden.

\subsubsection{Improving Peer Support and Crowd-Powered Platforms}

In the past, peer support and crowd-powered platforms have shown promising potential in providing wellbeing support, yet they face several inherent challenges that could be alleviated by an approach like ours. For example, the act of sharing and writing content can be daunting; LLMs can assist crowdworkers and peers by providing example stories and offering writing support to edit and restructure users' own experiences \cite{kim2024diarymate, reza2023abscribe, gatos2021hci}. Additionally, some individuals may be reluctant to open up to real people \cite{park2021wrote}. In these cases, the ability of LLMs to craft highly personalized narratives in a completely private manner — without the need to share personal details — could significantly increase their appeal, particularly for more inhibited users. Nonetheless, it is crucial to position LLMs as supportive tools that enhance the human-driven process and keep the essential human element in peer support intact.



\subsubsection{\revision{Considerations for People's Attitude Toward AI}}

\revision{While our study did not specifically measure people’s attitudes toward AI, participants’ comments suggest that such attitudes may influence their perceptions of LLM-enhanced stories. Many participants appeared open to the idea of AI-based tools providing support for managing negative thoughts, while others expressed resistance, reflecting a wide range of expectations regarding AI’s role in personal and emotional contexts~\cite{kaya2024roles, liu2022will}. This variation highlights the importance of considering how perceptions may shift depending on whether users are aware that a story is AI-generated or human-written — particularly since, in real-world deployments, authorship may not be masked.} 

\revision{As such, future research could benefit from directly measuring users’ attitudes toward AI~\cite{grassini2023development} and systematically investigating how knowledge of a story’s origin influences engagement, perceived helpfulness, and impact on negative thoughts and emotions.
In addition, future DMH tools could allow users to express their attitudes toward AI during onboarding in order to adapt to those views. Gradually building rapport, rather than offering support immediately, may also help accommodate diverse expectations and comfort levels~\cite{o2022building}.}

\subsubsection{Considerations for Longitudinal Deployments}

\DIS{While our study explored user perceptions of narrative interventions as an SSI, deploying similar interventions across multiple sessions will introduce additional design considerations. Future longitudinal studies are essential to investigate these implications in greater depth, but our observations hint at several aspects that could be important to consider.}

\DIS{Repeated engagement with narratives, even when the content differs, may result in habituation \cite{rankin2009habituation} if the stories consistently follow similar styles or structures, thereby reducing their effectiveness. To maintain user engagement, LLM-enhanced narratives should dynamically vary in format, length, tone, and writing style across sessions. Achieving this diversity may require insights from the literature on just-in-time adaptive interventions \cite{bhattacharjee2023investigating, kabir2022ask}, which emphasizes the role of contextual factors such as time, mood, emotions, and social context in shaping user engagement with DMH content. AI-driven systems could leverage this literature to create adaptive guidelines for LLMs, enabling the generation of contextually tailored stories. For instance, users might prefer shorter narratives during a busy workday but may engage more deeply with longer, reflective stories during a relaxed weekend \cite{bhattacharjee2023design}. By integrating such dynamic adjustments, longitudinal interventions may be able to maintain user engagement and provide meaningful experiences that adapt to their changing needs and circumstances.}

\subsection{Limitations \revision{and Future Work}}


Our study has several limitations that merit attention. First, our study population was confined to young adults. While some findings from this demographic might be generalizable to other groups, conducting similar studies with different populations could reveal new insights into their preferences and responses. Future studies should also consider including individuals with clinically elevated levels of depression and anxiety symptoms, keeping in mind the heightened need for risk management processes and oversight in such research.


\revision{Second, we did not embed specific action strategies within the narratives; rather, our goal was to communicate psychological principles for managing negative thoughts. Some participants mentioned that the stories inspired them to adopt a different mindset or consider taking small steps forward, but we did not systematically track whether or how participants acted on the stories. Future research should explore what kinds of actions, if any, individuals are motivated to take after engaging with such narratives. While our intervention did not include explicit suggestions, we recognize that introducing more action-oriented content in future systems may require careful consideration of timing and user readiness, as individuals may vary in their willingness or capacity to engage with such content depending on their context \cite{bhattacharjee2023investigating}.}

Finally, \revision{our study relied primarily on user perceptions of the LLM-enhanced stories — their views on various qualities of the stories and their effectiveness as interventions. That is, we did not evaluate the objective qualities of the content itself. While we constrained the LLMs to work within the framework of previously established narratives supporting evidence-based psychological strategies, future research should assess adherence to these strategies and other inappropriate content that may appear in generated stories, potentially involving clinicians in the evaluation process to ensure the accuracy and appropriateness of the content.} 

%% file: Section/conclusion.tex
\section{Conclusion}

LLMs offer promising capabilities to dynamically adapt and generate stories that can resonate with the unique challenges individuals face. Motivated by this potential, our study leveraged LLM to enhance and personalize existing human-written stories from past DMH interventions, comparing the LLM-enhanced versions with their original counterparts. We conducted an experiment with young adults, including crowdworkers and students, and observed that LLM stories were perceived to do better than human-written ones in conveying key takeaways, promoting reflection, and reducing belief in negative thoughts. These narratives were not only perceived as more relatable but also maintained a level of authenticity comparable to human-written stories. Our findings illuminate potential areas for future work on LLM-enhanced narrative interventions, such as balancing relatability with avoiding implausibility and refining the AI’s tone and writing style. We believe our work marks a significant step towards enhancing DMH interventions with emerging AI technology, enabling them to dynamically adapt based on user challenges.

\begin{acks}

We express our gratitude to Mary Czerwinski (Microsoft Research), Javier Hernandez (Microsoft Research), Haochen Song (University of Toronto), and Benjamin Kaveladze (Northwestern University) for their suggestions on relevant literature and their feedback on this work. This work was supported by grants from the National Institute of Mental Health (K01MH125172, R34MH124960), the Office of Naval Research (N00014-18-1-2755, N00014-21-1-2576), the Natural Sciences and Engineering Research Council of Canada (RGPIN-2019-06968), and the National Science Foundation (2209819). In addition, we acknowledge a gift from the Microsoft AI for Accessibility program to the Center for Behavioral Intervention Technologies that, in part, supported this work (\url{http://aka.ms/ai4a}). Ananya Bhattacharjee would also like to acknowledge support from the Inlight Research Fellowship, the Schwartz Reisman Institute Graduate Fellowship, and the Wolfond Scholarship in Wireless Information Technology.

\end{acks}

%% file: Section/appendix.tex
\appendix

\onecolumn
\section{Non-LLM Stories and Corresponding LLM-Enhanced Examples}
\label{sec: examples}
\input{Table/appendix-story-table}
\clearpage
\section{Narrative Qualities and Intervention Outcomes Across All Challenging Thoughts}
\label{sec: app_evaluation}

\begin{figure*}[htbp]
    \centering
    \includegraphics[width=0.95\textwidth]{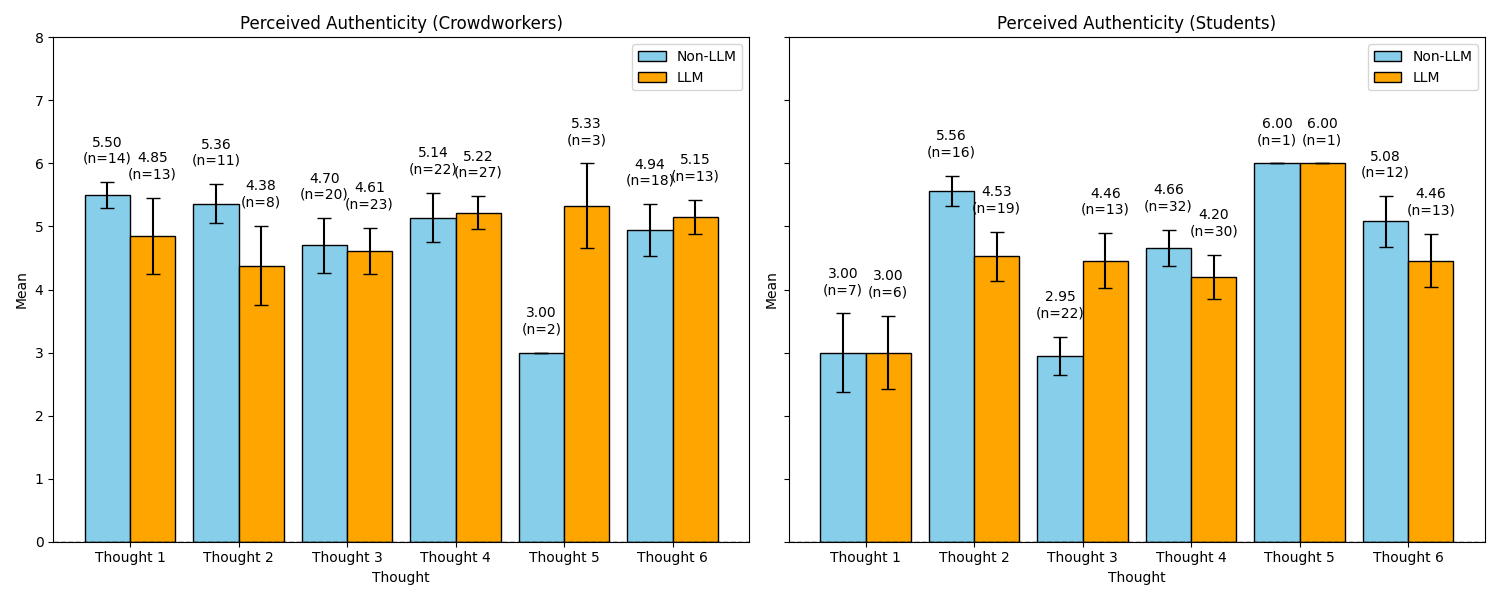}
    \caption{Mean perceived authenticity across Non-LLM and LLM stories addressing six challenging thoughts, shown separately for crowdworkers (left) and students (right). Error bars represent the standard error of the mean, with sample sizes indicated above each bar.}
    \label{fig:authenticity}
        \Description{Mean perceived authenticity across Non-LLM and LLM stories addressing six challenging thoughts, shown separately for crowdworkers (left) and students (right). Error bars represent the standard error of the mean, with sample sizes indicated above each bar.}
\end{figure*}

\begin{figure*}[htbp]
    \centering
    \includegraphics[width=0.95\textwidth]{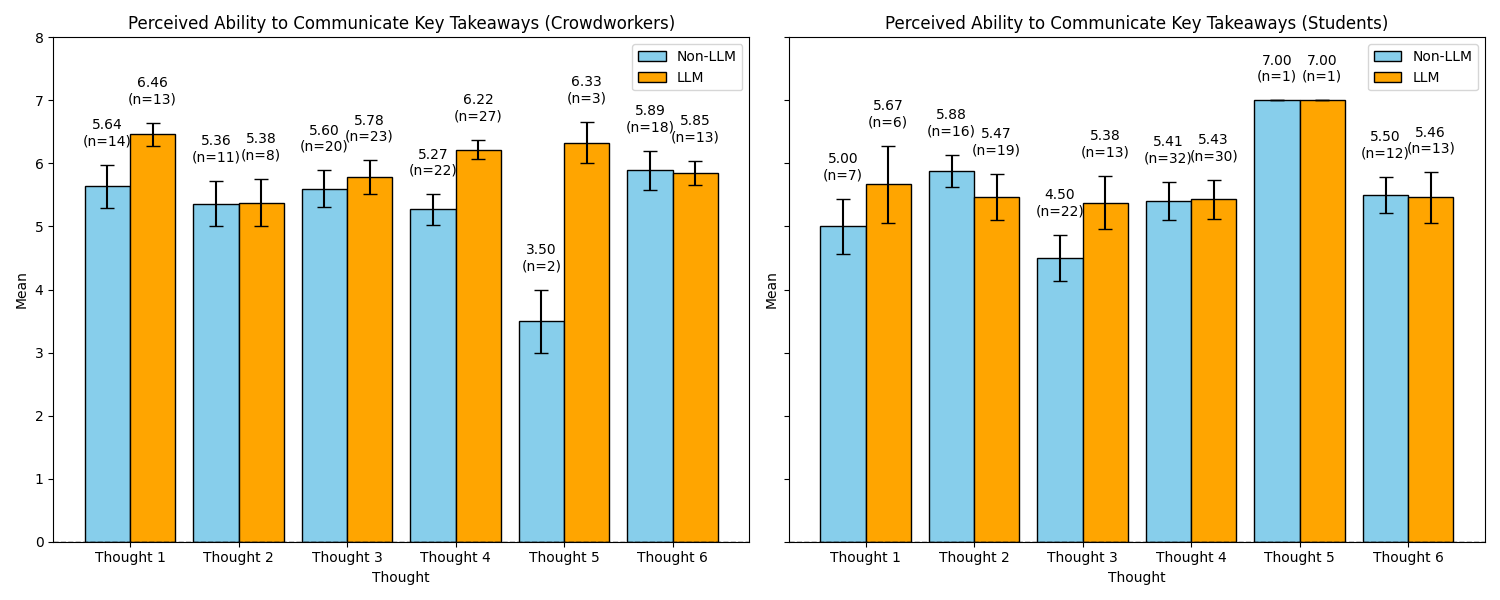}
    \caption{Mean perceived ability to communicate key takeaways across Non-LLM and LLM stories addressing six challenging thoughts, shown separately for crowdworkers (left) and students (right). Error bars represent the standard error of the mean, with sample sizes indicated above each bar.}
    \label{fig:key_message}
        \Description{Mean perceived ability to communicate key takeaways across Non-LLM and LLM stories addressing six challenging thoughts, shown separately for crowdworkers (left) and students (right). Error bars represent the standard error of the mean, with sample sizes indicated above each bar.}
\end{figure*}

\begin{figure*}[htbp]
    \centering
    \includegraphics[width=0.95\textwidth]{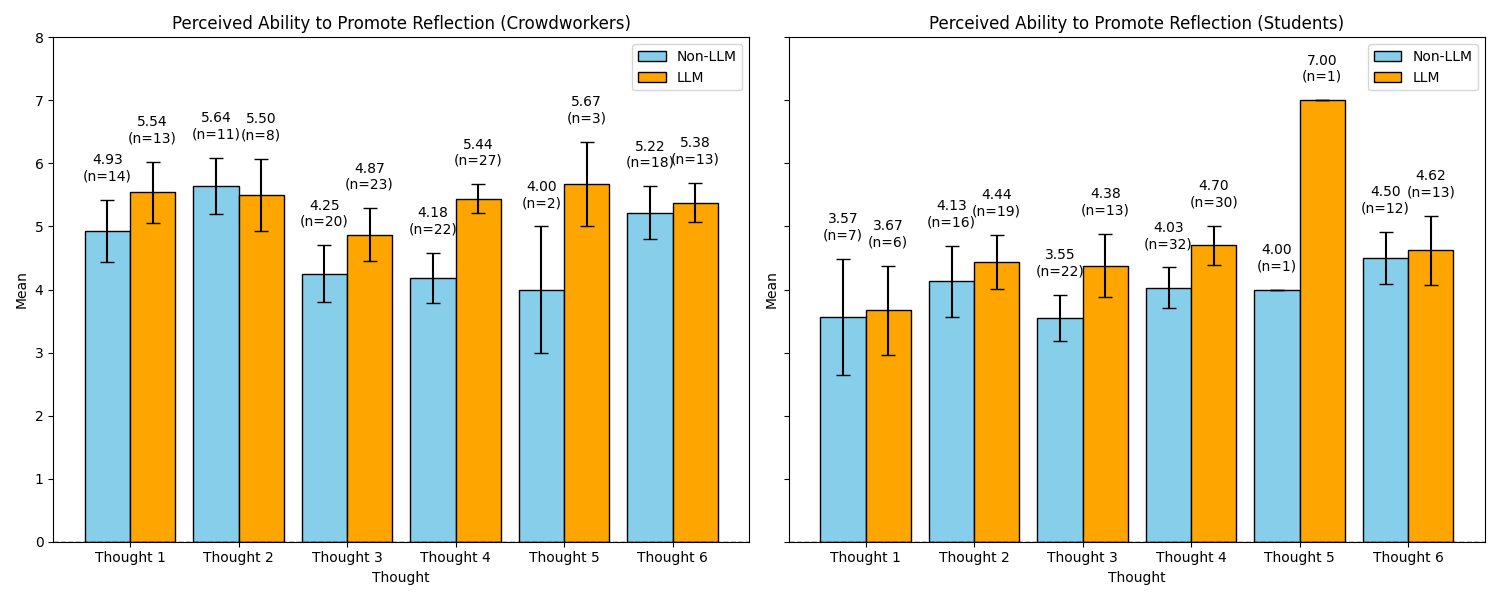}
    \caption{Mean perceived ability to promote reflection across Non-LLM and LLM stories addressing six challenging thoughts, shown separately for crowdworkers (left) and students (right). Error bars represent the standard error of the mean, with sample sizes indicated above each bar.}
    \label{fig:reflection}
    \Description{Mean perceived ability to promote reflection across Non-LLM and LLM stories addressing six challenging thoughts, shown separately for crowdworkers (left) and students (right). Error bars represent the standard error of the mean, with sample sizes indicated above each bar.}
\end{figure*}

\begin{figure*}[htbp]
    \centering
    \includegraphics[width=0.95\textwidth]{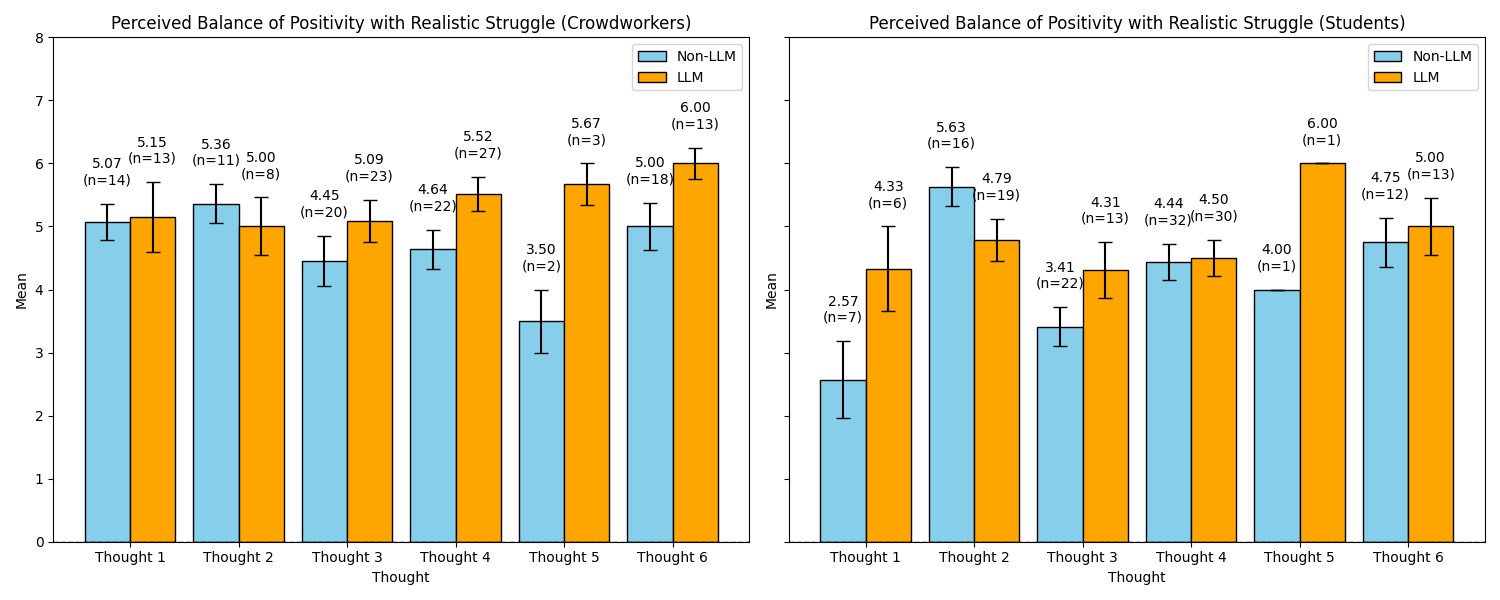}
    \caption{Mean perceived balance of positivity with realistic struggles across Non-LLM and LLM stories addressing six challenging thoughts, shown separately for crowdworkers (left) and students (right). Error bars represent the standard error of the mean, with sample sizes indicated above each bar.}
    \label{fig:balance}
    \Description{Mean perceived balance of positivity with realistic struggles across Non-LLM and LLM stories addressing six challenging thoughts, shown separately for crowdworkers (left) and students (right). Error bars represent the standard error of the mean, with sample sizes indicated above each bar.}
\end{figure*}

\begin{figure*}[htbp]
    \centering
    \includegraphics[width=0.95\textwidth]{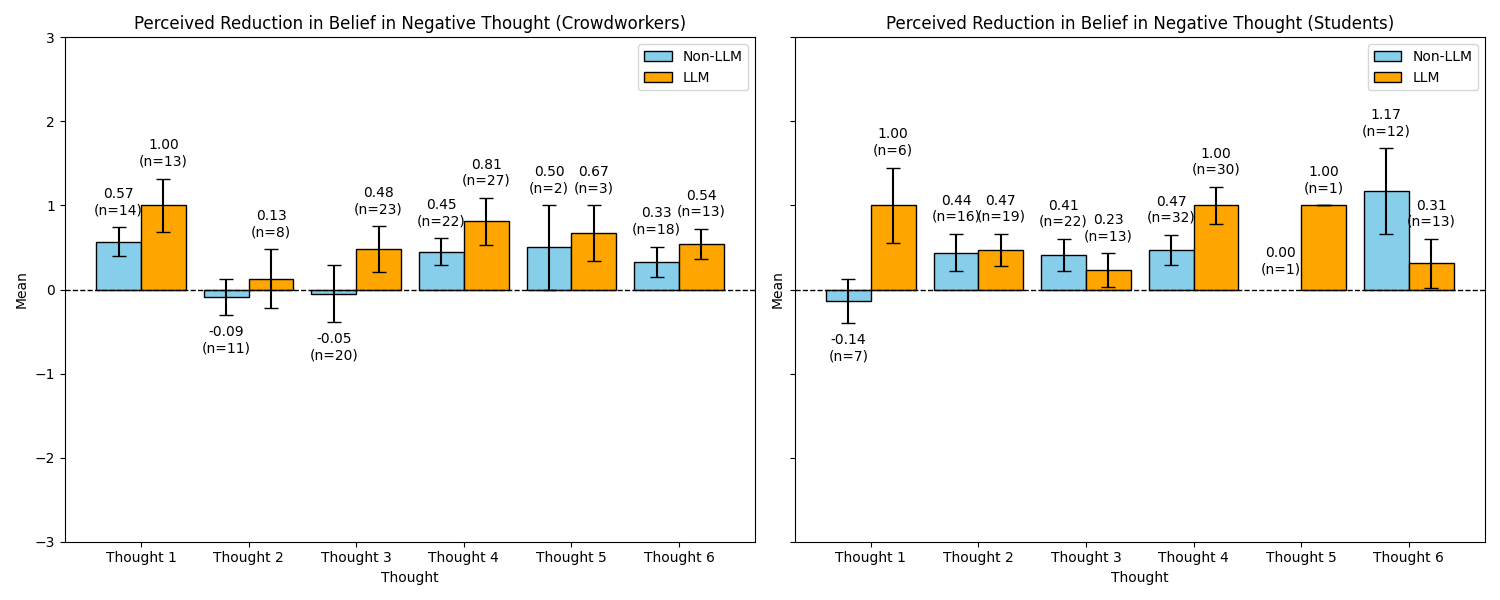}
    \caption{Mean perceived reduction in belief in negative thoughts across Non-LLM and LLM stories addressing six challenging thoughts, shown separately for crowdworkers (left) and students (right). Error bars represent the standard error of the mean, with sample sizes indicated above each bar.}
    \label{fig:change_negative_belief}
    \Description{Mean perceived reduction in belief in negative thoughts across Non-LLM and LLM stories addressing six challenging thoughts, shown separately for crowdworkers (left) and students (right). Error bars represent the standard error of the mean, with sample sizes indicated above each bar.}
\end{figure*}

\begin{figure*}[htbp]
    \centering
    \includegraphics[width=0.95\textwidth]{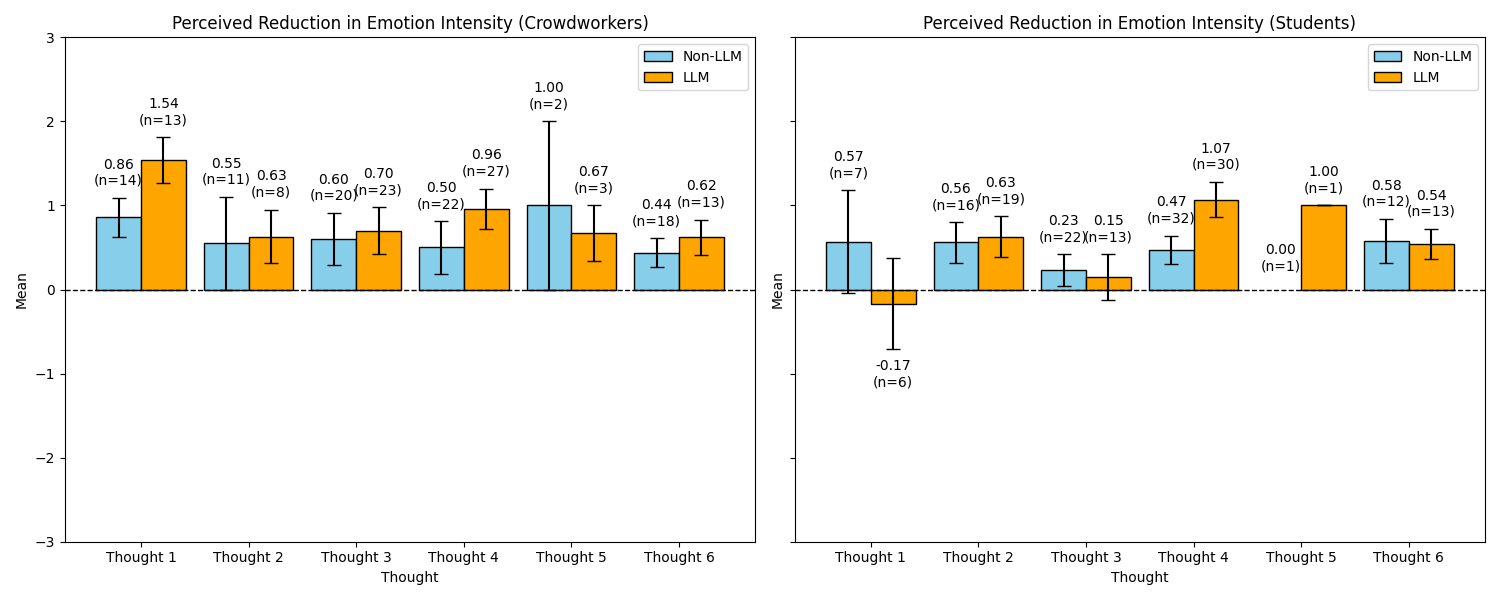}
    \caption{Mean perceived reduction in emotion intensity across Non-LLM and LLM stories addressing six challenging thoughts, shown separately for crowdworkers (left) and students (right). Error bars represent the standard error of the mean, with sample sizes indicated above each bar.}
    \label{fig:change_emotion}
    \Description{Mean perceived reduction in emotion intensity across Non-LLM and LLM stories addressing six challenging thoughts, shown separately for crowdworkers (left) and students (right). Error bars represent the standard error of the mean, with sample sizes indicated above each bar.}
\end{figure*}

\begin{figure*}[htbp]
    \centering
    \includegraphics[width=0.95\textwidth]{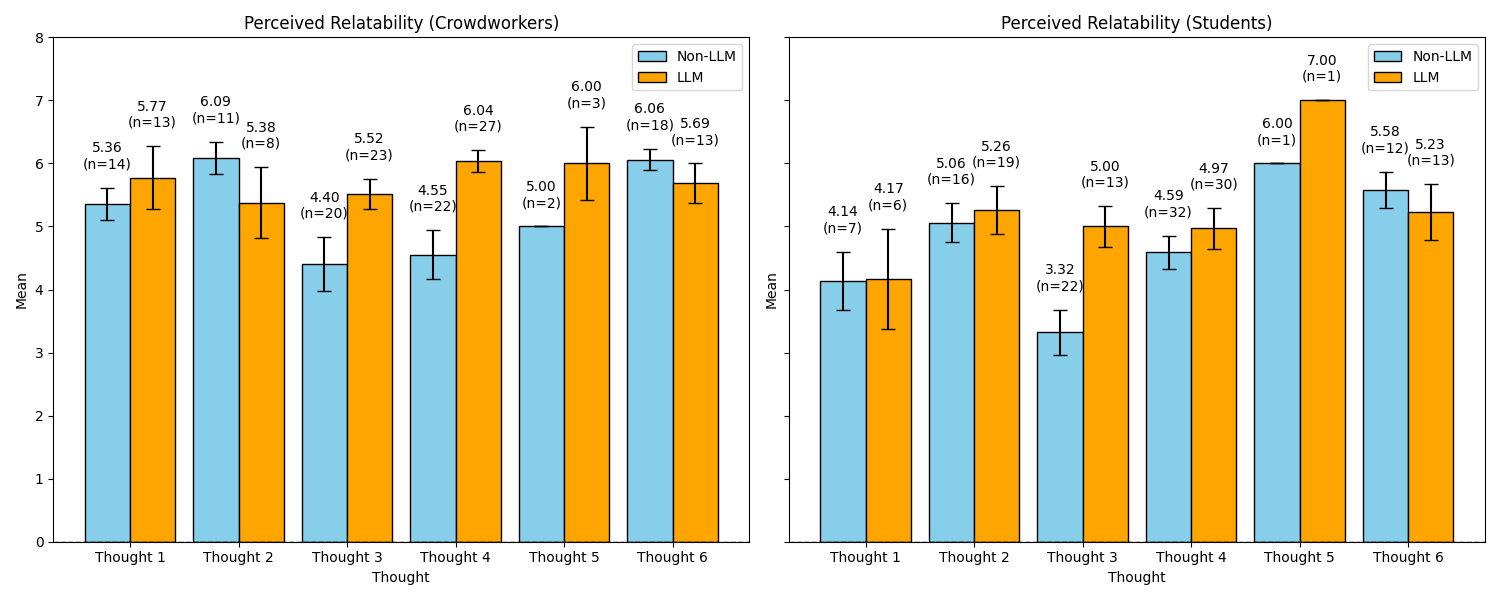}
    \caption{Mean perceived relatability across Non-LLM and LLM stories addressing six challenging thoughts, shown separately for crowdworkers (left) and students (right). Error bars represent the standard error of the mean, with sample sizes indicated above each bar.}
    \label{fig:relatability}
    \Description{Mean perceived relatability across Non-LLM and LLM stories addressing six challenging thoughts, shown separately for crowdworkers (left) and students (right). Error bars represent the standard error of the mean, with sample sizes indicated above each bar.}
\end{figure*}

\begin{figure*}[htbp]
    \centering
    \includegraphics[width=0.95\textwidth]{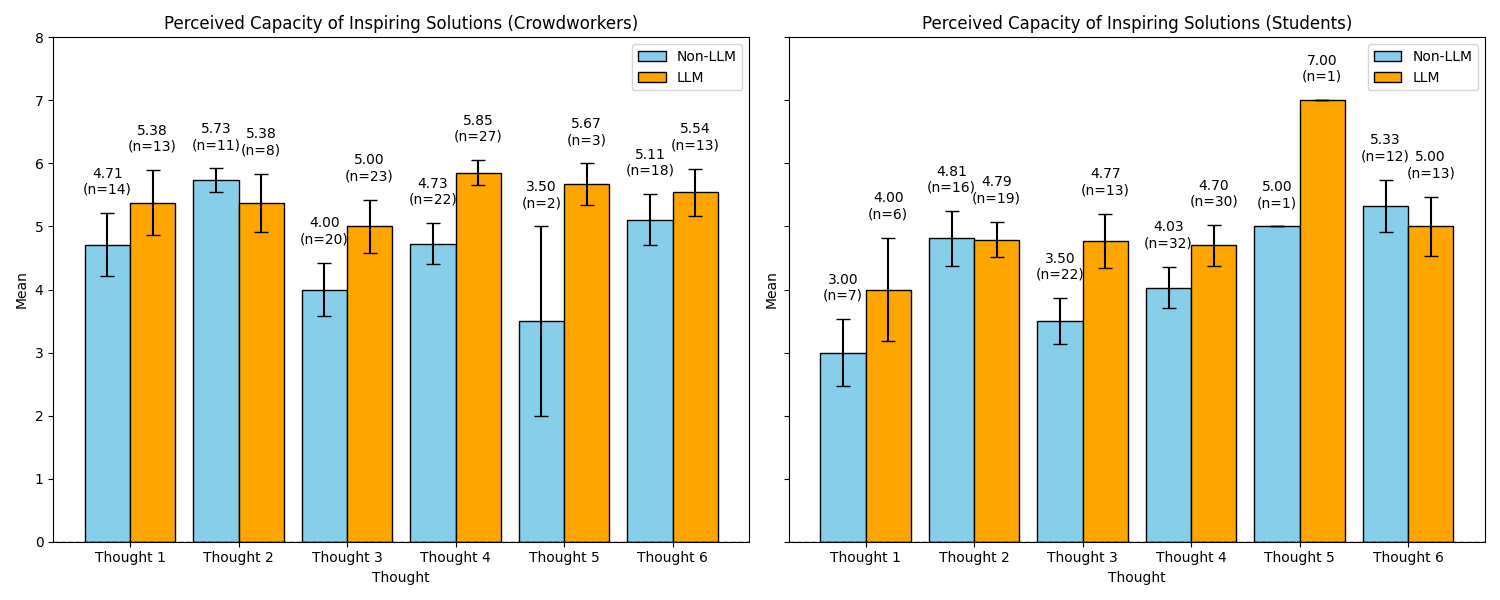}
    \caption{Mean perceived capacity of inspiring solutions across Non-LLM and LLM stories addressing six challenging thoughts, shown separately for crowdworkers (left) and students (right). Error bars represent the standard error of the mean, with sample sizes indicated above each bar.}
    \label{fig:inspiring_solutions}
    \Description{Mean perceived capacity of inspiring solutions across Non-LLM and LLM stories addressing six challenging thoughts, shown separately for crowdworkers (left) and students (right). Error bars represent the standard error of the mean, with sample sizes indicated above each bar.}
\end{figure*}

\begin{figure*}[htbp]
    \centering
    \includegraphics[width=0.95\textwidth]{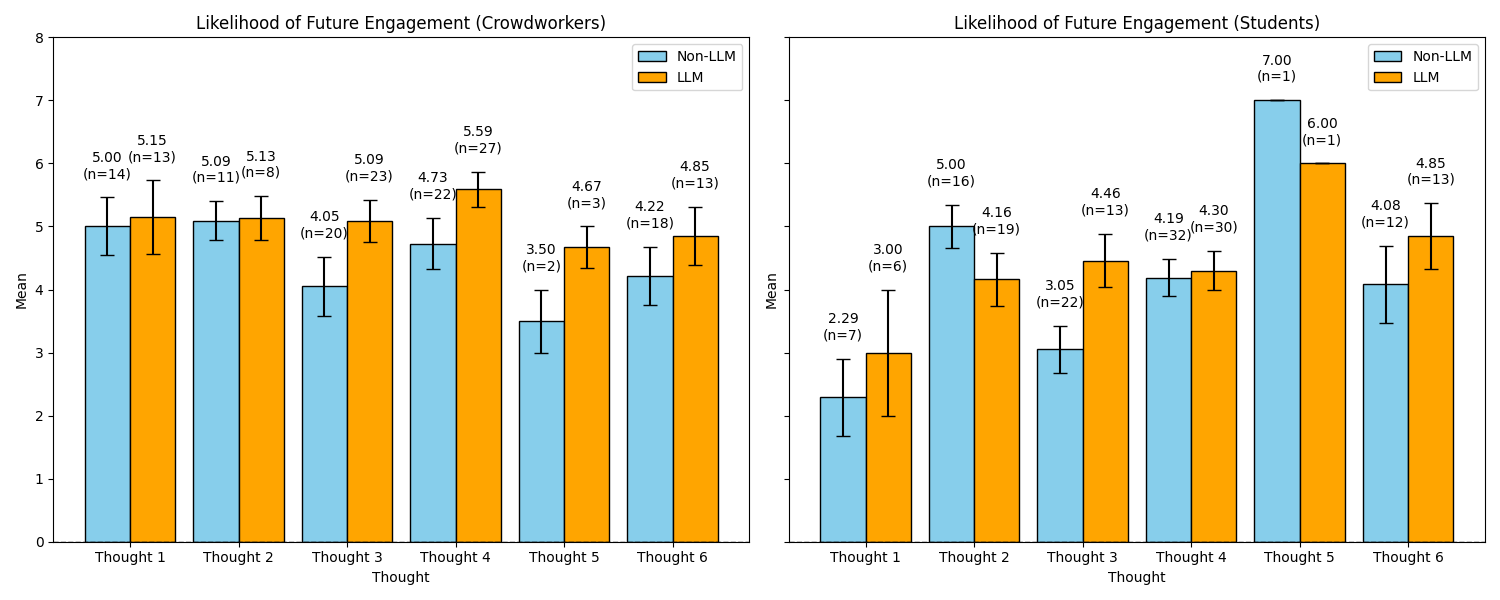}
    \caption{Mean likelihood of future engagement across Non-LLM and LLM stories addressing six challenging thoughts, shown separately for crowdworkers (left) and students (right). Error bars represent the standard error of the mean, with sample sizes indicated above each bar.}
    \label{fig:future_engagement}
    \Description{Mean likelihood of future engagement across Non-LLM and LLM stories addressing six challenging thoughts, shown separately for crowdworkers (left) and students (right). Error bars represent the standard error of the mean, with sample sizes indicated above each bar.}
\end{figure*}

\begin{figure*}[htbp]
    \centering
    \includegraphics[width=0.95\textwidth]{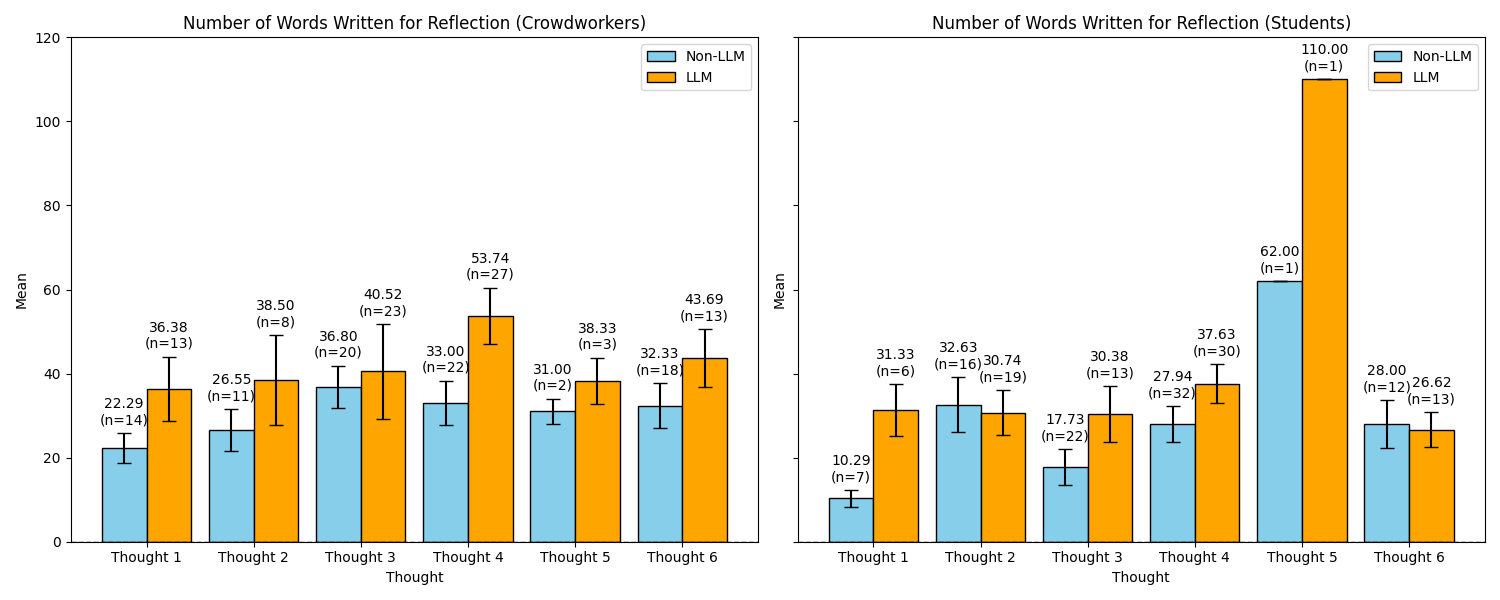}
    \caption{Mean number of words written in response to the reflection question across Non-LLM and LLM stories addressing six challenging thoughts, shown separately for crowdworkers (left) and students (right). Error bars represent the standard error of the mean, with sample sizes indicated above each bar.}
    \label{fig:reflection_word_count}
    \Description{Mean number of words written in response to the reflection question across Non-LLM and LLM stories addressing six challenging thoughts, shown separately for crowdworkers (left) and students (right). Error bars represent the standard error of the mean, with sample sizes indicated above each bar.}
\end{figure*}

In this section, we present participants' evaluations of stories from both the LLM and Non-LLM conditions, categorized by which of the six challenging thoughts they addressed. Figures \ref{fig:authenticity} through \ref{fig:balance} illustrate their qualities as narratives (measures related to RQ1), and Figures \ref{fig:change_negative_belief} through \ref{fig:reflection_word_count} focus on intervention outcomes (measures related to RQ2). Each figure displays results separately for the crowdworker and student groups, with error bars representing the standard error of the mean and sample sizes indicated above each bar. In each of the figures, the six challenging thoughts correspond to the following:

\begin{itemize}
    \item Thought 1: I always feel really down on myself.
    \item Thought 2: A lot of things could go wrong in the future.
    \item Thought 3: I am struggling more than I should.
    \item Thought 4: I haven’t achieved as much as other people have.
    \item Thought 5: The people I have counted on have let me down.
    \item Thought 6: I feel too shy or anxious when I’m in social situations.
\end{itemize}

%% file: Table/appendix-story-table.tex


\newcommand*{\tabindent}{\hspace{3mm}}

\begin{longtable}{|>{\raggedright\arraybackslash}p{0.1\textwidth}|p{0.2\textwidth}|p{0.30\textwidth}|p{0.30\textwidth}|}
    \captionsetup{justification=centering, singlelinecheck=true} 
    \caption{Non-LLM stories accompanied by LLM-enhanced versions} 
    \label{tab:appendix-story-table}
    \Description{Examples of Non-LLM stories and their LLM-enhanced counterparts. Each entry features a user description from various negative experience categories and the corresponding LLM-enhanced story generated.}
    \\
	\hline
	\textbf{User's Negative Thought}                        & \textbf{User's Provided Description} & \textbf{Non-LLM Story} & \textbf{LLM Story} \\
	\hline
	I always feel really down on myself
	                                                        &
	I feel disappointed in myself. There are many goals I set that I have not
	reached such as getting on the Dean's List scholar and making new friends
	in my first year. University has been a really tough transition for me
	since I moved abroad and I feel super lonely most of the time. I want to
	focus on my grades, but then I won't have enough time to make friends in
	class or through extracurricular activities. I am also quite introverted so it's hard for me to talk to new people. I envisioned a different university life. I just often feel down that none of my expectations are met.
	                                                        &
	I am really really hard on myself sometimes. I just keep fixating
	on and thinking about things I have done wrong.

	\bigskip

	But I am starting to be a bit more forgiving with myself because
	this is not how I want to feel and not how I want to treat myself. One of
	the ways I've been doing this is by telling myself a few positive
	affirmations every day. I'll tell myself over and over again things like
	``it's okay to make mistakes,'' ``it's okay not to feel happy,'' ``it's my right
	to be able to tell others no,'' or ``I am trying my best.''

	\bigskip

	I know it seems super silly and can feel a little awkward at times,
	but I've noticed that sometimes I'll say one of the affirmations in
	response to some of my bad thoughts about myself, so it's become more like
	a habit. They're just simple, true things that I sometimes lose sight of.

	\bigskip

	Being deliberate about the things we tell ourselves can make a
	difference in how we feel.

	\bigskip

	Please answer this reflective question: What simple, true
	things do you sometimes lose sight of?
	                                                        &
	Life hit hard when I moved to university, in a city miles away from
	home. Just like you, I was quite introverted and struggled with balancing
	academics and social life, multiply that with the pressure of getting on the
	Dean's list scholar. Loneliness became my consistent companion.

	\bigskip

	Then a friend told me about the power of positive affirmations. I
	dismissed it at first, but as I grew lonelier, I thought, why not give it a
	try? So, I started practicing. I would tell myself things like 'it's okay to
	make mistakes,' 'it's okay to feel lonely' or 'I'm not failing, I'm learning.'
	It felt awkward and silly in the beginning, but slowly, I noticed a shift in
	how I felt about myself.

	\bigskip

	Every time I started to feel down on myself, a self-affirmation would
	pop into my mind. I realized it was not about being perfect, but about focusing
	on my growth, my journey. This simple act of self-love still helps me during my
	bad days at university, and it's worth trying.

	\bigskip

	Please answer this reflective question: What kind phrases can
	you say to yourself during moments of self-disapproval or disappointment?
	\\
	\hline

	A lot of things could go wrong in the future
	                                                        &
	While things are fine in my life right now, I can't stop thinking about how
	a small incident could make me lose everything. I could make a mistake at
	work that gets me fired, and lose the ability to provide for my family. How
	will I feed myself or my daughter and make rent?
	                                                        &

	Currently, I am in the process of applying to grad school and I am
	worried about not getting in. I feel that if I don't get in, I will be very
	disappointed in myself. I find myself frequently going through various
	scenarios about all the things that could go wrong.

	\bigskip

	I know this is only hurting me, so I am trying to be more
	self-compassionate. I am trying to be as nice to myself as I would be to a
	friend who was applying to grad school.

	\bigskip

	If I were talking to a friend, I'd probably say something like, ``You
	are smart and successful already. I hope it works out, but if it doesn't
	you can reapply or find another path that will work for you.'' It's
	something I'm trying to remind myself so I can relax a little.

	\bigskip

	Self-compassion is all about giving the same kindness to ourselves we give to others.

    \bigskip
    
 	Please answer this reflective question: How can you be kind to
	yourself next time you're caught up in worry?
	                                                        &
	As I lay awake one night consumed by a whirlpool of thoughts about
	the potential problems I might face at work the next day, I suddenly
	realized how much I’d been tormenting myself with the 'what ifs'. I was so
	enveloped in the fear that a minor mistake could unravel everything, it
	took over my ability to live in the present.

	\bigskip

	That's when I decided to start treating myself with more
	compassion, as though I were a dear friend stuck in the same situation. I
	imagined what I'd say to them, 'You are competent and have tackled
	obstacles before, surely you can do it again. Even if the unforeseen
	happens, you'll find a way to pull through it.' I began repeating this to
	myself, becoming my own advocate.

	\bigskip

	Over time, I learned to cope better with my fears. Instead of
	showcasing my life's fragility, these anxieties became reminders that even
	if things go wrong, I have the capacity to bounce back.

	\bigskip

	Remember, self-compassion is giving the same kindness to ourselves
	that we extend to others. It's not an overnight transition, but eventually,
	it eases the burden of constant worry.
    
	\bigskip
    
	Please answer this reflective question: How can you apply
	self-compassion in your daily life when facing the fear of uncertain future
	outcomes?
	\\

	\hline

	I am struggling more than I should
	                                                        &
    I've been struggling financially, and it feels like I should have it figured out by now. I'm turning 25 this year, and it's hard to shake the feeling that I'm too old to still be barely making enough to cover rent. I wish I could be on top of my finances, instead of constantly worrying about money and just scraping by.
	                                                        &

	I have always struggled in life: struggled to fit in, struggled to
	sleep, struggled to read. It just seemed like in all the ways that people
	are successful, I struggled. I really started to feel down about my
	struggles in high school but it wasn't until my undergraduate years that I
	started seeing a professional.

	\bigskip

	I was lucky to be in school and to have counseling options available
	to me. I know there are some stigmas attached to seeing a professional, but
	I just needed someone unbiased, professional, and who could perhaps really
	help me understand myself better. The best thing that came out of that
	experience is that I continued to see professionals after college, whenever
	I felt stuck or was struggling, or just needed someone to listen. I don’t
	see a therapist every week anymore, but I know I can find help if I need
	it, and it's a relief.

	\bigskip

	Sometimes, we need to reach out to others to get a handle on
	things. Whether that's a professional or someone else.

    \bigskip
    
	Please answer this reflective question: Can you think of a time when reaching
	out to someone helped you break out of a rut?
	                                                        &
	As a bright-eyed 24-year-old, life felt like a jigsaw puzzle I was yet to solve, and finances, ironically the key corner piece. As days turned into weeks, and weeks into months, my wallet seemed to grow lighter, and the weight of my worries heavier. I was on the brink of turning 25 and the thought of barely making ends meet felt heavier than ever.

    \bigskip

    One day, out of the blue, I remembered how eager I was in college to learn and grow from my peers and professors. It was all about sharing and collaboration then. Why had I stopped reaching out? Was it fear? Was it the stigma? These questions pushed me beyond my comfort zone and I found myself venturing into the world of professional help.

    \bigskip

    I won't purport that overnight my troubles were resolved. But seeking help acted as a beacon of hope, gradually illuminating the darker corners of my finance woes. Years later, I still reach out to professionals when I find myself in a rut - not with the expectation of instant resolutions, but to gain perspective and feel understood.

    \bigskip

    Relying on others may sometimes make you feel vulnerable, but sometimes it is in this vulnerability that we find our greatest strengths. There's no shame in seeking help, after all, it's okay to not have all the pieces together. We're all a work in progress.

    \bigskip

    Please answer this reflective question: Does this narrative prompt you to reach out for professional help, or remind you of a time when doing so was the best decision you took?
	\\

	\hline

	I haven't achieved as much as other people have
	                                                        &
	I feel as though I haven't accomplished much compared to other 23 year
	olds, and it sometimes makes me feel horrible and depressed when I compare
	myself to those who live different lifestyles.
	                                                        &

	Ever since graduating high school 6 years ago and seeing everyone
	that I grew up with accomplish their goals and start to have their own
	families, I compare where my life is at in regards to where they're at. I
	tend to be really hard on myself and never feel like I'm doing well enough
	and am good enough.

	\bigskip

	When the pandemic started, I started working from home, and with
	places being closed, I was home 24/7. With no way of really ignoring my
	feelings, I hit an extremely low point. I made a Facebook post that was
	essentially a cry for help because I just didn't know what to do.
	Ultimately, I'm glad that I did.

    \bigskip

    A very dear friend of mine reached our privately to express he cares
	about me. I have been talking to my friend more often, who shared with me
	that he has gone through similar things as well. It's helped me see that
	acknowledging how I feel can help.

     \bigskip

    Our conversations have been a soothing balm, a beacon of shared
	experiences, providing comfort and perspective. So, remember, you're not
	alone in this tumultuously difficult journey. We all have stories, often
	unspoken, all challenging and unique.
 
 \bigskip
 
     If we only see the good sides of people's lives that they share,
	it's easy to forget how much other people struggle too.

	\bigskip

	Please answer this reflective question: What do you wish
	people would be more open about?

	                                                        &

	Six years post-high school and I kept grading myself by my friends'
	life charts, their flourishing careers, their growing families. I had
	painted a grueling picture of my life, always falling short of what they
	had achieved.

	\bigskip

	When the world quietened due to the pandemic, and I was confined at
	home, these comparisons became deafening echoes in my solitude. One harsh
	day, I succumbed to it and shared an undeclared plea for help on Facebook,
	a confession of my perceived underachievement. I was at the edge of
	despair.

	\bigskip

	Then occurred a shift I hadn't anticipated. 

    \bigskip

       A dear friend reached
	out with kind words, expressing his concern and shared his battles with
	similar circumstances. His candidness evoked an undeniable truth - our
	social media lives are mostly highlight reels, the failures and struggles
	often invisible. We, unknowingly, harshly measure our entire lives to
	others' well-curated feeds.
 
 \bigskip
 
	Please answer this reflective question: Have you reached out to
	a friend when you felt overwhelmed with self-doubt and how did that
	conversation make you feel about your own journey?

	\\
	\hline

	The people I have counted on have let me down
	                                                        &
    I feel really let down by my partner. I thought I could count on them, but when I needed them most, they weren't there. It's hard to shake the feeling of betrayal, like I don’t matter as much to them as they do to me.
	                                                        &
	I had to become independent at a very young age as I realized there
	was no one I could depend on. This has made it hard for my partners in
	relationships, as I don't know how to really let people in. I want to, but
	I've seen what happens when you let someone in. It can be devastating. The
	first time I completely trusted and depended on someone, I was left
	shattered for 7 years.

	\bigskip

    	Over that time, I suffered from chronic depression. I pushed
	people away in an attempt to protect myself.

\bigskip

	I have come to accept that we can't hate people for not sticking
	with us when they want to do something else. Negative experiences are a
	part of life. We can either let them swallow us, or we can push forward. I
	still cry sometimes, and some situations get the best of me.

	\bigskip

	There will be hard times, but you've gotten through them before and
	will again.

	\bigskip

	Please answer this reflective question: What helps you get through
	it when people let you down?

&
	
    I was thrust into independence at an early age, having realized that relying on others often led to disappointment. This developed a hardened shell around me, especially in romantic relationships where the fear of vulnerability reigned. I yearned for connection, yet the thought of opening up was paralyzing. Once I did open up, the outcome was a heartbreak that took me seven years to recover from.

    During those years, a shroud of chronic depression enveloped me. I pushed everyone away, building an even thicker protective barrier around myself.
    
    \bigskip
    
    Over time, acceptance found its way into my thought process. 

     \bigskip   
    
    I acknowledged that life is a blend of positives and negatives, and people's decisions are governed by their own desires. We can't bind them by our expectations. Yes, there are moments of weakness when I still surrender to tears.

    \bigskip

     But with every sunrise, there's an opportunity to pick myself up and march on. Sometimes, surviving the whole day seems daunting, but taking it one-minute at a time is manageable. Through all life's trials and tribulations, remember this: you've faced hard times before and emerged victorious.
     
     \bigskip

    Please answer this reflective question: What strategies do you find useful to cope when you feel let down by those you count on?
   
	\\
	\hline

	I feel too shy or anxious when I'm in social situations &

    I feel overwhelmed and awkward in social situations, like I don't know how to fit in or what to say. The anxiety just takes over, and I end up feeling invisible or out of place.

    \bigskip
    
     It’s exhausting, and I wish I could relax and be myself around others.
	                                                        &
	I am a shy person. It's part of who I am. I sometimes feel bad about
	this, like I have lost out on experiences because of being shy and not
	showing my true self to the world.

    While I know I am as worthy of having a
	voice as anyone else, there's this worry in my head that others will judge me.
 
    \bigskip
 
	What has helped me is to argue against my worries when I have them.
	I try to push back when the voice in my head says ``they're judging you''.
	
 \bigskip
    
    Plus, I have a lot of positive qualities. When people get to know me, they usually like me. My worries are not based on the reality of how
    people think. Remembering this has helped me feel a little more confident
    when I'm meeting someone new.
 
    \bigskip
    
    It's easy to forget that other people are often pretty focused on
	themselves, and don't spend as much time noticing and judging us as we
	might think.

    \bigskip

 	Please answer this reflective question: What strategies work for
	you when you feel shy or socially anxious?

	              &
    	I've often found myself tangled in the webs of social anxiety, like a silent character adrift in a sea of voices. There's this persistent whisper in my psyche, convincing me that everyone in the room is scrutinizing each slip, each stumble, and each stutter. 
	    \bigskip
 
   It’s a burden, and all I yearn for is to let go, to be the authentic ``me'' amidst the crowd.
    \bigskip
    
    I found solace in challenging this distressing voice in my head. Every time it chanted ``they're evaluating you'', I would confront it, argue against it.

    \bigskip

    I started to realize that people seldom ponder about my awkwardness as much as I imagine. Let's face it, we each have our own worlds to worry about.

    \bigskip

    Interestingly, I’ve come to embrace my introverted self as part of my identity. I’ve learned that this part of me possesses its own unique charm, and as people venture closer, they often appreciate my depth and catalogue of thought. This understanding has fostered a spark of confidence when meeting new faces.
    
    \bigskip
    
    In the vast universe of their own thoughts and worries, people rarely fixate on our quirks or mishaps. Embracing this fact has brought me a step closer to accepting my social struggles and actively working on overcoming them.

	\bigskip

    Please answer this reflective question: How do you combat your worries when you find yourself feeling socially anxious or shy? Are there certain affirmations or thought patterns that seem to help you most?
	\\
	\hline
\end{longtable}
\twocolumn

%% file: main.bbl

\begin{thebibliography}{117}


\ifx \showCODEN    \undefined \def \showCODEN     #1{\unskip}     \fi
\ifx \showDOI      \undefined \def \showDOI       #1{#1}\fi
\ifx \showISBNx    \undefined \def \showISBNx     #1{\unskip}     \fi
\ifx \showISBNxiii \undefined \def \showISBNxiii  #1{\unskip}     \fi
\ifx \showISSN     \undefined \def \showISSN      #1{\unskip}     \fi
\ifx \showLCCN     \undefined \def \showLCCN      #1{\unskip}     \fi
\ifx \shownote     \undefined \def \shownote      #1{#1}          \fi
\ifx \showarticletitle \undefined \def \showarticletitle #1{#1}   \fi
\ifx \showURL      \undefined \def \showURL       {\relax}        \fi
\providecommand\bibfield[2]{#2}
\providecommand\bibinfo[2]{#2}
\providecommand\natexlab[1]{#1}
\providecommand\showeprint[2][]{arXiv:#2}

\bibitem[Abd-Alrazaq et~al\mbox{.}(2023)]%
        {abd2023large}
\bibfield{author}{\bibinfo{person}{Alaa Abd-Alrazaq}, \bibinfo{person}{Rawan AlSaad}, \bibinfo{person}{Dari Alhuwail}, \bibinfo{person}{Arfan Ahmed}, \bibinfo{person}{Padraig~Mark Healy}, \bibinfo{person}{Syed Latifi}, \bibinfo{person}{Sarah Aziz}, \bibinfo{person}{Rafat Damseh}, \bibinfo{person}{Sadam~Alabed Alrazak}, \bibinfo{person}{Javaid Sheikh}, {et~al\mbox{.}}} \bibinfo{year}{2023}\natexlab{}.
\newblock \showarticletitle{Large language models in medical education: opportunities, challenges, and future directions}.
\newblock \bibinfo{journal}{\emph{JMIR Medical Education}} \bibinfo{volume}{9}, \bibinfo{number}{1} (\bibinfo{year}{2023}), \bibinfo{pages}{e48291}.
\newblock


\bibitem[Abd-Alrazaq et~al\mbox{.}(2019)]%
        {abd2019overview}
\bibfield{author}{\bibinfo{person}{Alaa~A Abd-Alrazaq}, \bibinfo{person}{Mohannad Alajlani}, \bibinfo{person}{Ali~Abdallah Alalwan}, \bibinfo{person}{Bridgette~M Bewick}, \bibinfo{person}{Peter Gardner}, {and} \bibinfo{person}{Mowafa Househ}.} \bibinfo{year}{2019}\natexlab{}.
\newblock \showarticletitle{An overview of the features of chatbots in mental health: A scoping review}.
\newblock \bibinfo{journal}{\emph{International journal of medical informatics}}  \bibinfo{volume}{132} (\bibinfo{year}{2019}), \bibinfo{pages}{103978}.
\newblock


\bibitem[Adler et~al\mbox{.}(2021)]%
        {adler2021identifying}
\bibfield{author}{\bibinfo{person}{Daniel~A Adler}, \bibinfo{person}{Vincent W-S Tseng}, \bibinfo{person}{Gengmo Qi}, \bibinfo{person}{Joseph Scarpa}, \bibinfo{person}{Srijan Sen}, {and} \bibinfo{person}{Tanzeem Choudhury}.} \bibinfo{year}{2021}\natexlab{}.
\newblock \showarticletitle{Identifying mobile sensing indicators of stress-resilience}.
\newblock \bibinfo{journal}{\emph{Proceedings of the ACM on interactive, mobile, wearable and ubiquitous technologies}} \bibinfo{volume}{5}, \bibinfo{number}{2} (\bibinfo{year}{2021}), \bibinfo{pages}{1--32}.
\newblock


\bibitem[Aesop(1998)]%
        {aesopfable}
\bibfield{author}{\bibinfo{person}{Aesop}.} \bibinfo{year}{1998}\natexlab{}.
\newblock \bibinfo{booktitle}{\emph{Aesop: The Complete Fables}}.
\newblock \bibinfo{publisher}{Penguin Classics. Penguin Books, London, UK}.
\newblock


\bibitem[Aguilera et~al\mbox{.}(2020)]%
        {aguilera2020mhealth}
\bibfield{author}{\bibinfo{person}{Adrian Aguilera}, \bibinfo{person}{Caroline~A Figueroa}, \bibinfo{person}{Rosa Hernandez-Ramos}, \bibinfo{person}{Urmimala Sarkar}, \bibinfo{person}{Anupama Cemballi}, \bibinfo{person}{Laura Gomez-Pathak}, \bibinfo{person}{Jose Miramontes}, \bibinfo{person}{Elad Yom-Tov}, \bibinfo{person}{Bibhas Chakraborty}, \bibinfo{person}{Xiaoxi Yan}, {et~al\mbox{.}}} \bibinfo{year}{2020}\natexlab{}.
\newblock \showarticletitle{mHealth app using machine learning to increase physical activity in diabetes and depression: clinical trial protocol for the DIAMANTE Study}.
\newblock \bibinfo{journal}{\emph{BMJ open}} \bibinfo{volume}{10}, \bibinfo{number}{8} (\bibinfo{year}{2020}), \bibinfo{pages}{e034723}.
\newblock


\bibitem[Akram et~al\mbox{.}(2020)]%
        {akram2020exploratory}
\bibfield{author}{\bibinfo{person}{Umair Akram}, \bibinfo{person}{Jennifer Drabble}, \bibinfo{person}{Glhenda Cau}, \bibinfo{person}{Frayer Hershaw}, \bibinfo{person}{Ashileen Rajenthran}, \bibinfo{person}{Mollie Lowe}, \bibinfo{person}{Carissa Trommelen}, {and} \bibinfo{person}{Jason~G Ellis}.} \bibinfo{year}{2020}\natexlab{}.
\newblock \showarticletitle{Exploratory study on the role of emotion regulation in perceived valence, humour, and beneficial use of depressive internet memes in depression}.
\newblock \bibinfo{journal}{\emph{Scientific reports}} \bibinfo{volume}{10}, \bibinfo{number}{1} (\bibinfo{year}{2020}), \bibinfo{pages}{899}.
\newblock


\bibitem[Allen and Leary(2010)]%
        {allen2010self}
\bibfield{author}{\bibinfo{person}{Ashley~Batts Allen} {and} \bibinfo{person}{Mark~R Leary}.} \bibinfo{year}{2010}\natexlab{}.
\newblock \showarticletitle{Self-Compassion, stress, and coping}.
\newblock \bibinfo{journal}{\emph{Social and personality psychology compass}} \bibinfo{volume}{4}, \bibinfo{number}{2} (\bibinfo{year}{2010}), \bibinfo{pages}{107--118}.
\newblock


\bibitem[Anderson and Wallace(2015)]%
        {anderson2015digital}
\bibfield{author}{\bibinfo{person}{Kim Anderson} {and} \bibinfo{person}{Beatriz Wallace}.} \bibinfo{year}{2015}\natexlab{}.
\newblock \showarticletitle{Digital storytelling as a trauma narrative intervention for children exposed to domestic violence}.
\newblock \bibinfo{journal}{\emph{Film and video-based therapy}} (\bibinfo{year}{2015}), \bibinfo{pages}{95--107}.
\newblock


\bibitem[Ardenghi et~al\mbox{.}(2024)]%
        {ardenghi2024supporting}
\bibfield{author}{\bibinfo{person}{Stefano Ardenghi}, \bibinfo{person}{Selena Russo}, \bibinfo{person}{Marco Bani}, \bibinfo{person}{Giulia Rampoldi}, {and} \bibinfo{person}{Maria~Grazia Strepparava}.} \bibinfo{year}{2024}\natexlab{}.
\newblock \showarticletitle{Supporting students with empathy: the association between empathy and coping strategies in pre-clinical medical students}.
\newblock \bibinfo{journal}{\emph{Current Psychology}} \bibinfo{volume}{43}, \bibinfo{number}{2} (\bibinfo{year}{2024}), \bibinfo{pages}{1879--1889}.
\newblock


\bibitem[Ariyaratne et~al\mbox{.}(2023)]%
        {ariyaratne2023comparison}
\bibfield{author}{\bibinfo{person}{Sisith Ariyaratne}, \bibinfo{person}{Karthikeyan~P Iyengar}, \bibinfo{person}{Neha Nischal}, \bibinfo{person}{Naparla Chitti~Babu}, {and} \bibinfo{person}{Rajesh Botchu}.} \bibinfo{year}{2023}\natexlab{}.
\newblock \showarticletitle{A comparison of ChatGPT-generated articles with human-written articles}.
\newblock \bibinfo{journal}{\emph{Skeletal radiology}} \bibinfo{volume}{52}, \bibinfo{number}{9} (\bibinfo{year}{2023}), \bibinfo{pages}{1755--1758}.
\newblock


\bibitem[Baumel et~al\mbox{.}(2019)]%
        {baumel2019objective}
\bibfield{author}{\bibinfo{person}{Amit Baumel}, \bibinfo{person}{Frederick Muench}, \bibinfo{person}{Stav Edan}, {and} \bibinfo{person}{John~M Kane}.} \bibinfo{year}{2019}\natexlab{}.
\newblock \showarticletitle{Objective user engagement with mental health apps: systematic search and panel-based usage analysis}.
\newblock \bibinfo{journal}{\emph{Journal of medical Internet research}} \bibinfo{volume}{21}, \bibinfo{number}{9} (\bibinfo{year}{2019}), \bibinfo{pages}{e14567}.
\newblock


\bibitem[Baumel et~al\mbox{.}(2018)]%
        {baumel2018digital}
\bibfield{author}{\bibinfo{person}{Amit Baumel}, \bibinfo{person}{Amanda Tinkelman}, \bibinfo{person}{Nandita Mathur}, {and} \bibinfo{person}{John~M Kane}.} \bibinfo{year}{2018}\natexlab{}.
\newblock \showarticletitle{Digital peer-support platform (7Cups) as an adjunct treatment for women with postpartum depression: feasibility, acceptability, and preliminary efficacy study}.
\newblock \bibinfo{journal}{\emph{JMIR mHealth and uHealth}} \bibinfo{volume}{6}, \bibinfo{number}{2} (\bibinfo{year}{2018}), \bibinfo{pages}{e9482}.
\newblock


\bibitem[Bhattacharjee et~al\mbox{.}(2024a)]%
        {bhattacharjee2024exploring}
\bibfield{author}{\bibinfo{person}{Ananya Bhattacharjee}, \bibinfo{person}{Pan Chen}, \bibinfo{person}{Abhijoy Mandal}, \bibinfo{person}{Anne Hsu}, \bibinfo{person}{Katie O'Leary}, \bibinfo{person}{Alex Mariakakis}, \bibinfo{person}{Joseph~Jay Williams}, {et~al\mbox{.}}} \bibinfo{year}{2024}\natexlab{a}.
\newblock \showarticletitle{Exploring User Perspectives on Brief Reflective Questioning Activities for Stress Management: Mixed Methods Study}.
\newblock \bibinfo{journal}{\emph{JMIR Formative Research}} \bibinfo{volume}{8}, \bibinfo{number}{1} (\bibinfo{year}{2024}), \bibinfo{pages}{e47360}.
\newblock


\bibitem[Bhattacharjee et~al\mbox{.}(2024b)]%
        {bhattacharjee2024actually}
\bibfield{author}{\bibinfo{person}{Ananya Bhattacharjee}, \bibinfo{person}{Zichen Gong}, \bibinfo{person}{Bingcheng Wang}, \bibinfo{person}{Timothy~James Luckcock}, \bibinfo{person}{Emma Watson}, \bibinfo{person}{Elena~Allica Abellan}, \bibinfo{person}{Leslie Gutman}, \bibinfo{person}{Anne Hsu}, {and} \bibinfo{person}{Joseph~Jay Williams}.} \bibinfo{year}{2024}\natexlab{b}.
\newblock \showarticletitle{" Actually I Can Count My Blessings": User-Centered Design of an Application to Promote Gratitude Among Young Adults}.
\newblock \bibinfo{journal}{\emph{Proceedings of the ACM on Human-Computer Interaction}} \bibinfo{volume}{8}, \bibinfo{number}{CSCW2} (\bibinfo{year}{2024}), \bibinfo{pages}{1--29}.
\newblock


\bibitem[Bhattacharjee et~al\mbox{.}(2023a)]%
        {bhattacharjee2023design}
\bibfield{author}{\bibinfo{person}{Ananya Bhattacharjee}, \bibinfo{person}{Jiyau Pang}, \bibinfo{person}{Angelina Liu}, \bibinfo{person}{Alex Mariakakis}, {and} \bibinfo{person}{Joseph~Jay Williams}.} \bibinfo{year}{2023}\natexlab{a}.
\newblock \showarticletitle{Design implications for one-way text messaging services that support psychological wellbeing}.
\newblock \bibinfo{journal}{\emph{ACM Transactions on Computer-Human Interaction}} \bibinfo{volume}{30}, \bibinfo{number}{3} (\bibinfo{year}{2023}), \bibinfo{pages}{1--29}.
\newblock


\bibitem[Bhattacharjee et~al\mbox{.}(2022)]%
        {bhattacharjee2022kind}
\bibfield{author}{\bibinfo{person}{Ananya Bhattacharjee}, \bibinfo{person}{Joseph~Jay Williams}, \bibinfo{person}{Karrie Chou}, \bibinfo{person}{Justice Tomlinson}, \bibinfo{person}{Jonah Meyerhoff}, \bibinfo{person}{Alex Mariakakis}, {and} \bibinfo{person}{Rachel Kornfield}.} \bibinfo{year}{2022}\natexlab{}.
\newblock \showarticletitle{" I kind of bounce off it": translating mental health principles into real life through story-based text messages}.
\newblock \bibinfo{journal}{\emph{Proceedings of the ACM on Human-computer Interaction}} \bibinfo{volume}{6}, \bibinfo{number}{CSCW2} (\bibinfo{year}{2022}), \bibinfo{pages}{1--31}.
\newblock


\bibitem[Bhattacharjee et~al\mbox{.}(2023b)]%
        {bhattacharjee2023investigating}
\bibfield{author}{\bibinfo{person}{Ananya Bhattacharjee}, \bibinfo{person}{Joseph~Jay Williams}, \bibinfo{person}{Jonah Meyerhoff}, \bibinfo{person}{Harsh Kumar}, \bibinfo{person}{Alex Mariakakis}, {and} \bibinfo{person}{Rachel Kornfield}.} \bibinfo{year}{2023}\natexlab{b}.
\newblock \showarticletitle{Investigating the role of context in the delivery of text messages for supporting psychological wellbeing}. In \bibinfo{booktitle}{\emph{Proceedings of the 2023 CHI Conference on Human Factors in Computing Systems}}. \bibinfo{pages}{1--19}.
\newblock


\bibitem[Bhattacharjee et~al\mbox{.}(2024c)]%
        {bhattacharjee2024understanding}
\bibfield{author}{\bibinfo{person}{Ananya Bhattacharjee}, \bibinfo{person}{Yuchen Zeng}, \bibinfo{person}{Sarah~Yi Xu}, \bibinfo{person}{Dana Kulzhabayeva}, \bibinfo{person}{Minyi Ma}, \bibinfo{person}{Rachel Kornfield}, \bibinfo{person}{Syed~Ishtiaque Ahmed}, \bibinfo{person}{Alex Mariakakis}, \bibinfo{person}{Mary~P Czerwinski}, \bibinfo{person}{Anastasia Kuzminykh}, {et~al\mbox{.}}} \bibinfo{year}{2024}\natexlab{c}.
\newblock \showarticletitle{Understanding the Role of Large Language Models in Personalizing and Scaffolding Strategies to Combat Academic Procrastination}. In \bibinfo{booktitle}{\emph{Proceedings of the CHI Conference on Human Factors in Computing Systems}}. \bibinfo{pages}{1--18}.
\newblock


\bibitem[Bilandzic and Busselle(2013)]%
        {bilandzic2013narrative}
\bibfield{author}{\bibinfo{person}{Helena Bilandzic} {and} \bibinfo{person}{Rick Busselle}.} \bibinfo{year}{2013}\natexlab{}.
\newblock \showarticletitle{Narrative persuasion}.
\newblock \bibinfo{journal}{\emph{The SAGE handbook of persuasion: Developments in theory and practice}}  \bibinfo{volume}{2} (\bibinfo{year}{2013}), \bibinfo{pages}{200--219}.
\newblock


\bibitem[Burgess et~al\mbox{.}(2019)]%
        {burgess2019think}
\bibfield{author}{\bibinfo{person}{Eleanor~R Burgess}, \bibinfo{person}{Kathryn~E Ringland}, \bibinfo{person}{Jennifer Nicholas}, \bibinfo{person}{Ashley~A Knapp}, \bibinfo{person}{Jordan Eschler}, \bibinfo{person}{David~C Mohr}, {and} \bibinfo{person}{Madhu~C Reddy}.} \bibinfo{year}{2019}\natexlab{}.
\newblock \showarticletitle{" I think people are powerful" The Sociality of Individuals Managing Depression}.
\newblock \bibinfo{journal}{\emph{Proceedings of the ACM on Human-Computer Interaction}} \bibinfo{volume}{3}, \bibinfo{number}{CSCW} (\bibinfo{year}{2019}), \bibinfo{pages}{1--29}.
\newblock


\bibitem[Carpenter et~al\mbox{.}(2016)]%
        {carpenter2016seeing}
\bibfield{author}{\bibinfo{person}{Jordan Carpenter}, \bibinfo{person}{Patrick Crutchley}, \bibinfo{person}{Ran~D Zilca}, \bibinfo{person}{H~Andrew Schwartz}, \bibinfo{person}{Laura~K Smith}, \bibinfo{person}{Angela~M Cobb}, {and} \bibinfo{person}{Acacia~C Parks}.} \bibinfo{year}{2016}\natexlab{}.
\newblock \showarticletitle{Seeing the “big” picture: big data methods for exploring relationships between usage, language, and outcome in internet intervention data}.
\newblock \bibinfo{journal}{\emph{Journal of Medical Internet Research}} \bibinfo{volume}{18}, \bibinfo{number}{8} (\bibinfo{year}{2016}), \bibinfo{pages}{e241}.
\newblock


\bibitem[Chan et~al\mbox{.}(2023)]%
        {chan2023mango}
\bibfield{author}{\bibinfo{person}{Szeyi Chan}, \bibinfo{person}{Jiachen Li}, \bibinfo{person}{Bingsheng Yao}, \bibinfo{person}{Amama Mahmood}, \bibinfo{person}{Chien-Ming Huang}, \bibinfo{person}{Holly Jimison}, \bibinfo{person}{Elizabeth~D Mynatt}, {and} \bibinfo{person}{Dakuo Wang}.} \bibinfo{year}{2023}\natexlab{}.
\newblock \showarticletitle{" Mango Mango, How to Let The Lettuce Dry Without A Spinner?'': Exploring User Perceptions of Using An LLM-Based Conversational Assistant Toward Cooking Partner}.
\newblock \bibinfo{journal}{\emph{arXiv preprint arXiv:2310.05853}} (\bibinfo{year}{2023}).
\newblock


\bibitem[Chen et~al\mbox{.}(2021)]%
        {chen2021scaffolding}
\bibfield{author}{\bibinfo{person}{Tianying Chen}, \bibinfo{person}{Kristy Zhang}, \bibinfo{person}{Robert~E Kraut}, {and} \bibinfo{person}{Laura Dabbish}.} \bibinfo{year}{2021}\natexlab{}.
\newblock \showarticletitle{Scaffolding the online peer-support experience: novice supporters' strategies and challenges}.
\newblock \bibinfo{journal}{\emph{Proceedings of the ACM on Human-Computer Interaction}} \bibinfo{volume}{5}, \bibinfo{number}{CSCW2} (\bibinfo{year}{2021}), \bibinfo{pages}{1--30}.
\newblock


\bibitem[Cheng et~al\mbox{.}(2022)]%
        {cheng2022human}
\bibfield{author}{\bibinfo{person}{Xusen Cheng}, \bibinfo{person}{Xiaoping Zhang}, \bibinfo{person}{Jason Cohen}, {and} \bibinfo{person}{Jian Mou}.} \bibinfo{year}{2022}\natexlab{}.
\newblock \showarticletitle{Human vs. AI: Understanding the impact of anthropomorphism on consumer response to chatbots from the perspective of trust and relationship norms}.
\newblock \bibinfo{journal}{\emph{Information Processing \& Management}} \bibinfo{volume}{59}, \bibinfo{number}{3} (\bibinfo{year}{2022}), \bibinfo{pages}{102940}.
\newblock


\bibitem[Clark(2013)]%
        {clark2013cognitive}
\bibfield{author}{\bibinfo{person}{David~A Clark}.} \bibinfo{year}{2013}\natexlab{}.
\newblock \showarticletitle{Cognitive restructuring}.
\newblock \bibinfo{journal}{\emph{The Wiley handbook of cognitive behavioral therapy}} (\bibinfo{year}{2013}), \bibinfo{pages}{1--22}.
\newblock


\bibitem[Clarke and Braun(2017)]%
        {clarke2017thematic}
\bibfield{author}{\bibinfo{person}{Victoria Clarke} {and} \bibinfo{person}{Virginia Braun}.} \bibinfo{year}{2017}\natexlab{}.
\newblock \showarticletitle{Thematic analysis}.
\newblock \bibinfo{journal}{\emph{The journal of positive psychology}} \bibinfo{volume}{12}, \bibinfo{number}{3} (\bibinfo{year}{2017}), \bibinfo{pages}{297--298}.
\newblock


\bibitem[Collins et~al\mbox{.}(2022)]%
        {collins2022covid}
\bibfield{author}{\bibinfo{person}{Christopher Collins}, \bibinfo{person}{Simone Arbour}, \bibinfo{person}{Nathan Beals}, \bibinfo{person}{Shawn Yama}, \bibinfo{person}{Jennifer Laffier}, {and} \bibinfo{person}{Zixin Zhao}.} \bibinfo{year}{2022}\natexlab{}.
\newblock \showarticletitle{Covid connect: Chat-driven anonymous story-sharing for peer support}. In \bibinfo{booktitle}{\emph{Proceedings of the 2022 ACM Designing Interactive Systems Conference}}. \bibinfo{pages}{301--318}.
\newblock


\bibitem[Davis et~al\mbox{.}(2020)]%
        {davis2020understand}
\bibfield{author}{\bibinfo{person}{Hilary Davis}, \bibinfo{person}{Darren~C Fisher}, {and} \bibinfo{person}{Ivana Randjelovic}.} \bibinfo{year}{2020}\natexlab{}.
\newblock \showarticletitle{" I Understand, Mate" A Co-designed Comic-based Digital Story from'Down Under'}. In \bibinfo{booktitle}{\emph{Proceedings of the 2020 ACM Designing Interactive Systems Conference}}. \bibinfo{pages}{243--254}.
\newblock


\bibitem[Delgado(1990)]%
        {delgado1990story}
\bibfield{author}{\bibinfo{person}{Richard Delgado}.} \bibinfo{year}{1990}\natexlab{}.
\newblock \showarticletitle{When a story is just a story: Does voice really matter?}
\newblock \bibinfo{journal}{\emph{Virginia Law Review}} (\bibinfo{year}{1990}), \bibinfo{pages}{95--111}.
\newblock


\bibitem[Demszky et~al\mbox{.}(2023)]%
        {demszky2023using}
\bibfield{author}{\bibinfo{person}{Dorottya Demszky}, \bibinfo{person}{Diyi Yang}, \bibinfo{person}{David~S Yeager}, \bibinfo{person}{Christopher~J Bryan}, \bibinfo{person}{Margarett Clapper}, \bibinfo{person}{Susannah Chandhok}, \bibinfo{person}{Johannes~C Eichstaedt}, \bibinfo{person}{Cameron Hecht}, \bibinfo{person}{Jeremy Jamieson}, \bibinfo{person}{Meghann Johnson}, {et~al\mbox{.}}} \bibinfo{year}{2023}\natexlab{}.
\newblock \showarticletitle{Using large language models in psychology}.
\newblock \bibinfo{journal}{\emph{Nature Reviews Psychology}} \bibinfo{volume}{2}, \bibinfo{number}{11} (\bibinfo{year}{2023}), \bibinfo{pages}{688--701}.
\newblock


\bibitem[Denny et~al\mbox{.}(2023)]%
        {denny2023conversing}
\bibfield{author}{\bibinfo{person}{Paul Denny}, \bibinfo{person}{Viraj Kumar}, {and} \bibinfo{person}{Nasser Giacaman}.} \bibinfo{year}{2023}\natexlab{}.
\newblock \showarticletitle{Conversing with copilot: Exploring prompt engineering for solving cs1 problems using natural language}. In \bibinfo{booktitle}{\emph{Proceedings of the 54th ACM Technical Symposium on Computer Science Education V. 1}}. \bibinfo{pages}{1136--1142}.
\newblock


\bibitem[Dimond et~al\mbox{.}(2013)]%
        {dimond2013hollaback}
\bibfield{author}{\bibinfo{person}{Jill~P Dimond}, \bibinfo{person}{Michaelanne Dye}, \bibinfo{person}{Daphne LaRose}, {and} \bibinfo{person}{Amy~S Bruckman}.} \bibinfo{year}{2013}\natexlab{}.
\newblock \showarticletitle{Hollaback! The role of storytelling online in a social movement organization}. In \bibinfo{booktitle}{\emph{Proceedings of the 2013 conference on Computer supported cooperative work}}. \bibinfo{pages}{477--490}.
\newblock


\bibitem[Dyer(1982)]%
        {dyer1982depth}
\bibfield{author}{\bibinfo{person}{Michael~George Dyer}.} \bibinfo{year}{1982}\natexlab{}.
\newblock \bibinfo{booktitle}{\emph{In-Depth Understanding. A Computer Model of Integrated Processing for Narrative Comprehension.}}
\newblock \bibinfo{type}{{T}echnical {R}eport}. \bibinfo{institution}{Yale University, New Haven, Connecticut}.
\newblock


\bibitem[Dym et~al\mbox{.}(2019)]%
        {dym2019coming}
\bibfield{author}{\bibinfo{person}{Brianna Dym}, \bibinfo{person}{Jed~R Brubaker}, \bibinfo{person}{Casey Fiesler}, {and} \bibinfo{person}{Bryan Semaan}.} \bibinfo{year}{2019}\natexlab{}.
\newblock \showarticletitle{" Coming Out Okay" Community Narratives for LGBTQ Identity Recovery Work}.
\newblock \bibinfo{journal}{\emph{Proceedings of the ACM on Human-Computer Interaction}} \bibinfo{volume}{3}, \bibinfo{number}{CSCW} (\bibinfo{year}{2019}), \bibinfo{pages}{1--28}.
\newblock


\bibitem[Ekin(2023)]%
        {ekin2023prompt}
\bibfield{author}{\bibinfo{person}{Sabit Ekin}.} \bibinfo{year}{2023}\natexlab{}.
\newblock \showarticletitle{Prompt engineering for ChatGPT: a quick guide to techniques, tips, and best practices}.
\newblock \bibinfo{journal}{\emph{Authorea Preprints}} (\bibinfo{year}{2023}).
\newblock


\bibitem[Evans-Lacko et~al\mbox{.}(2014)]%
        {evans2014effect}
\bibfield{author}{\bibinfo{person}{Sara Evans-Lacko}, \bibinfo{person}{Elizabeth Corker}, \bibinfo{person}{Paul Williams}, \bibinfo{person}{Claire Henderson}, {and} \bibinfo{person}{Graham Thornicroft}.} \bibinfo{year}{2014}\natexlab{}.
\newblock \showarticletitle{Effect of the Time to Change anti-stigma campaign on trends in mental-illness-related public stigma among the English population in 2003--13: an analysis of survey data}.
\newblock \bibinfo{journal}{\emph{The Lancet Psychiatry}} \bibinfo{volume}{1}, \bibinfo{number}{2} (\bibinfo{year}{2014}), \bibinfo{pages}{121--128}.
\newblock


\bibitem[Ferguson and Musheno(2000)]%
        {ferguson2000teaching}
\bibfield{author}{\bibinfo{person}{Jennifer~L Ferguson} {and} \bibinfo{person}{Michael Musheno}.} \bibinfo{year}{2000}\natexlab{}.
\newblock \showarticletitle{Teaching with stories: Engaging students in critical self-reflection about policing and in/justice}.
\newblock \bibinfo{journal}{\emph{Journal of Criminal Justice Education}} \bibinfo{volume}{11}, \bibinfo{number}{1} (\bibinfo{year}{2000}), \bibinfo{pages}{149--165}.
\newblock


\bibitem[Fitzpatrick et~al\mbox{.}(2017)]%
        {fitzpatrick2017delivering}
\bibfield{author}{\bibinfo{person}{Kathleen~Kara Fitzpatrick}, \bibinfo{person}{Alison Darcy}, {and} \bibinfo{person}{Molly Vierhile}.} \bibinfo{year}{2017}\natexlab{}.
\newblock \showarticletitle{Delivering cognitive behavior therapy to young adults with symptoms of depression and anxiety using a fully automated conversational agent (Woebot): a randomized controlled trial}.
\newblock \bibinfo{journal}{\emph{JMIR mental health}} \bibinfo{volume}{4}, \bibinfo{number}{2} (\bibinfo{year}{2017}), \bibinfo{pages}{e7785}.
\newblock


\bibitem[Garrig{\'o}s et~al\mbox{.}(2003)]%
        {garrigos2003modelling}
\bibfield{author}{\bibinfo{person}{Irene Garrig{\'o}s}, \bibinfo{person}{Jaime G{\'o}mez}, {and} \bibinfo{person}{Cristina Cachero}.} \bibinfo{year}{2003}\natexlab{}.
\newblock \showarticletitle{Modelling dynamic personalization in web applications}. In \bibinfo{booktitle}{\emph{International Conference on Web Engineering}}. Springer, \bibinfo{pages}{472--475}.
\newblock


\bibitem[Gatos et~al\mbox{.}(2021)]%
        {gatos2021hci}
\bibfield{author}{\bibinfo{person}{Do{\u{g}}a Gatos}, \bibinfo{person}{Asl{\i} G{\"u}nay}, \bibinfo{person}{G{\"u}ncel K{\i}rlang{\i}{\c{c}}}, \bibinfo{person}{Kemal Kuscu}, {and} \bibinfo{person}{Asim~Evren Yantac}.} \bibinfo{year}{2021}\natexlab{}.
\newblock \showarticletitle{How HCI bridges health and design in online health communities: a systematic review}. In \bibinfo{booktitle}{\emph{Proceedings of the 2021 ACM Designing Interactive Systems Conference}}. \bibinfo{pages}{970--983}.
\newblock


\bibitem[Gibbs et~al\mbox{.}(2002)]%
        {gibbs2002aesop}
\bibfield{author}{\bibinfo{person}{Laura Gibbs} {et~al\mbox{.}}} \bibinfo{year}{2002}\natexlab{}.
\newblock \bibinfo{booktitle}{\emph{Aesop's fables}}.
\newblock \bibinfo{publisher}{OUP Oxford}.
\newblock


\bibitem[Grassini(2023)]%
        {grassini2023development}
\bibfield{author}{\bibinfo{person}{Simone Grassini}.} \bibinfo{year}{2023}\natexlab{}.
\newblock \showarticletitle{Development and validation of the AI attitude scale (AIAS-4): a brief measure of general attitude toward artificial intelligence}.
\newblock \bibinfo{journal}{\emph{Frontiers in psychology}}  \bibinfo{volume}{14} (\bibinfo{year}{2023}), \bibinfo{pages}{1191628}.
\newblock


\bibitem[Grimes et~al\mbox{.}(2008)]%
        {grimes2008eatwell}
\bibfield{author}{\bibinfo{person}{Andrea Grimes}, \bibinfo{person}{Martin Bednar}, \bibinfo{person}{Jay~David Bolter}, {and} \bibinfo{person}{Rebecca~E Grinter}.} \bibinfo{year}{2008}\natexlab{}.
\newblock \showarticletitle{EatWell: sharing nutrition-related memories in a low-income community}. In \bibinfo{booktitle}{\emph{Proceedings of the 2008 ACM conference on Computer supported cooperative work}}. \bibinfo{pages}{87--96}.
\newblock


\bibitem[Guingrich and Graziano(2023)]%
        {guingrich2023chatbots}
\bibfield{author}{\bibinfo{person}{Rose Guingrich} {and} \bibinfo{person}{Michael~SA Graziano}.} \bibinfo{year}{2023}\natexlab{}.
\newblock \showarticletitle{Chatbots as social companions: How people perceive consciousness, human likeness, and social health benefits in machines}.
\newblock \bibinfo{journal}{\emph{arXiv preprint arXiv:2311.10599}} (\bibinfo{year}{2023}).
\newblock


\bibitem[Haugeland et~al\mbox{.}(2022)]%
        {haugeland2022understanding}
\bibfield{author}{\bibinfo{person}{Isabel Kathleen~Fornell Haugeland}, \bibinfo{person}{Asbj{\o}rn F{\o}lstad}, \bibinfo{person}{Cameron Taylor}, {and} \bibinfo{person}{Cato~Alexander Bj{\o}rkli}.} \bibinfo{year}{2022}\natexlab{}.
\newblock \showarticletitle{Understanding the user experience of customer service chatbots: An experimental study of chatbot interaction design}.
\newblock \bibinfo{journal}{\emph{International Journal of Human-Computer Studies}}  \bibinfo{volume}{161} (\bibinfo{year}{2022}), \bibinfo{pages}{102788}.
\newblock


\bibitem[Hauser and Schwarz(2016)]%
        {hauser2016attentive}
\bibfield{author}{\bibinfo{person}{David~J Hauser} {and} \bibinfo{person}{Norbert Schwarz}.} \bibinfo{year}{2016}\natexlab{}.
\newblock \showarticletitle{Attentive Turkers: MTurk participants perform better on online attention checks than do subject pool participants}.
\newblock \bibinfo{journal}{\emph{Behavior research methods}}  \bibinfo{volume}{48} (\bibinfo{year}{2016}), \bibinfo{pages}{400--407}.
\newblock


\bibitem[Hedderich et~al\mbox{.}(2024)]%
        {hedderich2024piece}
\bibfield{author}{\bibinfo{person}{Michael~A Hedderich}, \bibinfo{person}{Natalie~N Bazarova}, \bibinfo{person}{Wenting Zou}, \bibinfo{person}{Ryun Shim}, \bibinfo{person}{Xinda Ma}, {and} \bibinfo{person}{Qian Yang}.} \bibinfo{year}{2024}\natexlab{}.
\newblock \showarticletitle{A Piece of Theatre: Investigating How Teachers Design LLM Chatbots to Assist Adolescent Cyberbullying Education}. In \bibinfo{booktitle}{\emph{Proceedings of the CHI Conference on Human Factors in Computing Systems}}. \bibinfo{pages}{1--17}.
\newblock


\bibitem[Hinyard and Kreuter(2007)]%
        {hinyard2007using}
\bibfield{author}{\bibinfo{person}{Leslie~J Hinyard} {and} \bibinfo{person}{Matthew~W Kreuter}.} \bibinfo{year}{2007}\natexlab{}.
\newblock \showarticletitle{Using narrative communication as a tool for health behavior change: a conceptual, theoretical, and empirical overview}.
\newblock \bibinfo{journal}{\emph{Health education \& behavior}} \bibinfo{volume}{34}, \bibinfo{number}{5} (\bibinfo{year}{2007}), \bibinfo{pages}{777--792}.
\newblock


\bibitem[Hornstein et~al\mbox{.}(2023)]%
        {hornstein2023personalization}
\bibfield{author}{\bibinfo{person}{Silvan Hornstein}, \bibinfo{person}{Kirsten Zantvoort}, \bibinfo{person}{Ulrike Lueken}, \bibinfo{person}{Burkhardt Funk}, {and} \bibinfo{person}{Kevin Hilbert}.} \bibinfo{year}{2023}\natexlab{}.
\newblock \showarticletitle{Personalization strategies in digital mental health interventions: a systematic review and conceptual framework for depressive symptoms}.
\newblock \bibinfo{journal}{\emph{Frontiers in digital health}}  \bibinfo{volume}{5} (\bibinfo{year}{2023}), \bibinfo{pages}{1170002}.
\newblock


\bibitem[Hospital(2022)]%
        {deconstructing}
\bibfield{author}{\bibinfo{person}{McLean Hospital}.} \bibinfo{year}{2022}\natexlab{}.
\newblock \bibinfo{title}{Deconstructing Stigma}.
\newblock
\newblock
\urldef\tempurl%
\url{https://deconstructingstigma.org/}
\showURL{%
\tempurl}
\newblock
\shownote{Last accessed: 2022-04-06}.


\bibitem[Jahedi et~al\mbox{.}(2024)]%
        {jahedi2024personalization}
\bibfield{author}{\bibinfo{person}{Farzad Jahedi}, \bibinfo{person}{Paul~W Fay~Henman}, {and} \bibinfo{person}{Jillian~C Ryan}.} \bibinfo{year}{2024}\natexlab{}.
\newblock \showarticletitle{Personalization in digital psychological interventions for young adults}.
\newblock \bibinfo{journal}{\emph{International Journal of Human--Computer Interaction}} \bibinfo{volume}{40}, \bibinfo{number}{9} (\bibinfo{year}{2024}), \bibinfo{pages}{2254--2264}.
\newblock


\bibitem[Jo et~al\mbox{.}(2023)]%
        {jo2023understanding}
\bibfield{author}{\bibinfo{person}{Eunkyung Jo}, \bibinfo{person}{Daniel~A Epstein}, \bibinfo{person}{Hyunhoon Jung}, {and} \bibinfo{person}{Young-Ho Kim}.} \bibinfo{year}{2023}\natexlab{}.
\newblock \showarticletitle{Understanding the benefits and challenges of deploying conversational AI leveraging large language models for public health intervention}. In \bibinfo{booktitle}{\emph{Proceedings of the 2023 CHI Conference on Human Factors in Computing Systems}}. \bibinfo{pages}{1--16}.
\newblock


\bibitem[Jo et~al\mbox{.}(2024)]%
        {jo2024understanding}
\bibfield{author}{\bibinfo{person}{Eunkyung Jo}, \bibinfo{person}{Yuin Jeong}, \bibinfo{person}{SoHyun Park}, \bibinfo{person}{Daniel~A Epstein}, {and} \bibinfo{person}{Young-Ho Kim}.} \bibinfo{year}{2024}\natexlab{}.
\newblock \showarticletitle{Understanding the Impact of Long-Term Memory on Self-Disclosure with Large Language Model-Driven Chatbots for Public Health Intervention}. In \bibinfo{booktitle}{\emph{Proceedings of the CHI Conference on Human Factors in Computing Systems}}. \bibinfo{pages}{1--21}.
\newblock


\bibitem[Kabir et~al\mbox{.}(2022)]%
        {kabir2022ask}
\bibfield{author}{\bibinfo{person}{Kazi~Sinthia Kabir}, \bibinfo{person}{Stacey~A Kenfield}, \bibinfo{person}{Erin~L Van~Blarigan}, \bibinfo{person}{June~M Chan}, {and} \bibinfo{person}{Jason Wiese}.} \bibinfo{year}{2022}\natexlab{}.
\newblock \showarticletitle{Ask the users: a case study of leveraging user-centered design for designing Just-in-Time Adaptive Interventions (JITAIs)}.
\newblock \bibinfo{journal}{\emph{Proceedings of the ACM on Interactive, Mobile, Wearable and Ubiquitous Technologies}} \bibinfo{volume}{6}, \bibinfo{number}{2} (\bibinfo{year}{2022}), \bibinfo{pages}{1--21}.
\newblock


\bibitem[Kaya et~al\mbox{.}(2024)]%
        {kaya2024roles}
\bibfield{author}{\bibinfo{person}{Feridun Kaya}, \bibinfo{person}{Fatih Aydin}, \bibinfo{person}{Astrid Schepman}, \bibinfo{person}{Paul Rodway}, \bibinfo{person}{Okan Yeti{\c{s}}ensoy}, {and} \bibinfo{person}{Meva Demir~Kaya}.} \bibinfo{year}{2024}\natexlab{}.
\newblock \showarticletitle{The roles of personality traits, AI anxiety, and demographic factors in attitudes toward artificial intelligence}.
\newblock \bibinfo{journal}{\emph{International Journal of Human--Computer Interaction}} \bibinfo{volume}{40}, \bibinfo{number}{2} (\bibinfo{year}{2024}), \bibinfo{pages}{497--514}.
\newblock


\bibitem[Khandkar(2009)]%
        {khandkar2009open}
\bibfield{author}{\bibinfo{person}{Shahedul~Huq Khandkar}.} \bibinfo{year}{2009}\natexlab{}.
\newblock \showarticletitle{Open coding}.
\newblock \bibinfo{journal}{\emph{University of Calgary}}  \bibinfo{volume}{23} (\bibinfo{year}{2009}), \bibinfo{pages}{2009}.
\newblock


\bibitem[Kim et~al\mbox{.}(2023)]%
        {kim2023mindfuldiary}
\bibfield{author}{\bibinfo{person}{Taewan Kim}, \bibinfo{person}{Seolyeong Bae}, \bibinfo{person}{Hyun~Ah Kim}, \bibinfo{person}{Su-woo Lee}, \bibinfo{person}{Hwajung Hong}, \bibinfo{person}{Chanmo Yang}, {and} \bibinfo{person}{Young-Ho Kim}.} \bibinfo{year}{2023}\natexlab{}.
\newblock \showarticletitle{MindfulDiary: Harnessing Large Language Model to Support Psychiatric Patients' Journaling}.
\newblock \bibinfo{journal}{\emph{arXiv preprint arXiv:2310.05231}} (\bibinfo{year}{2023}).
\newblock


\bibitem[Kim et~al\mbox{.}(2024)]%
        {kim2024diarymate}
\bibfield{author}{\bibinfo{person}{Taewan Kim}, \bibinfo{person}{Donghoon Shin}, \bibinfo{person}{Young-Ho Kim}, {and} \bibinfo{person}{Hwajung Hong}.} \bibinfo{year}{2024}\natexlab{}.
\newblock \showarticletitle{DiaryMate: Understanding User Perceptions and Experience in Human-AI Collaboration for Personal Journaling}. In \bibinfo{booktitle}{\emph{Proceedings of the CHI Conference on Human Factors in Computing Systems}}. \bibinfo{pages}{1--15}.
\newblock


\bibitem[Kobak et~al\mbox{.}(2024)]%
        {kobak2024delving}
\bibfield{author}{\bibinfo{person}{Dmitry Kobak}, \bibinfo{person}{Rita~Gonz{\'a}lez M{\'a}rquez}, \bibinfo{person}{Em{\H{o}}ke-{\'A}gnes Horv{\'a}t}, {and} \bibinfo{person}{Jan Lause}.} \bibinfo{year}{2024}\natexlab{}.
\newblock \showarticletitle{Delving into ChatGPT usage in academic writing through excess vocabulary}.
\newblock \bibinfo{journal}{\emph{arXiv preprint arXiv:2406.07016}} (\bibinfo{year}{2024}).
\newblock


\bibitem[Kocielnik et~al\mbox{.}(2018)]%
        {kocielnik2018reflection}
\bibfield{author}{\bibinfo{person}{Rafal Kocielnik}, \bibinfo{person}{Lillian Xiao}, \bibinfo{person}{Daniel Avrahami}, {and} \bibinfo{person}{Gary Hsieh}.} \bibinfo{year}{2018}\natexlab{}.
\newblock \showarticletitle{Reflection companion: a conversational system for engaging users in reflection on physical activity}.
\newblock \bibinfo{journal}{\emph{Proceedings of the ACM on Interactive, Mobile, Wearable and Ubiquitous Technologies}} \bibinfo{volume}{2}, \bibinfo{number}{2} (\bibinfo{year}{2018}), \bibinfo{pages}{1--26}.
\newblock


\bibitem[Kornfield et~al\mbox{.}(2022a)]%
        {kornfield2022meeting}
\bibfield{author}{\bibinfo{person}{Rachel Kornfield}, \bibinfo{person}{Jonah Meyerhoff}, \bibinfo{person}{Hannah Studd}, \bibinfo{person}{Ananya Bhattacharjee}, \bibinfo{person}{Joseph~Jay Williams}, \bibinfo{person}{Madhu Reddy}, {and} \bibinfo{person}{David~C Mohr}.} \bibinfo{year}{2022}\natexlab{a}.
\newblock \showarticletitle{Meeting Users Where They Are: User-centered Design of an Automated Text Messaging Tool to Support the Mental Health of Young Adults}. In \bibinfo{booktitle}{\emph{Proceedings of the 2022 CHI Conference on Human Factors in Computing Systems}}. \bibinfo{pages}{1--16}.
\newblock


\bibitem[Kornfield et~al\mbox{.}(2022b)]%
        {kornfield2022involving}
\bibfield{author}{\bibinfo{person}{Rachel Kornfield}, \bibinfo{person}{David~C Mohr}, \bibinfo{person}{Rachel Ranney}, \bibinfo{person}{Emily~G Lattie}, \bibinfo{person}{Jonah Meyerhoff}, \bibinfo{person}{Joseph~J Williams}, {and} \bibinfo{person}{Madhu Reddy}.} \bibinfo{year}{2022}\natexlab{b}.
\newblock \showarticletitle{Involving Crowdworkers with lived experience in content-development for push-based digital mental health tools: lessons learned from crowdsourcing mental health messages}.
\newblock \bibinfo{journal}{\emph{Proceedings of the ACM on Human-computer Interaction}} \bibinfo{volume}{6}, \bibinfo{number}{CSCW1} (\bibinfo{year}{2022}), \bibinfo{pages}{1--30}.
\newblock


\bibitem[Korngiebel and Mooney(2021)]%
        {korngiebel2021considering}
\bibfield{author}{\bibinfo{person}{Diane~M Korngiebel} {and} \bibinfo{person}{Sean~D Mooney}.} \bibinfo{year}{2021}\natexlab{}.
\newblock \showarticletitle{Considering the possibilities and pitfalls of Generative Pre-trained Transformer 3 (GPT-3) in healthcare delivery}.
\newblock \bibinfo{journal}{\emph{NPJ Digital Medicine}} \bibinfo{volume}{4}, \bibinfo{number}{1} (\bibinfo{year}{2021}), \bibinfo{pages}{93}.
\newblock


\bibitem[Le~Glaz et~al\mbox{.}(2021)]%
        {le2021machine}
\bibfield{author}{\bibinfo{person}{Aziliz Le~Glaz}, \bibinfo{person}{Yannis Haralambous}, \bibinfo{person}{Deok-Hee Kim-Dufor}, \bibinfo{person}{Philippe Lenca}, \bibinfo{person}{Romain Billot}, \bibinfo{person}{Taylor~C Ryan}, \bibinfo{person}{Jonathan Marsh}, \bibinfo{person}{Jordan Devylder}, \bibinfo{person}{Michel Walter}, \bibinfo{person}{Sofian Berrouiguet}, {et~al\mbox{.}}} \bibinfo{year}{2021}\natexlab{}.
\newblock \showarticletitle{Machine learning and natural language processing in mental health: systematic review}.
\newblock \bibinfo{journal}{\emph{Journal of medical Internet research}} \bibinfo{volume}{23}, \bibinfo{number}{5} (\bibinfo{year}{2021}), \bibinfo{pages}{e15708}.
\newblock


\bibitem[Lee and Hong(2017)]%
        {lee2017designing}
\bibfield{author}{\bibinfo{person}{Kwangyoung Lee} {and} \bibinfo{person}{Hwajung Hong}.} \bibinfo{year}{2017}\natexlab{}.
\newblock \showarticletitle{Designing for self-tracking of emotion and experience with tangible modality}. In \bibinfo{booktitle}{\emph{Proceedings of the 2017 Conference on Designing Interactive Systems}}. \bibinfo{pages}{465--475}.
\newblock


\bibitem[Lipsey et~al\mbox{.}(2020)]%
        {lipsey2020evaluation}
\bibfield{author}{\bibinfo{person}{Amanda~Faye Lipsey}, \bibinfo{person}{Amy~D Waterman}, \bibinfo{person}{Emily~H Wood}, {and} \bibinfo{person}{Wendy Balliet}.} \bibinfo{year}{2020}\natexlab{}.
\newblock \showarticletitle{Evaluation of first-person storytelling on changing health-related attitudes, knowledge, behaviors, and outcomes: a scoping review}.
\newblock \bibinfo{journal}{\emph{Patient education and counseling}} \bibinfo{volume}{103}, \bibinfo{number}{10} (\bibinfo{year}{2020}), \bibinfo{pages}{1922--1934}.
\newblock


\bibitem[Liu et~al\mbox{.}(2022)]%
        {liu2022will}
\bibfield{author}{\bibinfo{person}{Yihe Liu}, \bibinfo{person}{Anushk Mittal}, \bibinfo{person}{Diyi Yang}, {and} \bibinfo{person}{Amy Bruckman}.} \bibinfo{year}{2022}\natexlab{}.
\newblock \showarticletitle{Will AI console me when I lose my pet? Understanding perceptions of AI-mediated email writing}. In \bibinfo{booktitle}{\emph{Proceedings of the 2022 CHI conference on human factors in computing systems}}. \bibinfo{pages}{1--13}.
\newblock


\bibitem[Llewellyn-Beardsley et~al\mbox{.}(2020)]%
        {llewellyn2020not}
\bibfield{author}{\bibinfo{person}{Joy Llewellyn-Beardsley}, \bibinfo{person}{Stefan Rennick-Egglestone}, \bibinfo{person}{Simon Bradstreet}, \bibinfo{person}{Larry Davidson}, \bibinfo{person}{Donna Franklin}, \bibinfo{person}{Ada Hui}, \bibinfo{person}{Rose McGranahan}, \bibinfo{person}{Kate Morgan}, \bibinfo{person}{Kristian Pollock}, \bibinfo{person}{Amy Ramsay}, {et~al\mbox{.}}} \bibinfo{year}{2020}\natexlab{}.
\newblock \showarticletitle{Not the story you want? Assessing the fit of a conceptual framework characterising mental health recovery narratives}.
\newblock \bibinfo{journal}{\emph{Social psychiatry and psychiatric epidemiology}} \bibinfo{volume}{55}, \bibinfo{number}{3} (\bibinfo{year}{2020}), \bibinfo{pages}{295--308}.
\newblock


\bibitem[Llewellyn-Beardsley et~al\mbox{.}(2019)]%
        {llewellyn2019characteristics}
\bibfield{author}{\bibinfo{person}{Joy Llewellyn-Beardsley}, \bibinfo{person}{Stefan Rennick-Egglestone}, \bibinfo{person}{Felicity Callard}, \bibinfo{person}{Paul Crawford}, \bibinfo{person}{Marianne Farkas}, \bibinfo{person}{Ada Hui}, \bibinfo{person}{David Manley}, \bibinfo{person}{Rose McGranahan}, \bibinfo{person}{Kristian Pollock}, \bibinfo{person}{Amy Ramsay}, {et~al\mbox{.}}} \bibinfo{year}{2019}\natexlab{}.
\newblock \showarticletitle{Characteristics of mental health recovery narratives: systematic review and narrative synthesis}.
\newblock \bibinfo{journal}{\emph{PloS one}} \bibinfo{volume}{14}, \bibinfo{number}{3} (\bibinfo{year}{2019}), \bibinfo{pages}{e0214678}.
\newblock


\bibitem[Ma et~al\mbox{.}(2023)]%
        {ma2023contextbot}
\bibfield{author}{\bibinfo{person}{Yao Ma}, \bibinfo{person}{Tahir Abbas}, {and} \bibinfo{person}{Ujwal Gadiraju}.} \bibinfo{year}{2023}\natexlab{}.
\newblock \showarticletitle{ContextBot: Improving Response Consistency in Crowd-Powered Conversational Systems for Affective Support Tasks}. In \bibinfo{booktitle}{\emph{Proceedings of the 34th ACM Conference on Hypertext and Social Media}}. \bibinfo{pages}{1--14}.
\newblock


\bibitem[Ma et~al\mbox{.}(2024)]%
        {ma2024evaluating}
\bibfield{author}{\bibinfo{person}{Zilin Ma}, \bibinfo{person}{Yiyang Mei}, \bibinfo{person}{Yinru Long}, \bibinfo{person}{Zhaoyuan Su}, {and} \bibinfo{person}{Krzysztof~Z Gajos}.} \bibinfo{year}{2024}\natexlab{}.
\newblock \showarticletitle{Evaluating the Experience of LGBTQ+ People Using Large Language Model Based Chatbots for Mental Health Support}. In \bibinfo{booktitle}{\emph{Proceedings of the CHI Conference on Human Factors in Computing Systems}}. \bibinfo{pages}{1--15}.
\newblock


\bibitem[Mahmood et~al\mbox{.}(2014)]%
        {mahmood2014dynamic}
\bibfield{author}{\bibinfo{person}{Tariq Mahmood}, \bibinfo{person}{Ghulam Mujtaba}, {and} \bibinfo{person}{Adriano Venturini}.} \bibinfo{year}{2014}\natexlab{}.
\newblock \showarticletitle{Dynamic personalization in conversational recommender systems}.
\newblock \bibinfo{journal}{\emph{Information Systems and e-Business Management}} \bibinfo{volume}{12}, \bibinfo{number}{2} (\bibinfo{year}{2014}), \bibinfo{pages}{213--238}.
\newblock


\bibitem[Matthews and Rhodes-Maquaire(2024)]%
        {matthews2024personalisation}
\bibfield{author}{\bibinfo{person}{Paul Matthews} {and} \bibinfo{person}{Clemence Rhodes-Maquaire}.} \bibinfo{year}{2024}\natexlab{}.
\newblock \showarticletitle{Personalisation and Recommendation for Mental Health Apps: A Scoping Review}.
\newblock \bibinfo{journal}{\emph{Behaviour \& Information Technology}} (\bibinfo{year}{2024}), \bibinfo{pages}{1--16}.
\newblock


\bibitem[Meyerhoff et~al\mbox{.}(2024)]%
        {meyerhoff2024small}
\bibfield{author}{\bibinfo{person}{Jonah Meyerhoff}, \bibinfo{person}{Miranda Beltzer}, \bibinfo{person}{Sarah Popowski}, \bibinfo{person}{Chris~J Karr}, \bibinfo{person}{Theresa Nguyen}, \bibinfo{person}{Joseph~J Williams}, \bibinfo{person}{Charles~J Krause}, \bibinfo{person}{Harsh Kumar}, \bibinfo{person}{Ananya Bhattacharjee}, \bibinfo{person}{David~C Mohr}, {et~al\mbox{.}}} \bibinfo{year}{2024}\natexlab{}.
\newblock \showarticletitle{Small Steps over time: A longitudinal usability test of an automated interactive text messaging intervention to support self-management of depression and anxiety symptoms}.
\newblock \bibinfo{journal}{\emph{Journal of Affective Disorders}}  \bibinfo{volume}{345} (\bibinfo{year}{2024}), \bibinfo{pages}{122--130}.
\newblock


\bibitem[Morris et~al\mbox{.}(2018)]%
        {morris2018towards}
\bibfield{author}{\bibinfo{person}{Robert~R Morris}, \bibinfo{person}{Kareem Kouddous}, \bibinfo{person}{Rohan Kshirsagar}, {and} \bibinfo{person}{Stephen~M Schueller}.} \bibinfo{year}{2018}\natexlab{}.
\newblock \showarticletitle{Towards an artificially empathic conversational agent for mental health applications: system design and user perceptions}.
\newblock \bibinfo{journal}{\emph{Journal of medical Internet research}} \bibinfo{volume}{20}, \bibinfo{number}{6} (\bibinfo{year}{2018}), \bibinfo{pages}{e10148}.
\newblock


\bibitem[Morris and Picard(2014)]%
        {morris2014crowd}
\bibfield{author}{\bibinfo{person}{Robert~R Morris} {and} \bibinfo{person}{Rosalind Picard}.} \bibinfo{year}{2014}\natexlab{}.
\newblock \showarticletitle{Crowd-powered positive psychological interventions}.
\newblock \bibinfo{journal}{\emph{The Journal of Positive Psychology}} \bibinfo{volume}{9}, \bibinfo{number}{6} (\bibinfo{year}{2014}), \bibinfo{pages}{509--516}.
\newblock


\bibitem[Morris et~al\mbox{.}(2015)]%
        {morris2015efficacy}
\bibfield{author}{\bibinfo{person}{Robert~R Morris}, \bibinfo{person}{Stephen~M Schueller}, {and} \bibinfo{person}{Rosalind~W Picard}.} \bibinfo{year}{2015}\natexlab{}.
\newblock \showarticletitle{Efficacy of a web-based, crowdsourced peer-to-peer cognitive reappraisal platform for depression: randomized controlled trial}.
\newblock \bibinfo{journal}{\emph{Journal of medical Internet research}} \bibinfo{volume}{17}, \bibinfo{number}{3} (\bibinfo{year}{2015}), \bibinfo{pages}{e72}.
\newblock


\bibitem[Motahar et~al\mbox{.}(2024)]%
        {motahar2024toward}
\bibfield{author}{\bibinfo{person}{Tamanna Motahar}, \bibinfo{person}{Noelle Brown}, \bibinfo{person}{Eliane~Stampfer Wiese}, {and} \bibinfo{person}{Jason Wiese}.} \bibinfo{year}{2024}\natexlab{}.
\newblock \showarticletitle{Toward Building Design Empathy for People with Disabilities Using Social Media Data: A New Approach for Novice Designers}. In \bibinfo{booktitle}{\emph{Proceedings of the 2024 ACM Designing Interactive Systems Conference}}. \bibinfo{pages}{3145--3160}.
\newblock


\bibitem[{National Institute of Mental Health (NIMH)}(2020)]%
        {nimh2020mentalhealth}
\bibfield{author}{\bibinfo{person}{{National Institute of Mental Health (NIMH)}}.} \bibinfo{year}{2020}\natexlab{}.
\newblock \bibinfo{title}{Mental Illness}.
\newblock \bibinfo{howpublished}{\url{https://www.nimh.nih.gov/health/statistics/mental-illness}}.
\newblock


\bibitem[Nurser et~al\mbox{.}(2018)]%
        {nurser2018personal}
\bibfield{author}{\bibinfo{person}{Kate~P Nurser}, \bibinfo{person}{Imogen Rushworth}, \bibinfo{person}{Tom Shakespeare}, {and} \bibinfo{person}{Deirdre Williams}.} \bibinfo{year}{2018}\natexlab{}.
\newblock \showarticletitle{Personal storytelling in mental health recovery}.
\newblock \bibinfo{journal}{\emph{Mental Health Review Journal}} (\bibinfo{year}{2018}).
\newblock


\bibitem[O'Dell et~al\mbox{.}(2022)]%
        {o2022building}
\bibfield{author}{\bibinfo{person}{Bessie O'Dell}, \bibinfo{person}{Katherine Stevens}, \bibinfo{person}{Anneka Tomlinson}, \bibinfo{person}{Ilina Singh}, {and} \bibinfo{person}{Andrea Cipriani}.} \bibinfo{year}{2022}\natexlab{}.
\newblock \bibinfo{title}{Building trust in artificial intelligence and new technologies in mental health}.
\newblock , \bibinfo{numpages}{45--46}~pages.
\newblock


\bibitem[O'Leary et~al\mbox{.}(2017)]%
        {o2017design}
\bibfield{author}{\bibinfo{person}{Kathleen O'Leary}, \bibinfo{person}{Arpita Bhattacharya}, \bibinfo{person}{Sean~A Munson}, \bibinfo{person}{Jacob~O Wobbrock}, {and} \bibinfo{person}{Wanda Pratt}.} \bibinfo{year}{2017}\natexlab{}.
\newblock \showarticletitle{Design opportunities for mental health peer support technologies}. In \bibinfo{booktitle}{\emph{Proceedings of the 2017 ACM conference on computer supported cooperative work and social computing}}. \bibinfo{pages}{1470--1484}.
\newblock


\bibitem[O'Leary et~al\mbox{.}(2018)]%
        {o2018suddenly}
\bibfield{author}{\bibinfo{person}{Kathleen O'Leary}, \bibinfo{person}{Stephen~M Schueller}, \bibinfo{person}{Jacob~O Wobbrock}, {and} \bibinfo{person}{Wanda Pratt}.} \bibinfo{year}{2018}\natexlab{}.
\newblock \showarticletitle{“Suddenly, we got to become therapists for each other” Designing Peer Support Chats for Mental Health}. In \bibinfo{booktitle}{\emph{Proceedings of the 2018 CHI Conference on Human Factors in Computing Systems}}. \bibinfo{pages}{1--14}.
\newblock


\bibitem[Oppenheimer et~al\mbox{.}(2009)]%
        {oppenheimer2009instructional}
\bibfield{author}{\bibinfo{person}{Daniel~M Oppenheimer}, \bibinfo{person}{Tom Meyvis}, {and} \bibinfo{person}{Nicolas Davidenko}.} \bibinfo{year}{2009}\natexlab{}.
\newblock \showarticletitle{Instructional manipulation checks: Detecting satisficing to increase statistical power}.
\newblock \bibinfo{journal}{\emph{Journal of experimental social psychology}} \bibinfo{volume}{45}, \bibinfo{number}{4} (\bibinfo{year}{2009}), \bibinfo{pages}{867--872}.
\newblock


\bibitem[Park et~al\mbox{.}(2023)]%
        {park2023anthropomorphic}
\bibfield{author}{\bibinfo{person}{Gain Park}, \bibinfo{person}{Seyoung Lee}, {and} \bibinfo{person}{Jiyun Chung}.} \bibinfo{year}{2023}\natexlab{}.
\newblock \showarticletitle{Do anthropomorphic chatbots increase counseling satisfaction and reuse intention? The moderated mediation of social rapport and social anxiety}.
\newblock \bibinfo{journal}{\emph{Cyberpsychology, Behavior, and Social Networking}} \bibinfo{volume}{26}, \bibinfo{number}{5} (\bibinfo{year}{2023}), \bibinfo{pages}{357--365}.
\newblock


\bibitem[Park et~al\mbox{.}(2021)]%
        {park2021wrote}
\bibfield{author}{\bibinfo{person}{SoHyun Park}, \bibinfo{person}{Anja Thieme}, \bibinfo{person}{Jeongyun Han}, \bibinfo{person}{Sungwoo Lee}, \bibinfo{person}{Wonjong Rhee}, {and} \bibinfo{person}{Bongwon Suh}.} \bibinfo{year}{2021}\natexlab{}.
\newblock \showarticletitle{“I wrote as if I were telling a story to someone I knew.”: Designing Chatbot Interactions for Expressive Writing in Mental Health}. In \bibinfo{booktitle}{\emph{Proceedings of the 2021 ACM Designing Interactive Systems Conference}}. \bibinfo{pages}{926--941}.
\newblock


\bibitem[Park(2018)]%
        {park2018social}
\bibfield{author}{\bibinfo{person}{Sun~Young Park}.} \bibinfo{year}{2018}\natexlab{}.
\newblock \showarticletitle{Social support mosaic: Understanding mental health management practice on college campus}. In \bibinfo{booktitle}{\emph{Proceedings of the 2018 Designing Interactive Systems Conference}}. \bibinfo{pages}{121--133}.
\newblock


\bibitem[Payne(2006)]%
        {payne2006narrative}
\bibfield{author}{\bibinfo{person}{Martin Payne}.} \bibinfo{year}{2006}\natexlab{}.
\newblock \bibinfo{booktitle}{\emph{Narrative therapy}}.
\newblock \bibinfo{publisher}{Sage}.
\newblock


\bibitem[Pereira and D{\'\i}az(2019)]%
        {pereira2019using}
\bibfield{author}{\bibinfo{person}{Juanan Pereira} {and} \bibinfo{person}{{\'O}scar D{\'\i}az}.} \bibinfo{year}{2019}\natexlab{}.
\newblock \showarticletitle{Using health chatbots for behavior change: a mapping study}.
\newblock \bibinfo{journal}{\emph{Journal of medical systems}}  \bibinfo{volume}{43} (\bibinfo{year}{2019}), \bibinfo{pages}{1--13}.
\newblock


\bibitem[Picard(2000)]%
        {picard2000affective}
\bibfield{author}{\bibinfo{person}{Rosalind~W Picard}.} \bibinfo{year}{2000}\natexlab{}.
\newblock \bibinfo{booktitle}{\emph{Affective computing}}.
\newblock \bibinfo{publisher}{MIT press}.
\newblock


\bibitem[Posner et~al\mbox{.}(2008)]%
        {posner2008columbia}
\bibfield{author}{\bibinfo{person}{K Posner}, \bibinfo{person}{D Brent}, \bibinfo{person}{C Lucas}, \bibinfo{person}{M Gould}, \bibinfo{person}{B Stanley}, \bibinfo{person}{G Brown}, \bibinfo{person}{P Fisher}, \bibinfo{person}{J Zelazny}, \bibinfo{person}{A Burke}, \bibinfo{person}{MJNY Oquendo}, {et~al\mbox{.}}} \bibinfo{year}{2008}\natexlab{}.
\newblock \showarticletitle{Columbia-suicide severity rating scale (C-SSRS)}.
\newblock \bibinfo{journal}{\emph{New York, NY: Columbia University Medical Center}}  \bibinfo{volume}{10} (\bibinfo{year}{2008}).
\newblock


\bibitem[Rankin et~al\mbox{.}(2009)]%
        {rankin2009habituation}
\bibfield{author}{\bibinfo{person}{Catharine~H Rankin}, \bibinfo{person}{Thomas Abrams}, \bibinfo{person}{Robert~J Barry}, \bibinfo{person}{Seema Bhatnagar}, \bibinfo{person}{David~F Clayton}, \bibinfo{person}{John Colombo}, \bibinfo{person}{Gianluca Coppola}, \bibinfo{person}{Mark~A Geyer}, \bibinfo{person}{David~L Glanzman}, \bibinfo{person}{Stephen Marsland}, {et~al\mbox{.}}} \bibinfo{year}{2009}\natexlab{}.
\newblock \showarticletitle{Habituation revisited: an updated and revised description of the behavioral characteristics of habituation}.
\newblock \bibinfo{journal}{\emph{Neurobiology of learning and memory}} \bibinfo{volume}{92}, \bibinfo{number}{2} (\bibinfo{year}{2009}), \bibinfo{pages}{135--138}.
\newblock


\bibitem[Reza et~al\mbox{.}(2023)]%
        {reza2023abscribe}
\bibfield{author}{\bibinfo{person}{Mohi Reza}, \bibinfo{person}{Nathan Laundry}, \bibinfo{person}{Ilya Musabirov}, \bibinfo{person}{Peter Dushniku}, \bibinfo{person}{Zhi~Yuan Yu}, \bibinfo{person}{Kashish Mittal}, \bibinfo{person}{Tovi Grossman}, \bibinfo{person}{Michael Liut}, \bibinfo{person}{Anastasia Kuzminykh}, \bibinfo{person}{Joseph~Jay Williams}, {et~al\mbox{.}}} \bibinfo{year}{2023}\natexlab{}.
\newblock \showarticletitle{ABScribe: Rapid Exploration of Multiple Writing Variations in Human-AI Co-Writing Tasks using Large Language Models}.
\newblock \bibinfo{journal}{\emph{arXiv preprint arXiv:2310.00117}} (\bibinfo{year}{2023}).
\newblock


\bibitem[Rodr{\'\i}guez~Vega et~al\mbox{.}(2014)]%
        {rodriguez2014mindfulness}
\bibfield{author}{\bibinfo{person}{B Rodr{\'\i}guez~Vega}, \bibinfo{person}{C Bay{\'o}n~P{\'e}rez}, \bibinfo{person}{A PalaoTarrero}, {and} \bibinfo{person}{A Fern{\'a}ndez~Liria}.} \bibinfo{year}{2014}\natexlab{}.
\newblock \showarticletitle{Mindfulness-based narrative therapy for depression in cancer patients}.
\newblock \bibinfo{journal}{\emph{Clinical psychology \& psychotherapy}} \bibinfo{volume}{21}, \bibinfo{number}{5} (\bibinfo{year}{2014}), \bibinfo{pages}{411--419}.
\newblock


\bibitem[Rosenthal(1971)]%
        {rosenthal1971specificity}
\bibfield{author}{\bibinfo{person}{Paul~I Rosenthal}.} \bibinfo{year}{1971}\natexlab{}.
\newblock \showarticletitle{Specificity, verifiability, and message credibility}.
\newblock \bibinfo{journal}{\emph{Quarterly Journal of Speech}} \bibinfo{volume}{57}, \bibinfo{number}{4} (\bibinfo{year}{1971}), \bibinfo{pages}{393--401}.
\newblock


\bibitem[Saksono et~al\mbox{.}(2021)]%
        {saksono2021storymap}
\bibfield{author}{\bibinfo{person}{Herman Saksono}, \bibinfo{person}{Carmen Castaneda-Sceppa}, \bibinfo{person}{Jessica~A Hoffman}, \bibinfo{person}{Magy Seif El-Nasr}, {and} \bibinfo{person}{Andrea Parker}.} \bibinfo{year}{2021}\natexlab{}.
\newblock \showarticletitle{StoryMap: Using Social Modeling and Self-Modeling to Support Physical Activity Among Families of Low-SES Backgrounds}. In \bibinfo{booktitle}{\emph{Proceedings of the 2021 CHI Conference on Human Factors in Computing Systems}}. \bibinfo{pages}{1--14}.
\newblock


\bibitem[Saksono et~al\mbox{.}(2023)]%
        {saksono2023evaluating}
\bibfield{author}{\bibinfo{person}{Herman Saksono}, \bibinfo{person}{Vivien Morris}, \bibinfo{person}{Andrea~G Parker}, {and} \bibinfo{person}{Krzysztof~Z Gajos}.} \bibinfo{year}{2023}\natexlab{}.
\newblock \showarticletitle{Evaluating Similarity Variables for Peer Matching in Digital Health Storytelling}.
\newblock \bibinfo{journal}{\emph{Proceedings of the ACM on Human-Computer Interaction}} \bibinfo{volume}{7}, \bibinfo{number}{CSCW2} (\bibinfo{year}{2023}), \bibinfo{pages}{1--25}.
\newblock


\bibitem[Sarlej and Ryan(2012)]%
        {sarlej2012representing}
\bibfield{author}{\bibinfo{person}{Margaret Sarlej} {and} \bibinfo{person}{Malcolm Ryan}.} \bibinfo{year}{2012}\natexlab{}.
\newblock \showarticletitle{Representing morals in terms of emotion}. In \bibinfo{booktitle}{\emph{Proceedings of the AAAI Conference on Artificial Intelligence and Interactive Digital Entertainment}}, Vol.~\bibinfo{volume}{8}.
\newblock


\bibitem[Schleider et~al\mbox{.}(2020)]%
        {schleider2020acceptability}
\bibfield{author}{\bibinfo{person}{Jessica~Lee Schleider}, \bibinfo{person}{Mallory Dobias}, \bibinfo{person}{Jenna Sung}, \bibinfo{person}{Emma Mumper}, {and} \bibinfo{person}{Michael~C Mullarkey}.} \bibinfo{year}{2020}\natexlab{}.
\newblock \showarticletitle{Acceptability and utility of an open-access, online single-session intervention platform for adolescent mental health}.
\newblock \bibinfo{journal}{\emph{JMIR mental health}} \bibinfo{volume}{7}, \bibinfo{number}{6} (\bibinfo{year}{2020}), \bibinfo{pages}{e20513}.
\newblock


\bibitem[Schleider and Weisz(2017)]%
        {schleider2017little}
\bibfield{author}{\bibinfo{person}{Jessica~L Schleider} {and} \bibinfo{person}{John~R Weisz}.} \bibinfo{year}{2017}\natexlab{}.
\newblock \showarticletitle{Little treatments, promising effects? Meta-analysis of single-session interventions for youth psychiatric problems}.
\newblock \bibinfo{journal}{\emph{Journal of the American Academy of Child \& Adolescent Psychiatry}} \bibinfo{volume}{56}, \bibinfo{number}{2} (\bibinfo{year}{2017}), \bibinfo{pages}{107--115}.
\newblock


\bibitem[Seyama and Nagayama(2007)]%
        {seyama2007uncanny}
\bibfield{author}{\bibinfo{person}{Jun'ichiro Seyama} {and} \bibinfo{person}{Ruth~S Nagayama}.} \bibinfo{year}{2007}\natexlab{}.
\newblock \showarticletitle{The uncanny valley: Effect of realism on the impression of artificial human faces}.
\newblock \bibinfo{journal}{\emph{Presence}} \bibinfo{volume}{16}, \bibinfo{number}{4} (\bibinfo{year}{2007}), \bibinfo{pages}{337--351}.
\newblock


\bibitem[Sharma et~al\mbox{.}(2021)]%
        {sharma2021towards}
\bibfield{author}{\bibinfo{person}{Ashish Sharma}, \bibinfo{person}{Inna~W Lin}, \bibinfo{person}{Adam~S Miner}, \bibinfo{person}{David~C Atkins}, {and} \bibinfo{person}{Tim Althoff}.} \bibinfo{year}{2021}\natexlab{}.
\newblock \showarticletitle{Towards facilitating empathic conversations in online mental health support: A reinforcement learning approach}. In \bibinfo{booktitle}{\emph{Proceedings of the Web Conference 2021}}. \bibinfo{pages}{194--205}.
\newblock


\bibitem[Sharma et~al\mbox{.}(2023a)]%
        {sharma2023human}
\bibfield{author}{\bibinfo{person}{Ashish Sharma}, \bibinfo{person}{Inna~W Lin}, \bibinfo{person}{Adam~S Miner}, \bibinfo{person}{David~C Atkins}, {and} \bibinfo{person}{Tim Althoff}.} \bibinfo{year}{2023}\natexlab{a}.
\newblock \showarticletitle{Human--AI collaboration enables more empathic conversations in text-based peer-to-peer mental health support}.
\newblock \bibinfo{journal}{\emph{Nature Machine Intelligence}} \bibinfo{volume}{5}, \bibinfo{number}{1} (\bibinfo{year}{2023}), \bibinfo{pages}{46--57}.
\newblock


\bibitem[Sharma et~al\mbox{.}(2023b)]%
        {sharma2023cognitive}
\bibfield{author}{\bibinfo{person}{Ashish Sharma}, \bibinfo{person}{Kevin Rushton}, \bibinfo{person}{Inna~Wanyin Lin}, \bibinfo{person}{David Wadden}, \bibinfo{person}{Khendra~G Lucas}, \bibinfo{person}{Adam Miner}, \bibinfo{person}{Theresa Nguyen}, {and} \bibinfo{person}{Tim Althoff}.} \bibinfo{year}{2023}\natexlab{b}.
\newblock \showarticletitle{Cognitive Reframing of Negative Thoughts through Human-Language Model Interaction}. In \bibinfo{booktitle}{\emph{The 61st Annual Meeting Of The Association For Computational Linguistics}}.
\newblock


\bibitem[Shaw and Homewood(2015)]%
        {shaw2015effect}
\bibfield{author}{\bibinfo{person}{Laura-Kate Shaw} {and} \bibinfo{person}{Judi Homewood}.} \bibinfo{year}{2015}\natexlab{}.
\newblock \showarticletitle{The effect of eating disorder memoirs in individuals with self-identified eating pathologies}.
\newblock \bibinfo{journal}{\emph{The Journal of nervous and mental disease}} \bibinfo{volume}{203}, \bibinfo{number}{8} (\bibinfo{year}{2015}), \bibinfo{pages}{591--595}.
\newblock


\bibitem[Shoes(2022)]%
        {walkinourshoes}
\bibfield{author}{\bibinfo{person}{Walk In~Our Shoes}.} \bibinfo{year}{2022}\natexlab{}.
\newblock \bibinfo{title}{Walk In Our Shoes}.
\newblock
\newblock
\urldef\tempurl%
\url{https://walkinourshoes.org/}
\showURL{%
\tempurl}
\newblock
\shownote{Last accessed: 2022-04-06}.


\bibitem[Singhal and Rogers(2012)]%
        {singhal2012entertainment}
\bibfield{author}{\bibinfo{person}{Arvind Singhal} {and} \bibinfo{person}{Everett Rogers}.} \bibinfo{year}{2012}\natexlab{}.
\newblock \bibinfo{booktitle}{\emph{Entertainment-education: A communication strategy for social change}}.
\newblock \bibinfo{publisher}{Routledge}.
\newblock


\bibitem[Slov{\'a}k et~al\mbox{.}(2016)]%
        {slovak2016scaffolding}
\bibfield{author}{\bibinfo{person}{Petr Slov{\'a}k}, \bibinfo{person}{Kael Rowan}, \bibinfo{person}{Christopher Frauenberger}, \bibinfo{person}{Ran Gilad-Bachrach}, \bibinfo{person}{Mia Doces}, \bibinfo{person}{Brian Smith}, \bibinfo{person}{Rachel Kamb}, {and} \bibinfo{person}{Geraldine Fitzpatrick}.} \bibinfo{year}{2016}\natexlab{}.
\newblock \showarticletitle{Scaffolding the scaffolding: Supporting children's social-emotional learning at home}. In \bibinfo{booktitle}{\emph{Proceedings of the 19th ACM Conference on Computer-Supported Cooperative Work \& Social Computing}}. \bibinfo{pages}{1751--1765}.
\newblock


\bibitem[Smith et~al\mbox{.}(2021)]%
        {smith2021effective}
\bibfield{author}{\bibinfo{person}{C~Estelle Smith}, \bibinfo{person}{William Lane}, \bibinfo{person}{Hannah Miller~Hillberg}, \bibinfo{person}{Daniel Kluver}, \bibinfo{person}{Loren Terveen}, {and} \bibinfo{person}{Svetlana Yarosh}.} \bibinfo{year}{2021}\natexlab{}.
\newblock \showarticletitle{Effective strategies for crowd-powered cognitive reappraisal systems: A field deployment of the flip* doubt web application for mental health}.
\newblock \bibinfo{journal}{\emph{Proceedings of the ACM on Human-Computer Interaction}} \bibinfo{volume}{5}, \bibinfo{number}{CSCW2} (\bibinfo{year}{2021}), \bibinfo{pages}{1--37}.
\newblock


\bibitem[Stade et~al\mbox{.}(2024)]%
        {stade2024large}
\bibfield{author}{\bibinfo{person}{Elizabeth~C Stade}, \bibinfo{person}{Shannon~Wiltsey Stirman}, \bibinfo{person}{Lyle~H Ungar}, \bibinfo{person}{Cody~L Boland}, \bibinfo{person}{H~Andrew Schwartz}, \bibinfo{person}{David~B Yaden}, \bibinfo{person}{Jo{\~a}o Sedoc}, \bibinfo{person}{Robert~J DeRubeis}, \bibinfo{person}{Robb Willer}, {and} \bibinfo{person}{Johannes~C Eichstaedt}.} \bibinfo{year}{2024}\natexlab{}.
\newblock \showarticletitle{Large language models could change the future of behavioral healthcare: a proposal for responsible development and evaluation}.
\newblock \bibinfo{journal}{\emph{NPJ Mental Health Research}} \bibinfo{volume}{3}, \bibinfo{number}{1} (\bibinfo{year}{2024}), \bibinfo{pages}{12}.
\newblock


\bibitem[Torous et~al\mbox{.}(2018)]%
        {torous2018mental}
\bibfield{author}{\bibinfo{person}{John Torous}, \bibinfo{person}{Hannah Wisniewski}, \bibinfo{person}{Gang Liu}, \bibinfo{person}{Matcheri Keshavan}, {et~al\mbox{.}}} \bibinfo{year}{2018}\natexlab{}.
\newblock \showarticletitle{Mental health mobile phone app usage, concerns, and benefits among psychiatric outpatients: comparative survey study}.
\newblock \bibinfo{journal}{\emph{JMIR mental health}} \bibinfo{volume}{5}, \bibinfo{number}{4} (\bibinfo{year}{2018}), \bibinfo{pages}{e11715}.
\newblock


\bibitem[Vromans and Schweitzer(2011)]%
        {vromans2011narrative}
\bibfield{author}{\bibinfo{person}{Lynette~P Vromans} {and} \bibinfo{person}{Robert~D Schweitzer}.} \bibinfo{year}{2011}\natexlab{}.
\newblock \showarticletitle{Narrative therapy for adults with major depressive disorder: Improved symptom and interpersonal outcomes}.
\newblock \bibinfo{journal}{\emph{Psychotherapy research}} \bibinfo{volume}{21}, \bibinfo{number}{1} (\bibinfo{year}{2011}), \bibinfo{pages}{4--15}.
\newblock


\bibitem[Wang et~al\mbox{.}(2023)]%
        {wang2023metrics}
\bibfield{author}{\bibinfo{person}{Tony Wang}, \bibinfo{person}{Haard~K Shah}, \bibinfo{person}{Raj~Sanjay Shah}, \bibinfo{person}{Yi-Chia Wang}, \bibinfo{person}{Robert~E Kraut}, {and} \bibinfo{person}{Diyi Yang}.} \bibinfo{year}{2023}\natexlab{}.
\newblock \showarticletitle{Metrics for peer counseling: triangulating success outcomes for online therapy platforms}. In \bibinfo{booktitle}{\emph{Proceedings of the 2023 CHI Conference on Human Factors in Computing Systems}}. \bibinfo{pages}{1--17}.
\newblock


\bibitem[Weizenbaum(1966)]%
        {weizenbaum1966eliza}
\bibfield{author}{\bibinfo{person}{Joseph Weizenbaum}.} \bibinfo{year}{1966}\natexlab{}.
\newblock \showarticletitle{ELIZA—a computer program for the study of natural language communication between man and machine}.
\newblock \bibinfo{journal}{\emph{Commun. ACM}} \bibinfo{volume}{9}, \bibinfo{number}{1} (\bibinfo{year}{1966}), \bibinfo{pages}{36--45}.
\newblock


\bibitem[Wells(2000)]%
        {wells2000dialogic}
\bibfield{author}{\bibinfo{person}{Gordon Wells}.} \bibinfo{year}{2000}\natexlab{}.
\newblock \showarticletitle{Dialogic inquiry in education}.
\newblock \bibinfo{journal}{\emph{Vygotskian perspectives on literacy research}} (\bibinfo{year}{2000}), \bibinfo{pages}{51--85}.
\newblock


\bibitem[Williams et~al\mbox{.}(2018)]%
        {williams2018recovery}
\bibfield{author}{\bibinfo{person}{Anne Williams}, \bibinfo{person}{Ellie Fossey}, \bibinfo{person}{John Farhall}, \bibinfo{person}{Fiona Foley}, \bibinfo{person}{Neil Thomas}, {et~al\mbox{.}}} \bibinfo{year}{2018}\natexlab{}.
\newblock \showarticletitle{Recovery after psychosis: qualitative study of service user experiences of lived experience videos on a recovery-oriented website}.
\newblock \bibinfo{journal}{\emph{JMIR Mental Health}} \bibinfo{volume}{5}, \bibinfo{number}{2} (\bibinfo{year}{2018}), \bibinfo{pages}{e9934}.
\newblock


\bibitem[Williams and Moser(2019)]%
        {williams2019art}
\bibfield{author}{\bibinfo{person}{Michael Williams} {and} \bibinfo{person}{Tami Moser}.} \bibinfo{year}{2019}\natexlab{}.
\newblock \showarticletitle{The art of coding and thematic exploration in qualitative research}.
\newblock \bibinfo{journal}{\emph{International management review}} \bibinfo{volume}{15}, \bibinfo{number}{1} (\bibinfo{year}{2019}), \bibinfo{pages}{45--55}.
\newblock


\end{thebibliography}
